\documentclass[fleqn,usenatbib]{mnras}

\usepackage[T1]{fontenc}

\DeclareRobustCommand{\VAN}[3]{#2}
\let\VANthebibliography\thebibliography
\def\thebibliography{\DeclareRobustCommand{\VAN}[3]{##3}\VANthebibliography}

\usepackage{graphicx}	% Including figure files
\usepackage{amsmath}	% Advanced maths commands
\usepackage{amssymb}	% Extra maths symbols

\newcommand{\kms}{\,km\,s$^{-1}$} % kilometres per second
\newcommand{\kmspc}{\,km\,s$^{-1}$\,kpc$^{-1}$} % kilometres per second
\newcommand{\Ha}{H$\alpha$\,} % kilometres per second

\usepackage[dvipsnames]{xcolor} %% texto de colores
\usepackage{color,soul} %to highlight
\usepackage{rotating}
\usepackage{longtable}

\usepackage{svg}

\usepackage{lscape}
\usepackage{caption}        
\usepackage{float}
\usepackage[none]{hyphenat}
\usepackage{booktabs}

\defcitealias{sardaneta-2024}{Paper~I}

\usepackage{marvosym}

\usepackage{newtxtext,newtxmath}
\title[H$\alpha$ kinematics of isolated edge-on galaxies ] 
{Extraplanar emission in isolated edge-on late-type galaxies. \\ 
II. The H$\alpha$ kinematics.
}
\author[M. M. Sardaneta et al.]{
Minerva M. Sardaneta,$^{1, 2}$\thanks{E-mail: mminerva@astro.unam.mx}
Philippe Amram,$^{3}$
Roberto Rampazzo,$^{4}$
Margarita Rosado,$^{1}$
\newauthor
Isaura Fuentes-Carrera$^{5}$
and Soumavo Ghosh$^{6}$
\\
$^{1}$Universidad Nacional Aut\'onoma de M\'exico. Instituto de Astronom\'ia. A.P. 70-264, 04510 Ciudad de M\'exico, M\'exico \\
$^{2}$Departamento de Astronom\'ia, Universidad de La Serena, Avda. Ra\'ul Bitr\'an 1305, La Serena, Chile\\
$^{3}$Aix Marseille Univ, CNRS, CNES, LAM, Marseille, France\\
$^{4}$INAF-Osservatorio Astrofisico di Asiago,Via dell’Osservatorio 8, 36012 Asiago, Italy\\
$^{5}$Escuela Superior de F\'isica y Matem\'aticas, Instituto Polit\'ecnico Nacional, U.P. Adolfo L\'opez Mateos, C.P.07738, Ciudad de M\'exico, M\'exico\\
$^{6}$Department of Astronomy, Astrophysics and Space Engineering, Indian Institute of Technology Indore, India - 453552
}

\date{Accepted 2025 October 20. Received 2025 October 13; in original form 2024 December 3}
\pubyear{2015}

\begin{document}
\label{firstpage}
\pagerange{\pageref{firstpage}--\pageref{lastpage}}
\maketitle

\begin{abstract}
Isolated galaxies are rare yet invaluable for studying secular evolution, as their physical properties can remain largely unaffected by external influences for several billion years, primarily shaped by internal evolutionary processes. This study focuses on a representative sample of nearly edge-on ($i\geq80^{\circ}$) late-type galaxies selected from the Catalogue of Isolated Galaxies (CIG). We analyse the \Ha kinematics derived from Fabry-Perot data and integrate these findings with a comprehensive examination of the UV, optical, and FIR properties of these galaxies to study their dynamic evolutionary processes. We investigate the individual kinematics by computing rotation curves and dynamical masses for each galaxy in the sample. The accuracy of our kinematic results is confirmed through comparisons with \ion{H}{i} data and by applying the B- and K-band Tully-Fisher relationships. Among the galaxies studied, we observed a rotational lag along the $z$-axis in half of the total sample (7 out of 14 cases), with an average lag ($\Delta V/\Delta z=32.0\pm10.6$~\kmspc) consistent with previous research findings. 
Notably, not all galaxies exhibiting measurable lag display morphological extraplanar components, leading us to conclude that this cannot serve as a definitive marker for the extraplanar Diffuse Ionized Gas (eDIG) component. While we found no significant correlations between rotation lag and overall galaxy properties, there seems to be a potential correlation with tidal strength. 
Based on the kinematic characteristics observed in our sample, we suggest that that the extended disc gas likely originates from interactions with the Circumgalactic Medium (CGM) rather than arising internally within the galaxies themselves. 

\end{abstract}

\begin{keywords}
galaxies: kinematics and dynamics -- 
galaxies: haloes -- 
galaxies: evolution -- 
galaxies: ISM 
\end{keywords}

%%%%%%%%%%%%%%%%%%%%%%%%%%%%%%%%%%%%%%%%%%%%%%%%%%

%%%%%%%%%%%%%%%%% BODY OF PAPER %%%%%%%%%%%%%%%%%%

\section{Introduction}

In order to understand the effects of the environment on fundamental galactic properties it is necessary to test models of the formation and evolution of galaxies  down to representative samples residing in low density environments \citep[e.g.][]{Rampazzo2020}.  At the extreme of these poor galaxy environments are placed the so called isolated galaxies. Several techniques have been adopted to identify such galaxies \citep[e.g.][]{Karachentseva-1973, verley-2007, hernandez-toledo-2010,  Argudo-F-2013}. Their high degree of isolation  ensures that such objects have not been appreciably affected by their nearest neighbours in the last 1-3 Gyr \citep{Karachentseva-2009}, i.e. that for a non-negligible part of their existence their physical properties, as e.g star formation, have been driven by internal evolutionary processes.

In \cite{sardaneta-2024} (hereafter  \citetalias{sardaneta-2024}),  we presented a sample of 14 high inclined ($i \geq 80^{\circ}$) late-type galaxies selected from the  \textit{Catalogue of Isolated Galaxies} \citep[CIG,][]{Karachentseva-1973}, we studied their ionized gas component, mapped by monochromatic H$\alpha$  emission, versus the old and young stellar populations, traced by near-infrared  (\textit{NIR}) and Ultraviolet (\textit{UV}) imaging. Investigating their  vertical structure, we found that 11 out of 14  isolated galaxies present the extraplanar diffuse ionized gas (eDIG) component.  Comparing our sample with general samples of highly inclined galaxies, the \citetalias{sardaneta-2024} study showed that it is unlikely that the eDIG component may be generated by environmental influence. This, however, must be confirmed using the most powerful tool to study the  perturbation of the gaseous components and, consequently, possible of extraplanar effects  \citep[e.g.][]{veilleux-2005}, i.e. kinematic signatures.

Edge-on ($i\simeq 90^{\circ}$) systems provide an optimal perspective for studying the vertical structure of  galactic discs, as they allow us to  distinguish different components along the vertical axis. In these galaxies, the disc and halo appear separated in projection, making it possible to study gas kinematics at various heights above and below the midplane \citep[e.g.][]{dettmar-1993, veilleux-1999}. Although motions perpendicular to the disc cannot be directly measured, their effects can be inferred from the observed gas kinematics at different vertical distances from the galactic mid-plane, providing insight into the vertical kinematics \citep[e.g.][]{burstein-1979, fraternali-2006, kamphuis-2007, rosado-2013, peters-2017}. A similar approach is also effective for the highly-inclined galaxies.

Late-type spiral galaxies having relatively high star formation rates per unit area  \citep[$L_{\rm FIR}/D_{25}^{2}\geq 3.2\times10^{40}\,\,\,\mathrm{erg\,s^{-1}\,kpc^{-2}}$,][]{rossa-2003-ii}, 
present a major interstellar medium (ISM) phase formed by layers of extraplanar Diffuse Ionized Gas (eDIG) reaching several kiloparsecs above/below the disc \citep[e.g.][]{rossa-2003-i}. In our Galaxy, this Warm ionized Medium (WIM) is also called the \textit{Reynolds layer} \citep[following its discovery by][]{reynolds-1984}, consisting of gas with a scale height of $\sim1\,\,\,\mathrm{kpc}$, $T\sim 10 ^{4}$~K and $n_{e}\sim 0.1\,\mathrm{cm^{-3}}$ \citep[][]{reynolds-1993, rand-1996, rossa-2003-i, heald-2006}. In general, the eDIG is a major component of the galactic ISM  which can provide insights into the different physical processes affecting the gas in galaxies. However, there has been no agreement so far on the nature of the ionization sources that could explain the eDIG behaviour observed in late-type galaxies \citep[e.g][]{FloresFajardo-2011, tomicic-2021, rautio-2022}. Some studies have found regions mixed OB-shock and hot low-mass evolved stars (also known by the acronym HOLMES, for HOt Low-Mass Evolved Stars) as the primary driver of eDIG ionization \citep[e.g.][]{FloresFajardo-2011, jones-manga-2017, levy-2019, rautio-2022, gonzalezDiaz-2024-betis-ii}. HOLMES, which include post-asymptotic giant branch (also known by the acronym PAGB, for Post Asymptotic Giant Branch) stars and white dwarfs, are found abundantly in the thick discs and lower haloes of galaxies \citep[e.g][]{marino-2011-mnras}. Their considerable vertical distribution compared to OB stars make them suitable candidates to be an eDIG ionization source \citep[e.g.][]{FloresFajardo-2011, jones-manga-2017, levy-2019, rautio-2022}. Nevertheless, although the PAGB scenario has been deeply studied, recent integral field spectroscopy studies have agreed that the eDIG may rather be a result of  the Circum-galactic Medium (CGM) accretion \citep[][]{marasco-2019, levy-2019, tomicic-2021, bizyaev-2022}.

It was observed  that the extraplanar component presents kinematic properties different from those of the ionized gas bound to massive stars and distributed  in the thin galactic disc \citep[e.g][]{kamphuis-2007, rosado-2013, Zschaechner-2015-a, bizyaev-2017, levy-2019}. The kinematic information of edge-on galaxies allows to distinguish the inner-disc from the outer-disc gas \citep{rand-1998, heald-2006, heald-2007, bizyaev-2017, levy-2019}. Studies of the eDIG kinematics in both the neutral hydrogen component (radio) \citep[e.g.][]{swaters-1997, fraternali-2005, lee-2001, Zschaechner-2015-a, marasco-2019, li-fraternali-marasco-2023} and the warm ionized component (optical) \citep[e.g.][]{rand-2000, miller-2003, heald-2006, rosado-2013, ho-2016, bizyaev-2017, tomicic-2021, sardaneta-2022, rautio-2022}, have shown that the eDIG is in rotation like the thin disc. Nevertheless, the halo gas rotates more slowly than the gas in the plane, this vertical velocity gradient has been referred to as the \textit{rotation lag} of the disc.

A full analysis of how gaseous haloes rotate requires high spectral and spatial resolutions and two  dimensional spatial coverage \citep[e.g.][]{veilleux-1999, heald-2006, rosado-2013} that long-slit spectroscopy does not allow \citep{miller-2003, fraternali-2004, FloresFajardo-2011}. 
In this sense, in the optical wavelength, studies on samples of edge-on galaxies using integral field unit (IFU) spectroscopy data from SAMI\footnote{Sydney-AAO Multi-object Integral field spectrograph (SAMI, spectral resolution $R_{\rm opt}\sim 1750-4500$).} \citep{ho-2016}, MaNGA\footnote{Mapping Nearby Galaxies at Apache Point (MaNGA, spectral resolution $R\sim 2000$).} \citep[][]{jones-manga-2017, bizyaev-2017}, CALIFA\footnote{Calar-Alto Legacy Integral Field Area (CALIFA, spectral resolution $R\sim 1650$).} \citep{levy-2018, levy-2019} and MUSE\footnote{Multi Unit Spectrograph Explorer (MUSE, spectral resolution $R\sim3000$).} 
\citep{tomicic-2021, gonzalezDiaz-2024-betis-i, gonzalezDiaz-2024-betis-ii} were performed with the aim of understanding the interplay between kinematic properties of the extraplanar gas and integral properties of their host galaxies, but without still having achieved a conclusive result, partially due to their lower spectral resolution (the best resolution achived by those instruments is the one of SAMI and is only worth $R_{\rm opt}\sim4\,500$), their smaller field of view (FoV), and their sample selection criteria  \citepalias[see][]{sardaneta-2024}.

Optical scanning Fabry-Perot (FP) interferometry enables integral field spectroscopy as well, providing complete two-dimensional coverage of extended regions emitting in the optical wavelength range \citep[e.g.][]{veilleux-1999, veilleux-1999-ii, rosado-2008, sardaneta-2020}.  The FP data are especially sensitive to the monochromatic \Ha emission, offering a large field of view and high spectral resolution \citep[$R\sim10,000$,][]{amram-1991}. The high-resolution FP data also allow detailed study of extended objects by tracing both circular and non-circular gas motions across entire galaxies \citep[e.g.][]{fuentesc-2004,fuentes-2007, epinat-2008-ii, rosado-2013, sardaneta-2022}. 
Our goal is to determine the origin of the galactic halo by studying physical processes that affect the gas in galaxies allowing us to understand which are the main sources of ionization of the eDIG. 
In this work, we study the kinematics of the extraplanar gas of the 14 late-type highly inclined isolated galaxies of our sample using FP data \citepalias[see][]{sardaneta-2024}. 
We further emphasize the fact that our sample of isolated galaxies may enable us to shed light on the origin, either internal or external, of the phenomenon.

This paper is structured as follows. In Section~\ref{Sec:Obs}, we describe the data acquisition and the data reduction process. In Section~\ref{Sec:Data-analysis}, we explain how the FP H$\alpha$ data  are used to determine the galactic rotation curves and the rotation lag. In Section~\ref{Sec:Results}  we show our results. In Section \ref{Sec:Disc} we discuss our results in the light of the literature, and in Section \ref{Sec:Conc} we  draw our conclusions.  Throughout this work we assumed a constant Hubble $H_{0}=70\,\,\,\mathrm{km\,s^{-1}\, Mpc^{-1}}$ \citep[e.g.][]{ho-2016, Jones2018, levy-2019}.

\begin{table*}
\caption{General parameters of the highly inclined isolated galaxy sample}
\begin{center}
\begin{tabular}{ccccccccccccc}
\hline
CIG & Other & RA$_{\rm phot}$ & Dec$_{\rm phot}$ & V$_{\rm sys}$ & Distance & $i_{\rm Opt}$ & m$_{\rm B}$ & $A_{v}$ & $K_{s}$  & r$_{\rm K_{3\sigma}}$ & PA$_{\rm K_{3\sigma}}$ & $M_{\rm * (MIR)}$ \\ 
 Name & Name & (hh mm ss) & ($^{\circ}$ ' ") &  (\kms)  & (Mpc) & (deg) &  (mag) &  (mag) &  (mag) & (arcsec)   & (deg)  &  ($10^{9}\,M_{\odot}$)  \\ 
(1) & (2) & (3) & (4) & (5) & (6) & (7) & (8) & (9) & (10) & (11) & (12) & (13)\\ 
\hline
71 & UGC 01391 & 01 55 15.9 & +10 00 50.1 & 5901 & 84.3 & 83.8 & 15.35 & 0.236 & 10.8 & 18.7 & 178.8 & 14.1 \\ 
95 & UGC 01733 & 02 15 20.7 & +22 00 21.9  & 4418 & 63.1 & 86.5 & 15.48 & 0.282 & 11.6 & 12.0 & 126.2 & 3.6 \\ 
159 & UGC 03326 & 05 39 36.7 & +77 18 44.8 & 4121 & 58.9 & 85.4 & 15.3 & 0.352 & 9.6 & 42.0 & 66.2 & 3.8 \\ 
171 & UGC 03474 & 06 32 37.7 & +71 33 38.2 & 3634 & 51.9 & 84.0 & 15.4 & 0.603 & 10.1 & 44.4 & 159.7 & 1.4 \\ 
183 & UGC 03791 & 07 18 31.8 & +27 09 28.7 & 5090 & 72.7 & 80.4 & 15.38 & 0.194 & 11.4 & 13.5 & 16.6 & 8.1 \\ 
201 & UGC 03979 & 07 44 30.9 & +67 16 25.6 & 4061 & 58.0 & 80.9 & 14.78 & 0.147 & 10.6 & 20.5 & 151.7 & 5.1 \\ 
329 & UGC 05010 & 09 24 55.1 & +26 46 29.0 & 4096 & 58.5 & 81.3 & 13.77 & 0.085 & 9.3 & 35.9 & 115.4 & 31.9 \\ 
416 & UGC 05642 & 10 25 41.6 & +11 44 21.4 & 2322 & 33.2 & 81.1 & 14.68 & 0.091 & 11.6 & 11.9 & 95.0 & 2.6 \\ 
593 & UGC 08598 & 13 36 40.7 & +20 12 00.1 & 4909 & 70.1 & 83.2 & 14.94 & 0.054 & 10.8 & 14.1 & 165.5 & 12.1 \\ 
847 & UGC 11132 & 18 09 26.2 & +38 47 41.0 & 2837 & 40.5 & 81.2 & 15.38 & 0.095 & 10.7 & 20.5 & 140.2 & 11.9 \\ 
906 & UGC 11723 & 21 20 17.5 & -01 41 03.3 & 4899 & 70.0 & 80.9 & 14.7 & 0.131 & 10.0 & 31.3 & 32.8 & 5.1 \\ 
922 & UGC 11785 & 21 39 26.7 & +02 49 37.6 & 4074 & 58.2 & 84.2 & 15.18 & 0.231 & 11.0 & 19.6 & 60.0 & 4.3 \\ 
936 & UGC 11859 & 21 58 07.4 & +01 00 32.0 & 3011 & 43.0 & 85.7 & 15.16 & 0.147 & 11.4 & 12.3 & 61.6 & 1.0 \\ 
1003 & UGC 12304 & 23 01 08.3 & +05 39 14.3 & 3470 & 49.6 & 82.5 & 15.11 & 0.174 & 10.3 & 25.7 & 124.7 & 6.0 \\ 
\hline
\end{tabular}
\label{Tab:general}
\end{center}
Columns: 
(1) CIG galaxy name; (2) UGC galaxy name; 
(3) and (4) \textsc{WCS} coordinates (J2000) of the 2MASS $K_{s}$-band photometric centre  \citepalias[see][]{sardaneta-2024}; 
(5)~$V_{\rm sys}$: systemic velocity from NED; 
(6)~heliocentric distance to the galaxy;
(7)~$i_{\rm Opt}$:~inclination computed with the apparent optical major and minor axes from NED  assuming a flat disc \citepalias[see][]{sardaneta-2024}; 
(8)~apparent B-band magnitude from NED; 
(9)~the total absorption in magnitudes at V  from NED; 
(10)~apparent \textit{K$_{s}$}-band magnitude from NED; 
(11)~ position angle~(PA) of the isophotal level at 3$\sigma$ of the 2MASS $K_{s}$-band image  \citepalias[see][]{sardaneta-2024};  
(12)~stellar mass ($M_{*}$) computed using Mid-Infrared (MIR) photometric~data \citepalias[see][]{sardaneta-2024}. 
\end{table*}

\begin{figure*}
\centering

\includegraphics[width=0.24\hsize]{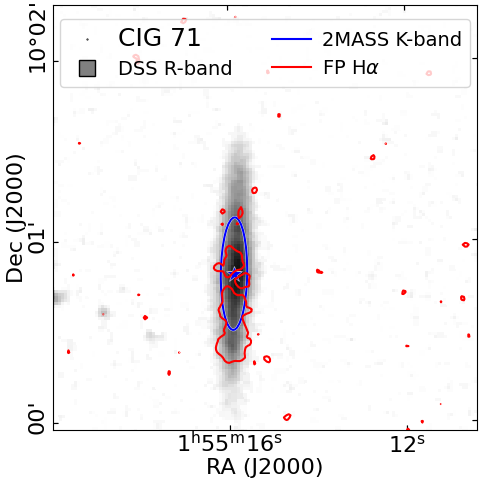}
\includegraphics[width=0.24\hsize]{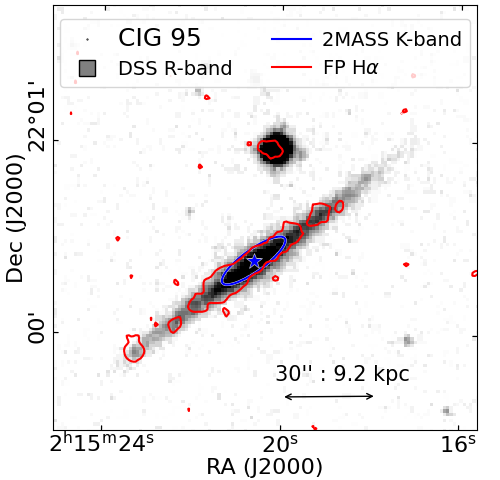}
\includegraphics[width=0.24\hsize]{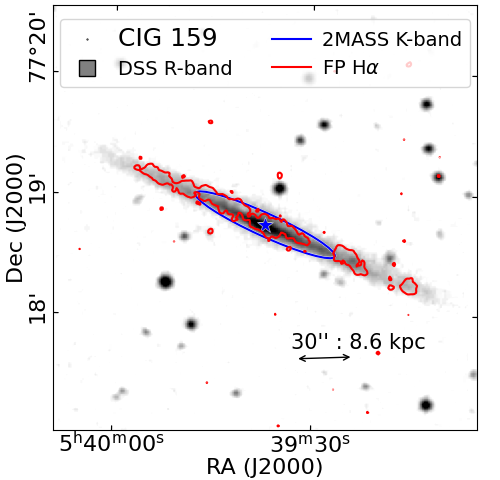}
\includegraphics[width=0.24\hsize]{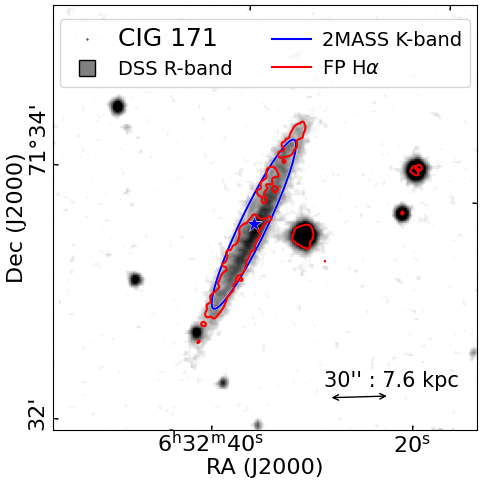}

\includegraphics[width=0.24\hsize]{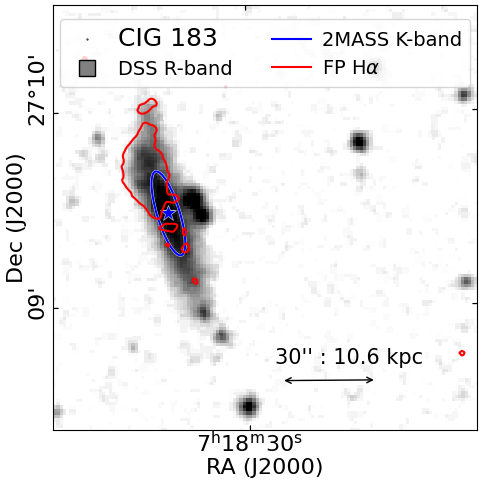}
\includegraphics[width=0.24\hsize]{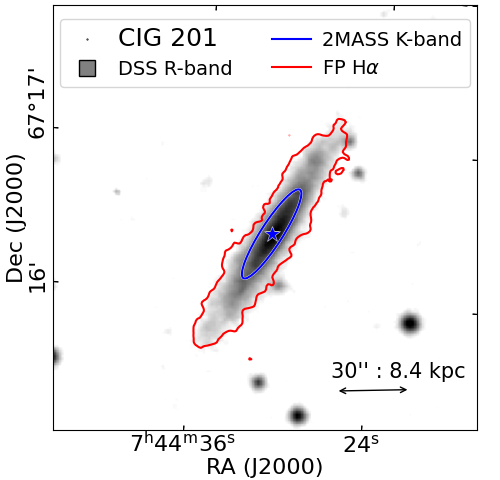}
\includegraphics[width=0.24\hsize]{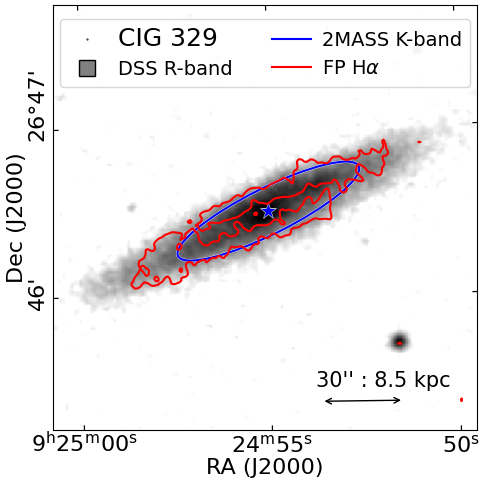}
\includegraphics[width=0.24\hsize]{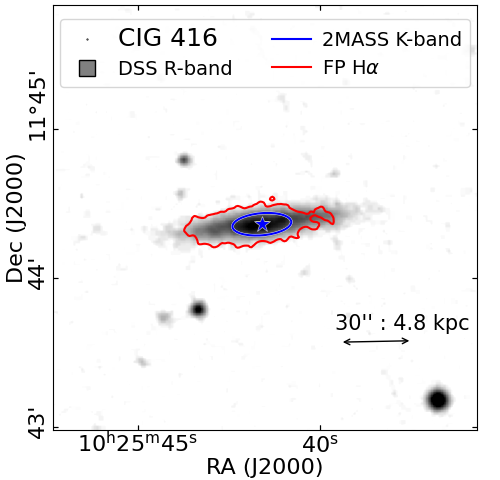}

\includegraphics[width=0.24\hsize]{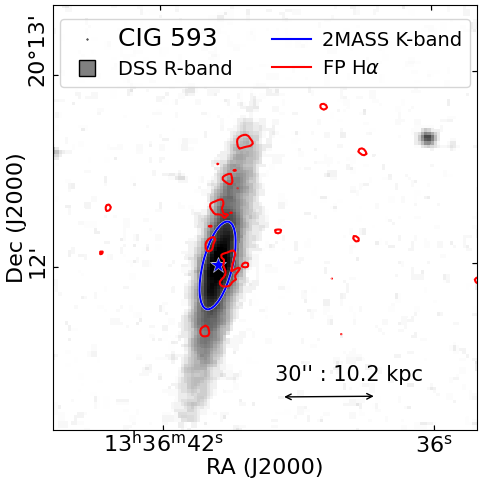}
\includegraphics[width=0.24\hsize]{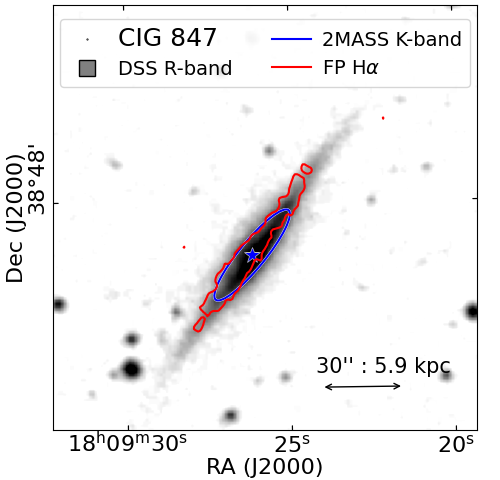}
\includegraphics[width=0.24\hsize]{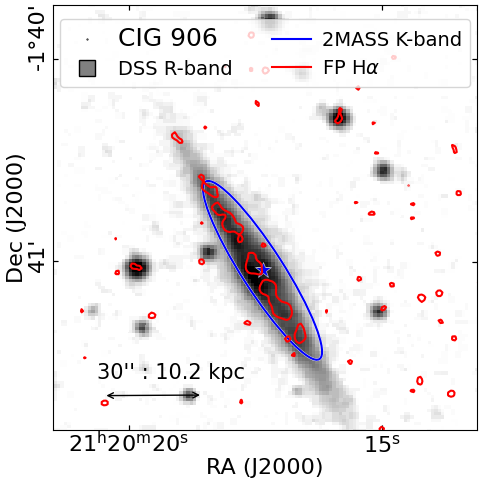}
\includegraphics[width=0.24\hsize]{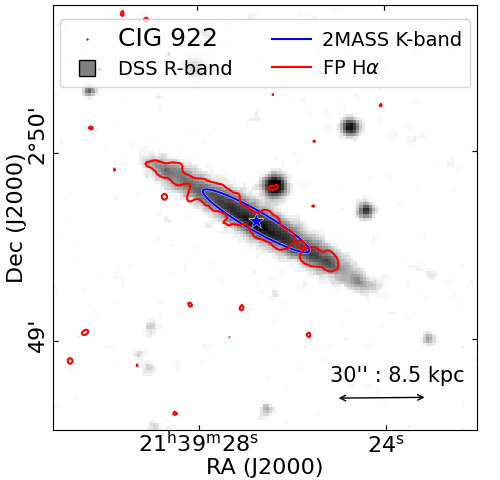}

\includegraphics[width=0.24\hsize]{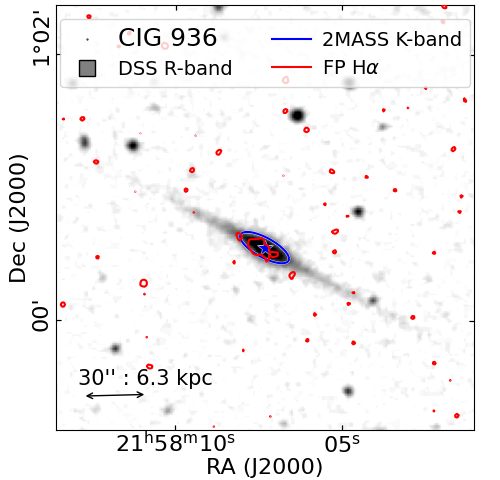}
\includegraphics[width=0.24\hsize]{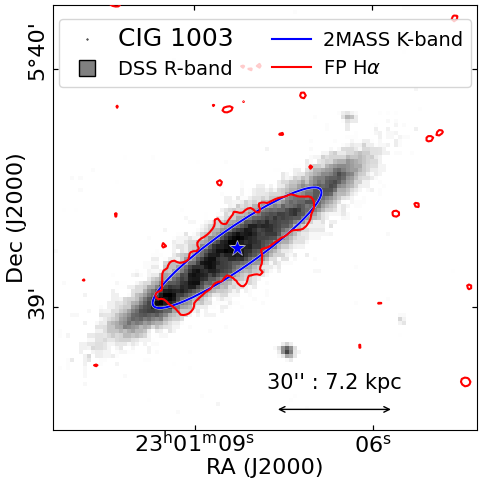}

\caption{
The R-band map from the Digitized Sky Survey (DSS) of galaxies in our sample. The ellipse tracing the $3\sigma$ surface brightness of the K-band image is overlaid on the maps to represent the stellar disc formed by the old stellar population. 
The contour of the lowest \Ha emission level corresponding to a signal-to-noise ratio SNR$\geq$5 is shown, illustrating the distribution of ionized gas relative to the old stellar disc and the optical emission.
}
\label{fig:R-dss}
\end{figure*}

\section{Sample, observations and data reduction}\label{Sec:Obs}

We aim to study the environmental influence on the occurrence of the ISM extraplanar component.  Assuming that over some billion years dynamically isolated galaxies have evolved by purely internal reasons \citep[e.g.][]{verdesm-2005, Karachentsev-2011, rampazzo-2016,rampazzo-2021}, we have selected a complete sample of 14 galaxies from the Catalogue of Isolated Galaxies \citep[CIG,][]{Karachentseva-1973}. Galaxies in our sample satisfy the following characteristics: (1)~late-type galaxies (S) influenced by the visually luminous Population~I stars; (2)~galaxies with high inclination ($i \geq80^{\circ}$); (3)~nearby galaxies, with $z\leq0.02$, to ensure that the emission of the H$\alpha$ line redshifted fits in the optical wavelength; and (4)~galaxies with apparent K-band magnitude $K_{s}\leq 12$~mag, because the near-infrared (NIR) is sensitive to nuclear rings and large-scale bars, which might be a critical component to fuelling active nuclei \citep[see e.g.][]{kormendy-1982, Eskridge-2002, jarrett-2003}. 
Spectral information is available for only 3 out of the 14 galaxies in our sample, identifying two as star-forming (CIG~416 and 936) and one as a LINER (CIG~329) \citep[see][]{sabater-2012}. 
The selection criteria and general properties of the sample, as well as the morphology of each individual galaxy, were described extensively in  \citetalias{sardaneta-2024}. However, for the sake of synthesis and clarity, we present the general parameters of galaxies in  our sample in Table~\ref{Tab:general}.

Observations of the H$\alpha$ emission line were performed using the scanning Fabry-Perot (FP) interferometer, GHASP, attached at the Cassegrain focus of the 1.93~m telescope at the \textit{Observatoire de Haute-Provence} (OHP). All the details related with the telescope and the instrument are described in  \citetalias{sardaneta-2024}. The general data reduction process is  widely reported in \cite{daigle-2006-fp}. As summarized in \citetalias{sardaneta-2024}, the general data reduction process was performed with the packages \textsc{reducWizard} 
%\footnote{\url{http://www.astro.umontreal.ca/fantomm/reduction/}} 
and \textsc{ComputeEverything} which are based on  \textsc{IDL}\footnote{Interactive Data Language, ITT Visual  Information Solutions: \url{https://www.l3harrisgeospatial.com/Software-Technology/IDL}} routines.
In this paper, the instrumental and observational parameters are listed in Table~\ref{Tab:obs}.

\begin{table}
\caption{Instrumental and observational parameters}
\begin{center}
\begin{tabular}{cc}
\hline
Parameter  & Value \\ 
\hline
Telescope & 1.93~m OHP \\ 
Aperture ratio of the focal reducer & $f$/3.9 \\ 
Instrument & GHASP \\ 
Detector type & IPCS GaAs system \\ 
Detector size  & 512$\times$512 pix$^{2}$ \\ 
Image scale  & 0.68 arcsec pix$^{-1}$ \\ 
Field of view  & 5.9$\times$5.9 arcmin$^{2}$  \\ 
Interference order at \Ha & 798 \\ 
FSR at \Ha  & 8.23~\AA\, / 376 \kms \\ 
Finesse observed & $\sim$13 \\ 
Resolution & $\sim$10000 \\ 
Spectral sampling at \Ha  & $\sim$0.26~\AA\,  / $\sim$11.5  \kms  \\ 
Average seeing  & $\sim$2.9 arcsec \\ 
Exposure time  & $\sim$180 – 220 s \\ 
Calibration line  & $6598.95$~\AA\, (\ion{Ne}{} lamp)\\
\hline
\end{tabular}
\end{center}
\label{Tab:obs}
\end{table}

After a general data reduction process \citepalias[described in][]{sardaneta-2024}, the integrated data cubes were calibrated in wavelength, creating for each target a wavelength-sorted data cube by applying the phase map correction to the interferogram data cube. 
In \citetalias{sardaneta-2024}, we also explained  that a Voronoi tessellation was applied to the FP data cubes to guarantee a signal-to-noise ratio SNR$\geq$5 in the \Ha maps. Adaptive binning improves the SNR but may create large bins dominated by sky emission. To distinguish galaxy bins from sky bins, a velocity continuity criterion is applied, removing those with low flux or excessive size. A cut-off value, set to roughly one-tenth of the total velocity amplitude, helps distinguish genuine galactic emission from background noise. This  automatic cleaning process is thoroughly explained in \cite{epinat-2008-i}. 
Then, for each wavelength calibrated cube, a monochromatic, continuum, radial velocity and velocity dispersion map was computed. 
For each individual pixel , the momenta of the emission line (intensity, barycentre and width) were derived directly from the observed profiles using a selective intensity weighted mean algorithm extensively described in \cite{daigle-2006-fp}. The velocity was computed as the intensity-weighted centroid of the spectral bins within the emission line boundaries, with the continuum subtracted.

The free spectra range (FSR) determines the depth of the data cube and is a function of the observed wavelength ($\lambda$) and the interference order ($p$) at this wavelength: FSR$=\lambda/p$. The number of channels to be scanned within the FSR is linked to the finesse of the Airy function of the etalon (in fact, the number of channels respects the Shannon-Nyquist criteria: $n\geq 2 \times$\,the\,effective\, finesse) and has no relation to the FSR. 
In this work, we have used  a FSR of $\sim$8~\AA\, ($\sim$375~$\mathrm{km\, s^{-1}}$ at H$\alpha$) to cover most of the galaxy light-of-sight velocity gradients. 
If the radial velocity range of a galaxy, taking into account its inclination factor, exceeds the FSR, parts of the velocity field may appear shifted to the adjacent interference orders, $p-1$ and/or $p+1$, instead of the central order $p$. Filters placed in front of the etalon typically isolate the primary interference orders $p-1$, $p$, and $p+1$, blocking all others. Consequently, when abrupt transitions in the velocity field around the FSR are detected, these are corrected by adding or subtracting the FSR value as needed to restore continuity \citep[see e.g.][]{benoit-2008}.

Because of the radial velocity and velocity dispersion maps were extracted with a single emission-line detection algorithm, when two emission lines are spectrally close and have comparable amplitudes, they may be taken as a single one having a larger velocity dispersion \citep{daigle-2006-i}.  
Although a Gaussian fitting is not used to determine the velocity or velocity dispersion maps, we assume a Gaussian distribution when correcting the velocity dispersion and apply the standard correction:
$\sigma^{2}=\sigma^{2}_{\rm obs}-\sigma^{2}_{\rm inst}-\sigma^{2}_{\rm th}$,
where 
$\sigma_{\rm obs}$ is the velocity dispersion obtained from the data cube, $\sigma_{\rm inst}=\mathrm{FSR}/(\mathcal{F}\times 2\sqrt{2\mathrm{ln} 2})\simeq12.5\,\mathrm{km\,s^{-1}}$ is the correction to the emission-profile width due to the instrument in function of the FSR of the instrument and the finesse observed ($\mathcal{F}$) (see Table~\ref{Tab:obs}), 
$\sigma_{\rm th}=(kT_{e}/m_{H})^{1/2}=9.1\,\mathrm{km\,s^{-1}}$ is the correction for the thermal width, with electronic temperature of a \ion{H}{ii} region of $T_{e}=10^{4}\,\mathrm{K}$ \citep{osterbrock-2006}. 
Since turbulence is not spatially uniform across a galaxy \citep[e.g.][]{relano-beckman-2005, marasco-2019}, subtracting a uniform turbulent component ($\sigma_{\rm turb}\simeq25$~\kms, typical value for \ion{H}{ii} regions) from the entire velocity dispersion map would lead to unphysical values (e.g., null or negative dispersions) in several regions. Therefore, we did not apply a global correction to the dispersion map. Instead, turbulent broadening will be considered locally in regions of interest when analysing the kinematics in detail in a further work.

The astrometric information was attached to the processed files by using the \textsc{IDL} task \textsc{koords} from the \textsc{KARMA} package \citep{gooch-1996}  and, the data analysis was made with \textsc{ADHOCw}\footnote{`Analyse et Depouillement Homogene des
Observations Cigale for Windows'
\url{http://cesam.lam.fr/fabryperot/index/softwares} developed by J.
Boulesteix at Marseille Observatory in 2005.} program,
\textsc{IRAF}\footnote{`Image Reduction and Analysis Facility'
\url{http://iraf.noao.edu/}} tasks, 
the SAO Image DS9 software\footnote{SAO Image DS9. An image display and visualization tool for astronomical data \citep{ds9}  \url{https://sites.google.com/cfa.harvard.edu/saoimageds9}} 
and our own \textsc{Python} scripts.

In order to distinguish the extraplanar ionized gas in the galaxies of our sample, we defined the old stellar disc using the 3$\sigma$ surface brightness ellipse of the 2MASS K-band image. 
Once the \Ha emitting gas of each galaxy was identified, any gas outside this ellipse was considered to be extraplanar ionized gas (see \citetalias{sardaneta-2024}). 
In this work, Figure~\ref{fig:R-dss} shows the contour of the lowest \Ha emission level corresponding to a SNR$\geq$5 overlaid on the R-band map from the Digitized Sky Survey (DSS), alongside the K-band ellipse, outlining the spatial relationship between the ionized gas and the stellar disc.

\section{Data analysis}\label{Sec:Data-analysis}

\subsection{Position-velocity diagrams along the major-axis}\label{Sec:pvdmajor}

\begin{figure*}
\centering
\includegraphics[width=\hsize]{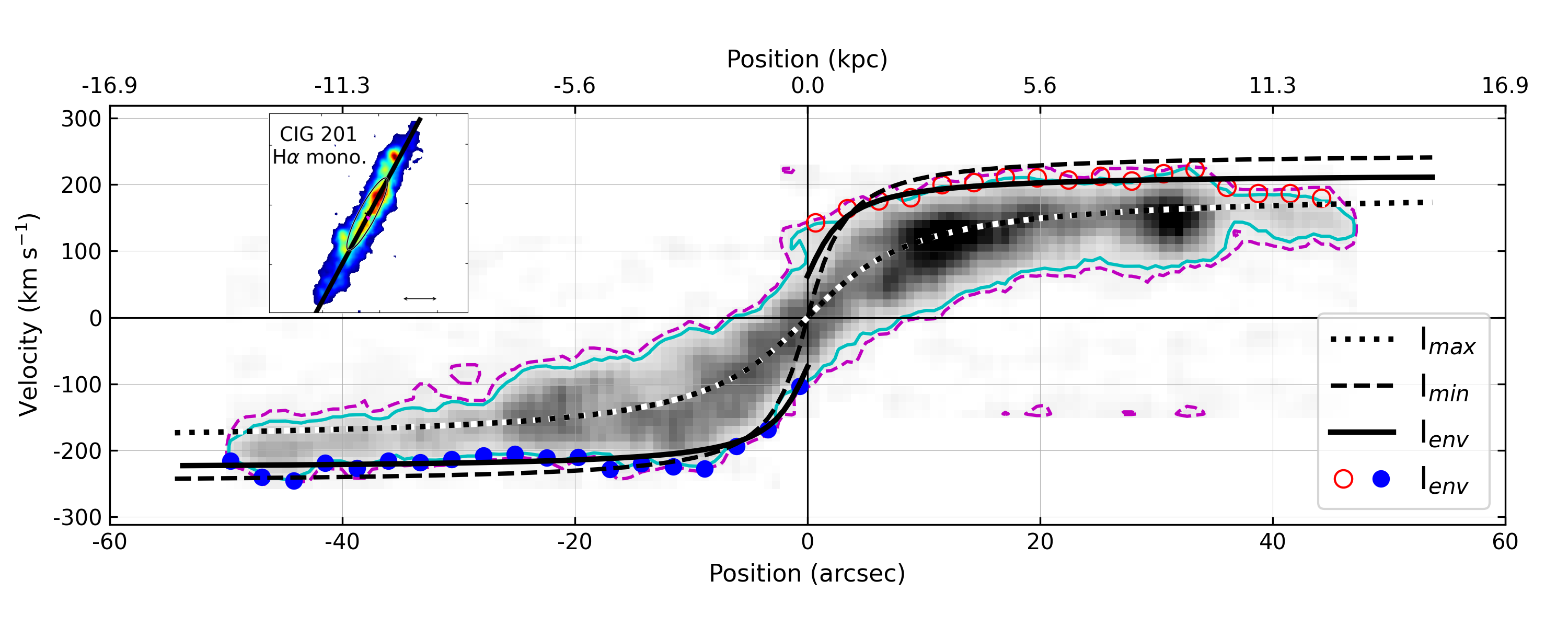}
\caption{
Position-velocity diagram (PVD) along the kinematic major axis ($z=0$) of the galaxy CIG~201, derived from the Voronoi-tessellated H$\alpha$ data cube including only emission with SNR$\geq5\sigma$. 
The ET method is applied to extract the rotation curve from the PVD. 
The maximum intensity ($I_{\rm max}$) is traced by a dotted line crossing the centre of the PVD and approximated by an $\arctan$ function. 
The minimum intensity ($I_{\rm min}$), corresponding to the $5\sigma$ emission contour, is traced by the dashed line and also fitted with an $\arctan$ function. 
The envelope intensity ($I_{\rm env}$), computed using the relation~\ref{equation:I_env} with $\eta=0.3$, is shown by the filled blue circles (approaching side) and empty red circles (receding side). The solid black line represents the best $\arctan$ fit to these $I_{\rm env}$ points. Contours corresponding to $I_{\rm min}$ (purple dashed line contour) and $I_{\rm env}$ (cyan solid line contour) are superimposed onto the PVD. The upper-left inset displays the location of the pseudo-slit overlaid on the H$\alpha$ monochromatic map of CIG~201.
}
\label{Fig:envelope}
\end{figure*}

In order to analyse the velocity distribution through the emission intensity, we built the Position-Velocity  Diagrams (PVD) from each  wavelength-sorted data cube after having extracted the continuum \citep[e.g.][]{benoit-2008, rosado-2013, sardaneta-2022}. First,  the data cube was rotated so that the position angle (PA) of the galaxy  aligns with the X-axis of the PVD. We defined a virtual slit with a width of two  pixels ($\sim$1.4~arcsec) along the galactic major axis, which was used to extract the \Ha emission intensity. In particular, only for CIG\,593 (Figure~\ref{Fig:c593}), a wider virtual slit of 5~pixels ($\sim$3.4~arcsec) was defined because its  \Ha-emitting gas is limited. Then, using the virtual slit centreed along the galactic major axis, we  extracted the intensity of the \Ha emission from the gas, projected onto the XY plane of the data cube (see e.g. Figure~\ref{Fig:envelope}).

The kinematic parameters of a galaxy, such as its centre, position angle (PA${\rm kin}$), and systemic velocity ($V{\rm sys}$), significantly influence the resulting PVDs. While variations in the position angle directly alter the shape of the PVD, changes in the galaxy’s centre and systemic velocity only shift the distribution horizontally or vertically without affecting its intrinsic shape, potentially introducing asymmetries and complicating the interpretation of gas kinematics (see Appendix~\ref{Appendix:pvd_symmetry}).

 Variations in the position angle directly alter the shape of the PVD, while changes in the galaxy's centre and systemic velocity shift the PVD horizontally or vertically without modifying its intrinsic shape. Such variations in any of these parameters can introduce asymmetries in the PVDs, increasing the difficulty of interpreting the gas motions within the galaxy (see Appendix~\ref{Appendix:pvd_symmetry}).

\begin{figure*}
\centering
\includegraphics[width=0.96\hsize]{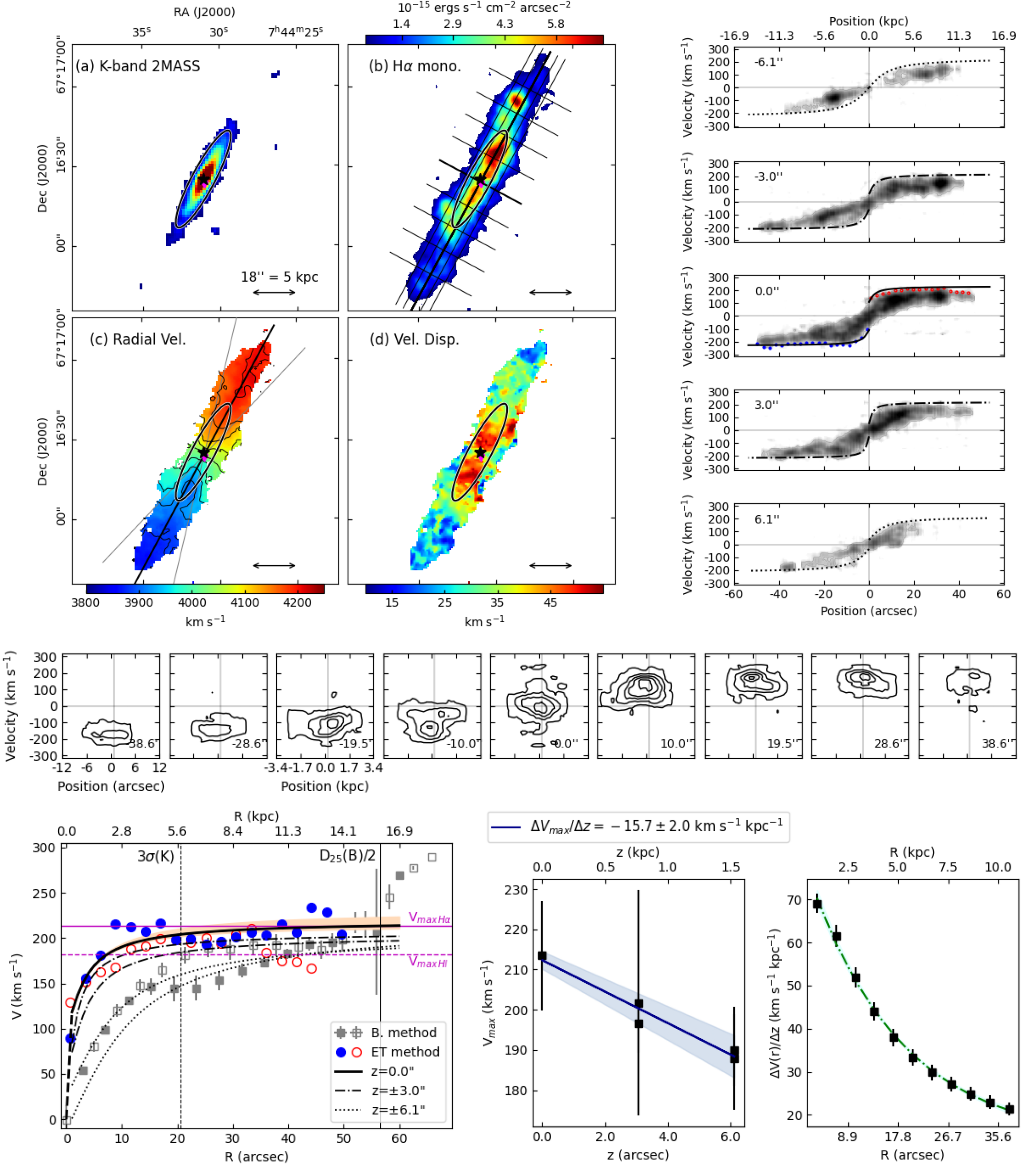}
\caption{
CIG~201~(UGC~3979). 
Example of the layout of the graphs and maps used for the kinematic analysis of each galaxy. 
Top left: 
(a)~2MASS~\textit{K$_s$}-band image, 
(b)~\Ha monochromatic map, 
(c)~radial velocity map and, 
(d)~velocity dispersion map,
with the ellipse fitted to the 3$\sigma$ surface brightness level of the 2MASS~\textit{K$_s$}-band image and the photometric (star, $\star$) and kinematic (diamond, $\diamond$) centres overlaid. 
Top right: PVDs along the major axis at various heights from the galaxy's kinematic major axis. 
The major-axis PVD shows the $I_{env}$ derived using $\eta = 0.3$, represented by filled blue circles for the receding side and empty red circles for the approaching side, these points were used to fit the $\arctan$ function (solid line). 
The $I_{env}$ using $\eta = 0.3$ at different heights from the kinematic major axis are plotted: dash-dotted lines correspond to heights of $\pm 3.0$ arcsec and dotted lines to heights of $\pm 6.1$ arcsec. 
Middle: PVDs along the minor axis at increasing distances from the kinematic center.
The kinematic major axis (black line) and pseudo-slits (gray lines) are shown on the \Ha map.
Bottom left: rotation curves derived from the intensity peak (filled squares for approaching, empty squares for receding) and the ET method (blue filled circles for approaching, red empty circles for receding). 
For the ET method, the inclination-corrected $\arctan$ fitting to I$_{\rm env}$ (with $\eta=0.3$) for the major-axis PVD is shown as a solid line, while the shaded area corresponds to the range $\eta = [0.2, 0.5]$. To measure the lag in rotation, inclination-corrected dashed and dotted lines from the upper panels are also included. 
Bottom middle: OLS fit of the relation between maximum rotation velocities at different heights and the height ($z$), with the confidence region shaded.
Bottom right: Lag in rotation vs. galactic radius, the slope of this plot could indicate the origin of the extraplanar component (see the text).
}
\label{Fig:e.g.-c201}
\end{figure*}

The kinematic parameters of a galaxy, such as its centre, position angle (PA$_{\rm kin}$) and systemic velocity ($V_{\rm sys}$), significantly influence the resulting PVD. While variations in the PA$_{\rm kin}$ directly alter the shape of the PVD, changes in the galaxy’s centre and $V_{\rm sys}$ only shift the distribution horizontally or vertically without affecting its intrinsic shape. However, variations in any of these parameters can introduce asymmetries in the PVDs, increasing the difficulty of interpreting the gas kinematics within the galaxy (see Appendix~\ref{Appendix:pvd_symmetry}). 
%% PA %%
We define the PA$_{\rm kin}$ as the counterclockwise angle relative to North. Varying the PA results in shorter PVDs with asymmetries along the position axis ($x$-axis). We initially extracted the PVD using the photometric PA$_{\rm K_{3\sigma}}$ and subsequently refined it to achieve the most extended and intensity-symmetric PVD (see Tables~\ref{Tab:general} and~\ref{Tab:kin}). In general, the difference between the photometric and kinematic PA of the galaxies of our sample does not exceed $\pm1.2^{\circ}$, except for CIG~183, 593, and 936, where the morphological differences between the K-band and \Ha images are more significant \citepalias[see][]{sardaneta-2024}. 
%% V_sys %%
On the other hand, modifying the $V_{\rm sys}$ alters the symmetry of the PVD along the velocity axis ($y$-axis). 
Here, the $V_{\rm sys}$ that gave as a result the most symmetric PVD agreed the one  published in the NED (see Table~\ref{Tab:general}), although an uncertainty up to $\pm$10-20~\kms may occur due to the difference in wavelength between the calibration and the galaxy wavelengths \citep[see e.g.][]{amram-1996}. %, $\Delta\simeq35\,\AA$ 
%% centre %% 
Finally, for most galaxies, the kinematic centre agreed within $\sim$2~asec with the peak of the light distribution of the K-band image. 
%% fin %% 
Thus, we selected the PVDs with the highest symmetry along both axes, as these best reflect the most physically consistent kinematic parameters.  
We computed the PVDs using the virtual slit and the established PA$_{\rm kin}$, $V_{\rm sys}$, and kinematic centre for each galaxy, ensuring that the captured \Ha emission  accurately represents the velocity distribution along the major axis (see e.g. Figure~\ref{Fig:e.g.-c201}).

%
% %

%

\subsection{Rotation curves}\label{Sec:rcs}

Rotation curves (RCs) are one of the most basic sources of information on the dynamics of galaxies \citep{sofue-2001} used to discuss super massive black holes, the dark matter distribution in the halo, the characteristics of a bulge in a disc, bar structure, and kinematic peculiarities \citep[e.g.][]{takamiya-2002, moiseev-2014}. 
The classical method for determining the  RC from a velocity profile is called the intensity-peak method (IPM, see Appendix~\ref{Appendix:intensity-peak}). It consists of fitting the profile to a theoretical function like a Gaussian or, alternatively, in directly computing the barycentre of the line, the goal being in both cases to measure the wavelength of the intensity peak that provides the mean velocity of the line \citep{sofue-2001}. However, in high-inclined systems, each line-of-sight (LoS) contains much more information on the kinematics of the galaxy, since it encompasses a larger portion of the disc and/or halo. At each position we receive the integrated signal from various radii, so we have to disentangle that information to get the radial density profile. 
A diagram demonstrating this effect is provided in \citet[][see their figure~2]{garcia-ruiz-2002}. Due to these complexities, alternative methods have been developed to calculate the RC in highly inclined galaxies \citep[e.g.][]{sancisi-1979, sofue-1998, garcia-ruiz-2002, heald-2006, heald-2006-b, heald-2007, rosado-2013, sardaneta-2022}.

\subsubsection{The envelope-tracing method}

In the case of edge-on galaxies, the classical method fails because the shape of the profile is  strongly affected by the previously mentioned LoS effects, which tend to underestimate the true rotation velocities. As a result, corrections must be applied to the estimated velocities to accurately map the rotation and infer the gravitational potential. These effects arise because, at each galactocentric distance along the major axis of an edge-on spiral, the LoS encompasses several different clouds of  gas with different projected velocities, spanning continuously (if the disc is uniform) from full rotation (for gas at the line of nodes) to the systemic velocity (for gas orbiting perpendicular to the LoS) \citep[e.g.][]{sancisi-1979, swaters-1997, garcia-ruiz-2002, heald-2006, heald-2006-b, heald-2007, rosado-2013}.

In the envelope-tracing method (ETM) described by \cite{sofue-1998}, the rotation velocity is defined by:
% %
\begin{equation}
V_{\mathrm{rot}}\sin i=(V_{env}-V_{\rm sys}) - (\sigma_{T}^{2}+\sigma_{\mathrm{LSF}}^{2})^{1/2},
\label{sofue-ec3}
\end{equation}
where $V_{\rm sys}$ is the systemic velocity, $i$ is the galactic inclination respect to the sky plane,  $\sigma_{\mathrm{LSF}}$ is the instrumental velocity resolution, $\sigma_{T}$ is the thermal broadening of the gas and,  $V_{env}$ is the terminal velocity characterised by the intensity of the envelope of an observed PVD along the line profile \citep[see][]{sardaneta-2022}. 
The intensity of the envelope is defined by 
\begin{equation}
I_{env}^{2}=(\eta I_{max})^{2}+(I_{min})^{2},
\label{equation:I_env}
\end{equation}
where $I_{max}$ is the maximum intensity in the line profile, $I_{min}$ is the minimum value of the contour, typically taken to be 3~times the root-main-square ($3\sigma$) noise in the PVD, and $\eta$ is a constant usually in range $0.2-0.5$ \cite[e.g.][]{sofue-2001, garcia-ruiz-2002, heald-2006, heald-2006-b, heald-2007, rosado-2013, sardaneta-2022}.

\subsubsection{Kinematic parameters}

The ETM uses the PVD traced along the major axis of the galaxy, obtained after simulating a pseudo-slit with a width equal to two pixels (see Section~\ref{Sec:pvdmajor}), and adopts all the kinematic parameters (centre, position angle, and systemic velocity; see Table~\ref{Tab:kin}) derived from it, with the exception of the inclination, which cannot be determined using this method.

In \citetalias{sardaneta-2024}, we explained that the galactic inclination $i$, computed using the apparent major and minor axes from optical wavelengths published in the NED, was a key selection parameter for the selection of our sample. Table~\ref{Tab:general} lists these inclination values. However, since optical inclinations can be affected by star-forming regions, we defined the stellar disc based on the 3$\sigma$ surface brightness ellipse from the 2MASS K-band image to measure morphological parameters of galaxies in our sample. In addition, \cite{epinat-2010} suggested that measuring inclination in the NIR rest frame minimizes contamination from star-forming regions. However, we identified particular cases (CIG~95, 416, 593, and 936) where a bright, centrally concentrated NIR emission, resembling a bulge, dominated the central region, affecting the measured inclination. Although NIR inclinations might suggest that these galaxies are not edge-on, visual inspection, as made by e.g. \cite{bizyaev-2017}, \cite{levy-2019} and \cite{kourkchi-2019}, confirms their edge-on nature through dust lanes in optical images (see Figure~\ref{fig:R-dss}).

In addition, in \citetalias{sardaneta-2024}, we discussed how physical parameters vary across wavelengths. For instance, NIR and MIR data may suggest that our sample is formed by faint dwarf galaxies, whereas young stellar discs indicate sizes and masses comparable to the Milky Way. 
In this sense, inclinations derived from different stellar population tracers revealed that discrepancies between wavelengths arise from real asymmetries in the old stellar disc and the presence of an extended UV disc and halo. The inclination measured from the 3$\sigma$ isophote in the NIR image was the most consistent with optical values. 
Even though NIR-derived inclinations might suggest that some galaxies in our sample are not highly inclined, this discrepancy arises from wavelength-dependent asymmetries rather than an actual low inclination. Since the NIR band provides a more reliable quantitative estimation of the stellar disc inclination, we adopt the inclination measured at the 3$\sigma$ isophote in the $K$-band image, listed in Table~\ref{Tab:kin}, to correct the LoS effects in the ETM.

\begin{table*}
\caption{Kinematic parameters}
\begin{center}
\begin{tabular}{ccccccccccc}
\hline 
CIG & RA$_{\rm kin}$ & Dec$_{\rm kin}$ & $D_{25_{B}}$ & $i_{K_{3\sigma}}$ & PA$_{\rm kin}$ & V$_{\mathrm{rot}}$ \ion{H}{i}    & V$_{\mathrm{rot}}$ \Ha & M$_{\rm dyn}$ \Ha & $\Delta V/\Delta z$ \Ha &  Height \Ha  \\
 Name & (hh mm ss) & (deg ‘ ‘’) &  (kpc)  & (deg) & (deg) &   (\kms)  & (\kms) & ($10^{10}\,M_{\odot}$) & (\kmspc) & (kpc) \\
(1) & (2) & (3) & (4) & (5) & (6) &   (7) & (8) & (9)   & (10)  & (11) \\ 
\hline 
71& 01 55 15.9 & +10 00 50.1 &35.4&78.6& 178.4& 180.7 $\pm$ 5.8  & 183.0 $\pm$ 7.4 & 13.8 & -21.9 $\pm$ 15.4  & 1.2 \\
95& 02 15 20.8 & +22 00 20.9 &33.4&75.0& 128.3& 131.8 $\pm$ 2.7$^{*}$  & 136.1 $\pm$ 13.6 & 7.2& ...   & 1.8$^{*}$  \\
159& 05 39 40.9 & +77 18 52.3 &60.8&87.0& 67.9& 248.8 $\pm$ 11.3  & 245.3 $\pm$ 28.0 & 42.6 & ...   & 0.5 \\
171& 06 32 37.6 & +71 33 41.5 &33.8&86.6&159.2& 170.1 $\pm$ 4.8  & 224.3 $\pm$ 16.4 & 19.8 & -41.5 $\pm$ 12.6   & ... \\
183& 07 18 32.4 & +27 09 42.0 &26.0&76.9&8.0& 152.0 $\pm$ 4.4$^{*}$   & 146.9 $\pm$ 29.7 & 6.5 & ...  &   2.8$^{*}$  \\
201& 07 44 31.1 & +67 16 24.5 &31.4&80.6&151.7& 182.4 $\pm$ 8.5  & 213.5 $\pm$ 13.6 & 16.7 & -15.7 $\pm$ 2.0  &  1.8  \\
329& 09 24 55.1 & +26 46 31.8 &42.8&79.1&115.4& 301.3 $\pm$ 1.4$^{*}$  & 364.9 $\pm$ 28.6 & 66.2 & -32.2 $\pm$ 11.0  &  1.7$^{*}$ \\
416& 10 25 41.7 & +11 44 21.2 &18.0&69.3&94.0& 115.3 $\pm$ 7.8  & 114.2 $\pm$ 15.1 &2.7& -44.5 $\pm$ 11.5  &  1.0 \\
593& 13 36 40.5 & +20 12 14.4 &35.5&72.3&170.1& 221.7 $\pm$ 4.9$^{*}$  & 231.4 $\pm$ 11.2 & 22.1 & ...  &  1.6$^{*}$  \\
847& 18 09 26.0 & +38 47 41.3 &24.6&79.2&142.8& 158.0 $\pm$ 5.3$^{*}$   & 169.3 $\pm$ 11.0 & 8.2 & -21.7$\pm$ 11.3 & 0.3$^{*}$  \\
906& 21 20 17.7 & -01 40 56.8 &37.9&81.3&31.0& 206.1 $\pm$ 8.6  & 187.1 $\pm$ 16.1 &15.4& -95.0 $\pm$ 24.3  & ... \\
922& 21 39 27.0 & +02 49 40.3 &29.4&84.6&58.4& 159.4 $\pm$ 4.2  & 195.8 $\pm$ 11.0 &13.1& ...  &  1.0 \\ 
936& 21 58 07.6 & +01 00 32.7 &38.7&70.2&56.4& 141.8 $\pm$ 4.1  & 159.7 $\pm$ 36.7 &11.5&  ...  & ... \\
1003& 23 01 08.3 & +05 39 13.6 &22.9&81.5&123.0& 115.9 $\pm$ 7.7  & 154.7 $\pm$ 8.8 &6.4& -46.7 $\pm$ 10.4  &  1.2 \\
\hline
\end{tabular}
\label{Tab:kin}
\end{center}
Columns: 
(1)~CIG galaxy name; 
(2) and (3)~\textsc{wcs} coordinates (J2000) of the kinematic centre; 
(4)~$D_{25_{B}}$:~optical diameter from NED; 
(5)~$i_{K_{3\sigma}}$~inclination computed taking into account the galactic thickness  \citepalias[see][]{sardaneta-2024} using the ellipse fitted to the 3$\sigma$ surface brightness level of the 2MASS $K_{s}$-band image; 
(6)~kinematic position angle; 
(7)~rotation velocity computed from the \ion{H}{i} line width from \cite{Jones2018}; (8)~rotation velocity derived from the ETM rotation curve computed with \Ha emission data;  
(9)~dynamical mass \citet{lequeux-1983}; 
(10)~mean rotational lag along the vertical axis measured at $D_{25_{B}}$;
(11)~averaged maximum vertical ($z$) distance reached by the extraplanar component at \Ha emission \citepalias[see table 7 in][]{sardaneta-2024}, the  ($^{*}$) marks galaxies strictly classified as isolated using \ion{H}{i} data by \cite{Jones2018}. 
\end{table*}

\subsubsection{RC model fitting}

Rotation curves can be fitted by a variety of parametric models, some of which are motivated by the physical expectation of contributions by a baryonic component with a mass distribution mimicking that of the light plus a dark matter spheroidal halo \citep[][]{persic-1996, giovalnelli-haynes-2002, valotto-2004, courteau-1997, catinella-2006}. 
\cite{courteau-1997} determined that the $\arctan$ function provides an adequate simple fit and provides an adequate match to most RCs.  Because this model reproduces fairly well the shape of the envelope of the PVD with the smallest number of arguments, we fitted this function to both the minimum  ($I_{min}$) and the maximum ($I_{max}$) contours of the intensity line profiles to obtain the terminal velocity  ($V_{env}$), and so, the galactic RC. The equation of the $\arctan$ model is given by:
%This function is given by:
%
\begin{equation}
V_{env}(r)= V_{sys} + V_{0}\, \tan^{-1} \left(\frac{r}{r_t}\right) + C,
\label{eq-vr-teorica}
\end{equation}
where 
$r_{t}$ is the transition radius between the rising and flat part of the RC, 
$V_{sys}$ is the systemic velocity, 
$V_{0}$ is the asymptotic velocity, and 
$C$ is the intercept on the $y-$axis \citep[see also e.g.][]{Drew-2018ApJ...869...58D, Zhao-2021ApJ...913..111Z, sardaneta-2022, UrrejolaM-2022ApJ...935...20U}.  
Simply-traced envelopes on the two sides of the nucleus have a discontinuity at the nucleus due to the finite beam width, to avoid this discontinuity, we connect both sides of the RC by a straight (solid-body like) line crossing the nucleus at zero velocity \citep[see][]{sofue-1997AJ....114.2428S}.

\subsubsection{Obtaining the RC}

By definition, a PVD represents the emission intensity as a function of position along a spatial axis. The highest intensity is typically found in the central regions, where the density of gas and stars is greatest, while the outer regions exhibit lower intensities due to decreasing density. From the PVD along the kinematic major axis of each galaxy, we approximated $I_{\rm max}$ by fitting an $\arctan$ function tracing the central region of the PVD. Since the data cube includes only Voronoi-tessellated H$\alpha$ emission with SNR$\geq5\sigma$, we extracted the $5\sigma$ isocontour of the PVD and defined $I_{\rm min}$ by fitting another $\arctan$ function to this contour. Using this approach, we computed the envelope intensity, $I_{\rm env}$, for $\eta=0.3$, following relation~\ref{equation:I_env}. 
To further refine this measurement, we extracted the points tracing the isocontour corresponding to the envelope intensity and fitted an $\arctan$ function to them. Figure~\ref{Fig:envelope} illustrates this process using the PVD along the major axis of CIG~201 as an example. Table~\ref{Tab:atan} presents the best-fitting parameters and the coefficient of determination obtained from the $\arctan$ model fitting to the $V_{env}$ (see relation~\ref{eq-vr-teorica}). The RCs computed with both the IPM and ETM methods are shown in Figures~\ref{Fig:c71} to~\ref{Fig:c1003} in Appendix~\ref{Appendix:Maps}, and the corresponding kinematic parameters are listed in Table~\ref{Tab:kin}.

Some studies have shown that the velocity dispersion in extraplanar ionized gas can be significantly enhanced due to non-thermal motions, i.e. 
turbulence \citep[e.g.][]{relano-beckman-2005, silchenko-2023}.  
These turbulent contributions, typically in the range  $\sigma_{\rm turb}\sim20-30$~\kms for \ion{H}{ii} regions \citep[e.g.][]{courtes-1989, relano-beckman-2005, moiseev-2008}, often dominate over the thermal broadening ($\sigma_{\rm T}\approx 9$\,km\,s$^{-1}$ at $T_e = 10^4$\,K) and the instrumental resolution (e.g., $\sigma_{\mathrm{LSF}} \approx 12.5$\,km\,s$^{-1}$), and are considered a major source of line broadening in ionized gas of disc galaxies. 
However, turbulence is not spatially uniform within a galaxy. It tends to be more prominent at large vertical heights ($z$), where the eDIG tends to show broader line profiles and more complex kinematic \citep[e.g.][]{relano-beckman-2005, marasco-2019,  silchenko-2023, gonzalezDiaz-2024-betis-ii}. 
In contrast, our RCs are derived from PVDs extracted along the kinematic major axis, where the \Ha  emission traces the central disc regions. These regions are dominated by ordered rotational motion and do not show the kinematic signatures associated with turbulent broadening. Therefore, applying a turbulent correction in this context would be physically unjustified, because the central disc does not exhibit the conditions under which such corrections are necessary.

\subsection{The lag in rotation}\label{Sec:Lag}

Studies on the eDIG kinematics either in radio \citep[e.g.][]{swaters-1997, lee-2001, fraternali-2005, marasco-2012, Zschaechner-2015-a, marasco-2019, li-fraternali-marasco-2023} or in optical \citep[][]{rand-2000, miller-2003, heald-2006, heald-2006-b, heald-2007, rosado-2013, bizyaev-2017, levy-2019, bizyaev-2022} wavelengths have determined that the halo gas rotates more slowly than the gas in the plane. This vertical velocity gradient, described as the lag of the disc rotation, 
has been interpreted within the framework of a disc-halo flow model, where gas is transported into the halo, moves radially outward, and slows down due to angular momentum conservation in a weaker gravitational potential \citep[e.g][]{bregman-1980, collins-2002}. 
However, this classical model alone cannot fully explain the observed rotation lag. Although the conservation of angular momentum causes the fountain clouds to lose rotational velocity, the discrepancy between predictions and measurements suggests the existence of additional physical processes. The interaction of the fountain clouds with the CGM, which has lower specific angular momentum, further slows their rotation and redirects their motion inward, highlighting the need for more complex physical processes to fully explain the lag \citep[][]{fraternali-binney-2008, marasco-2012, marasco-2019, li-fraternali-marasco-2023}.

Galactic winds, displayed as bipolar outflows, both disturb the symmetry of the extraplanar velocity field and increase the extraplanar emission line widths \citep[e.g.][]{Sharp-2010}. 
In cluster galaxies ongoing ram-pressure stripping (RPS), the extraplanar gas is oriented at a different angle from the disc plane, but showing a continuous velocity field between the disc and the extraplanar region \citep[e.g.][]{vollmer-2009}. 
In contrast, the eDIG follows the velocity field of the galaxy showing inconspicuous  asymmetry in the vertical gradient, i.e. the gas above and below the disc slows down equally with the $z-$axis \citep[e.g.][]{heald-2006, heald-2006-b, heald-2007, ho-2016, bizyaev-2017, bizyaev-2022}. 
However, it is likely that in some of the wind-dominated galaxies or galaxies suffering ongoing RPS, a portion of the pre-existing eDIG remains corotating with the discs while the outflowing gas interacts with the rest of the eDIG \citep[e.g.][]{ho-2016, tomicic-2021, sardaneta-2022}.

The technique typically used to measure the rotation lag from integral field spectroscopy data, including those obtained from FP, consists in measuring velocities in cuts parallel to the major axis \citep[e.g.][]{marasco-fraternali-2011A&A...525A.134M, rosado-2013, bizyaev-2017, levy-2019, sardaneta-2022}.
Some authors measure this lag at a specific radius of the galaxy, for example, where the maximum rotation velocity is observed \citep[e.g.][]{levy-2019, sardaneta-2022}.
In \citetalias{sardaneta-2024}, we detected extraplanar \Ha emission along the $z$-axis in 11 out of 14 galaxies (see their table7). Although not all galaxies show an extended or visually distinguishable eDIG structure, we intend here to measure the lag in the entire sample of 14 galaxies.

To do so, we extracted five PVDs along the major axis at different disc heights for all galaxies, which are plotted in the right panels of Figures\ref{Fig:c71} to~\ref{Fig:c1003}, except for CIG~593 due to its poor ionized gas emission. 
We applied the ET method to each PVD extracted parallel to the major axis at different disc heights. 
Since rotation curves often continue rising until the last observed radius, potentially underestimating the true maximum velocity, we follow \cite{torres-flores-2011} by adopting the rotation velocity measured at the optical radius ($D_{25}(B)/2$) as a reference. This provides a consistent basis for comparing rotation velocities across vertical heights. 
We then computed the average lag by fitting an ordinary least squares regression to the velocities obtained at the five heights (see bottom middle panel of Figure\ref{Fig:e.g.-c201}). 
In Appendix~\ref{App:tirific}, we present the analysis of a mock data cube replicating one of the sample galaxies, demonstrating that the ETM applied to each PVD successfully recovers the model’s input parameters, particularly the vertical rotation lag.

This method assumes that the gas at different heights traces the same underlying kinematics, aside from the expected lag. Although extraplanar gas can show enhanced velocity dispersion due to turbulence or outflows, our data do not reveal a systematic increase in line width with vertical height. In general, the radial velocity fields display symmetry above and below the plane, consistent with a lagging but corotating component. Moreover, the velocity dispersion maps reveal localized enhancements rather than a uniform trend with $z$. Therefore, we do not apply a turbulent broadening correction when estimating the vertical rotation lag. Such contributions will be analysed locally in a future work focused on the kinematics of specific extraplanar regions.

On the other hand, rotation along the z-axis cannot be detected in all galaxies, either because there is no variation or because the  \Ha emission is obscured by dust. In addition, there are cases where the emission at a large height becomes poor and may result in an inconsistent rotational lag value when the $\arctan$ function is fitted, as in the case of CIG~906 (see Figure~\ref{Fig:c906}). Regardless of whether it was possible to detect any variation in rotation along the $z$-axis, this measurement is listed in Table~\ref{Tab:kin}. We have found that in 7 out of 14 galaxies in our sample there is an averaged decrement in the rotation along the $z$-axis of $\Delta V/\Delta z=32.0\pm10.6$~\kmspc.

Using the same recipe for measuring the lag at the optical radius, $D_{25}(B)/2$, but at several radii, we computed the variation of the lag with respect to the radius of the galaxies. We observed that in all galaxies presenting lag in rotation the relation with the radius showed a clear trend, either continuously rising or falling. Thus, with an ordinary line square regression we fitted an exponential function to this curve finding that the amplitude of this fitted function overlaps within the margin of error of the previously lag measured (see bottom right panel of Figure~\ref{Fig:e.g.-c201}). The implications of the gradient of this fitting will be discussed later in Section~\ref{Sec:Results}.

\subsection{Dynamical mass}\label{Sec:dynMass}

We may assume that a galaxy is composed of a flat baryonic disc superimposed on a massive dark-matter halo, which could be either flat or spherical. 
In this sense, as a first approximation, the mass value of a late-type galaxy at radius $R$ lies between the mass value for a homogeneous sphere and that for an infinitesimally thin disc. 
In the case of the homogeneous sphere, the dynamical mass can be approximated by assuming a central point mass and calculated using the relation $M(R)=\alpha \ RV_{\mathrm{rot}}^{2}(R)/G$ from \cite{lequeux-1983}, where $G$ is the universal gravity constant. In this relation, the limiting values of $\alpha=(0.6,\, 1)$ correspond to the galaxy being dominated by a flat or spherical massive dark-matter halo, respectively. 
This is, for the infinitesimally thin disc, the mass value is about 0.6~times that of the homogeneous sphere at the same radius. 
We estimated the dynamical mass of the galaxies in our sample using the \Ha maximum rotational velocity and the radius derived from the optical diameter at the B-band ($D_{25_{B}}/2$, from NED). 
This estimation serves as a proxy for the total dynamical mass enclosed within the B-band optical radius. The corresponding parameters and dynamical mass for each galaxy are listed in Table~\ref{Tab:kin}.

\section{Results}\label{Sec:Results}

In this section, we present our results related to the measurements of $V_{\rm max}$ of the rotation curves obtained with the ET method. To begin, we assess the reliability of these measurements by exploring the fitting to the near-infrared (\textit{NIR}) and optical Tully-Fisher relations proposed by \cite{torres-flores-2011} and \cite{epinat-2008-ii}, respectively. Next, to provide insights into the alignment or divergence between the warm and cold gaseous components of galaxies in our sample, we compare $V_{\rm max}$ at the optical radius ($D_{25_{B}}$, see Table~\ref{Tab:general}) with the rotational velocity derived from \ion{H}{i} data \citep[data from][]{Jones2018}. 
Finally, we present results related to the galaxy rotation lag, analysing its correlation with various physical properties such as rotation velocity, velocity dispersion and galactic environmental features. We present in detail the individual results for each galaxy in our sample in Appendix~\ref{Sec:individual-notes}.

\subsection{Assessing the resulting $V_{\rm max}$.}

Spiral galaxies present significant differences in their brightness, size, rotational velocity and the shape of their rotation curves, nevertheless, they follow a consistent scaling correlation between luminosity (\textit{L}) and maximum rotational velocity ($V_{\rm max}$). This correlation is known as the Tully-Fisher (TF) relation \citep{tully-fisher-1977} which was originally defined to estimate distances to spiral galaxies \citep[see][and references therein]{mo}. The TF relation is also important for our understanding of galaxy formation and evolution since it defines a relationship between parameters that provide crucial information about the structure, composition and general characteristics of a galaxy, such as the luminosity and the dynamical mass (due to stars, gas and dark matter), related to the $V_{\rm max}$ value \citep[e.g.][]{epinat-2008-ii, torres-flores-2011, gomezl-2019}.

\begin{figure*}
\centering
\includegraphics[width=0.9\hsize]{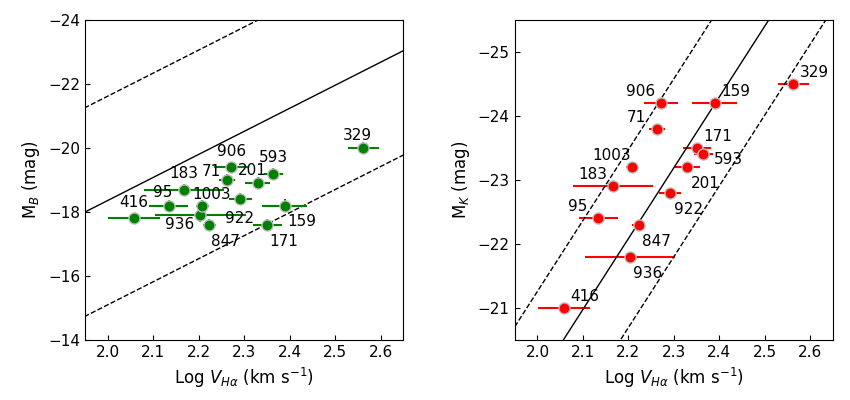}
\caption{
Comparison of the location of our isolated galaxy sample (circles) with the general galaxy population from the GHASP sample, in the Tully-Fisher relations for the B-band \citep[left panel, derived by][]{epinat-2008-ii} and the NIR \citep[right panel, K-band, derived by ][]{torres-flores-2011}. 
The solid line represents the fit to the GHASP, with the dashed lines outlining the uncertainty range (see the text). 
The consistency between the isolated galaxy sample and the GHASP relations serves to validate the accuracy of the rotation velocity measurements for the isolated galaxies.
}
\label{Fig:tully-fisher}
\end{figure*}

The Gassendi \Ha survey of SPirals \citep[GHASP project,][]{garrido-2002} is the largest sample of spiral and irregular galaxies observed to date using FP techniques, consisting of 203 galaxies, covering a large range of morphological types and absolute magnitudes. 
In Figure~\ref{Fig:tully-fisher}, we compare our isolated galaxies with the GHASP sample in both the B-band and the near-infrared (NIR) TF relations by using the linear fitting determined by 
\cite{epinat-2008-ii}, 
$M_{B} = (-7.2 \pm 1.2) \log V_{\rm max} - (3.9\pm 3.2)$, 
and by \cite{torres-flores-2011}, 
$M_{K} = (-11.07 \pm 0.63) \log(V_{\rm max}) + (2.27 \pm 1.39)$, 
for the GHASP sample in these different bands. Notably, the NIR TF relation offers advantages due to smaller extinction corrections compared to optical bands \citep[see e.g.][]{verheijen-2001}. However, both fittings align well with our sample.

In the NIR TF relation plotted in right panel of Figure~\ref{Fig:tully-fisher}, there is only one particular case.  
CIG~329 is the only one barred galaxy in our sample (see \citetalias{sardaneta-2024}) and defined as a LINER by \cite{sabater-2012} using the available SDSS spectrum. Its rotation curve exhibits a proper alignment between the tangent function fitting and the rotation curves calculated with the two methods, barycentre and envelope (see Figure~\ref{Fig:c329}). Therefore, the observed shift in the location of this galaxy within the \textit{K}-band TF relation can be attributed to the non-circular motions result of the stellar bar \citep[see e.g.][]{torres-flores-2011} and/or to the enhanced \textit{NIR} magnitude due to AGN activity.

\subsection{Relation between H$\alpha$ and \ion{H}{i} velocities}\label{Sec:Ha-Hi}

\begin{figure}
\centering
\includegraphics[width=1\hsize]{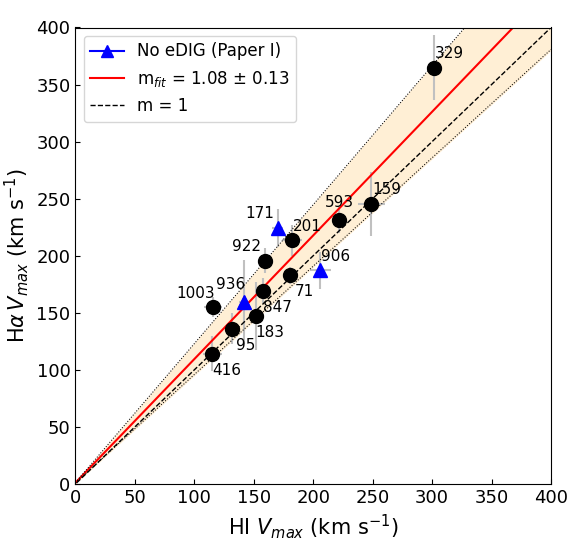}
\caption{
Comparison between the maximum rotation velocities derived from \ion{H}{i} \citep[from][]{Jones2018} and \Ha emission for galaxies in our sample. 
The red line represents the OLS best-fitting function ($V_{\rm max,H\alpha} = 1.08 \pm 0.13\,V_{\rm max, \ion{H}{i}}$) and the shaded area indicates the confidence region. As a reference, the dashed line traces the one-to-one \ion{H}{i} and H$\alpha$ $V_{\rm max}$ relation (slope $m=1$). We highlight three galaxies (triangles) which did not exhibit a morphological extraplanar component in \citetalias[][]{sardaneta-2024} and now they do not show any particular trend in this plot. The slight differences between \Ha and \ion{H}{i} velocities in these isolated galaxies are likely due to internal effects. 
}
\label{Fig:HIvsHa}
\end{figure}

\cite{Jones2018} compiled \ion{H}{i} properties of 844 isolated galaxies from the CIG Catalogue from their own  \ion{H}{i} observations with Arecibo, Effelsberg, Nan\c{c}ay and Green Bank Telescopes, as well as spectra collected from the literature in the frame of the AMIGA project. For each galaxy, they provide line widths $W_{\rm \ion{H}{i}}$ measured at 50\% of the peak flux per galaxy in $\mathrm{km\,s^{-1}}$.  We have computed the maximum \ion{H}{i} rotation velocity ($V_{\rm max, \ion{H}{i}}$) from the \ion{H}{i} line width assuming that $W_{\rm \ion{H}{i}}=2 V_{\rm max,\ion{H}{i}}\sin\,i$ \citep[e.g.][]{lee-2017, gomezl-2019}, using the photometric inclination  determined in \citetalias{sardaneta-2024}. Table~\ref{Tab:kin} shows the \ion{H}{i} rotation velocities corrected by inclination for each galaxy.  Figure~\ref{Fig:HIvsHa} shows the comparison between \ion{H}{i} and H$\alpha$ $V_{\rm max}$ values. We have performed an ordinary least square (OLS) estimation to find a correlation between \ion{H}{i} and H$\alpha$ $V_{\rm max}$ for galaxies in our sample obtaining $V_{\rm max,H\alpha}=(1.08\pm0.13)\,V_{\rm max, \ion{H}{i}}$. It is worth noticing that $V_{\rm max,H\alpha}$ is, in most cases, extrapolated from the rotation velocity points, as \Ha discs are not extended enough to reach the flat part of the rotation curve.

Since galaxies in our sample are assumed not to have interacted with the IGM, which could strip them of gas, the slight difference in \Ha and \ion{H}{i} velocities may primarily be attributed to internal effects rather than external ones. 
Specifically, internal kinematic disturbances, such as gas distribution asymmetries and non-circular motions,  can affect both the \Ha and \ion{H}{i} velocities. 
Asymmetries in the gas distribution create uneven peaks or offsets in the PVD, showing localized variations in velocity that affect the measured rotational velocities. While kinematic perturbations appear as deviations in the profile in the PVD, indicating areas where the gas may be moving inward or outward rather than along the galactic plane.

Moreover, the discrepancy between \ion{H}{i} and H$\alpha$ $V_{\rm max}$ values could be linked to the differences in the resolution of the data. 
The \ion{H}{i} data have relatively low spatial resolution \citep[Arecibo:~3.3 to 3.9~arcmin, Effelsberg:~8.8~arcmin, Nan\c{c}ay:~4~arcmin and Green Bank:~9 to 21~arcmin,][]{Jones2018} integrating emission over extensive areas within the galaxy. 
The low spatial resolution averages the velocities from different regions, combining the rotational motion of the gas at different radii and smoothing out local variations that the FP detects more accurately in \Ha emission. 
Indeed, some galaxies that deviate from the one-to-one \ion{H}{i} and H$\alpha$ $V_{\rm max}$ relationship  (dashed line in Figure~\ref{Fig:HIvsHa}), such as CIG~201, 922, and 1003, exhibit large bright regions emitting in \Ha along the disc. These regions, tracing recent star formation, contribute to the intensity variations that shape the PVD, influencing the determination of the maximum in the rotation curve. Meanwhile, the lower spatial resolution of the \ion{H}{i} data results in a velocity width that represents the gas motion averaged over a large area, potentially masking localized kinematic variations detected in \Ha emission. 
Nevertheless, the relationship factor observed of $m=1.08\pm0.13$ indicates that the discrepancy between  \ion{H}{i} and \Ha $V_{\rm max}$ values is not critical, suggesting that the two data sets remain comparable.

\subsection{Limitations of the $\arctan$ function fit }

Disk galaxies are complex systems comprising multiple structural components. A RC is obtained by approximating the full velocity field as a function of radius, averaging and smoothing rotation velocities, and neglecting small-scale variations \citep{sofue-2001}. 
The $\arctan$ function offers a simple fit to most RCs, though it does not account for non-circular motions, spiral arms, or sharp velocity peaks near the turnover radius. More sophisticated models may achieve better accuracy but often introduce strong covariances between parameters \citep{courteau-1997}. The envelope-tracing method remains one of the most practical approaches to deriving RCs in edge-on disk galaxies. However, it struggles to accurately determine velocities in central regions due to unresolved, rapidly rotating components, steep velocity gradients, and complex gas distributions \citep{takamiya-2002}. Additionally, observational factors such as seeing, slit width, beam size, and spectral resolution, as well as intrinsic properties like inclination, interstellar velocity dispersion, and gas distribution, further influence the shape of PVDs \citep{sofue-2001, takamiya-2002}.

We found that the $\arctan$ model may not adequately describe the observed rotation curves for several galaxies, particularly CIG~95, 183, 329, 416, and 1003. For $\eta=0.3$, the maximum coefficient of determination for the fit was only $R^{2}=0.6$, suggesting that the model does not fully reproduce reproduce the observed kinematics (see Table~\ref{Tab:atan}). 
These discrepancies arise because the PVDs of these galaxies exhibit significant deviations from a smooth S-shape, presenting asymmetries (e.g. CIG~95 and 183), intensity enhancements due to star-forming regions (e.g. CIG~416 and 1003), or distortions caused by non-circular motions (e.g. CIG~183).  
Moreover, if the gas distribution is ring-like or concentrated in specific regions (e.g. CIG~329), the ET method may overestimate or underestimate the actual velocities. These factors introduce substantial uncertainties in the RC determination, making a simple two-parameter model insufficient to capture the full kinematic complexity of these galaxies.

To obtain the maximum rotation velocity ($V_{max}$), we extended the $\arctan$ fit to the radius corresponding to $M_B$. Despite its limitations, the extrapolated RC gives as a result a $V_{max}$ consistent with the one derived using the barycenter method and with that derived from \ion{H}{i} observations (see Section~\ref{Sec:Ha-Hi}). The agreement between these independent tracers suggests that, although the ET method and the $\arctan$ model do not capture all the complexities of non-circular motions and structural asymmetries, they still provide a reasonable approximation of the overall galactic kinematics. Thus, the $\arctan$ function remains a useful tool for characterizing the global rotation properties of edge-on galaxies.

\subsection{Rotation lag and galaxy properties}

Using the same technique to compute the maximum rotation velocity, we assessed the rotation of the disc at different heights. Our results showed a reliable (average $\Delta V/\Delta z=32.0\pm10.6$~\kmspc) decreasing velocity with height (lag in rotation or lag) in 7 out of 14 galaxies (CIG~71, 171, 201, 329, 416, 847, 1003). This is  consistent with previous studies which have reported vertical gradients in azimuthal velocity of $10-30$~\kmspc\, in nearby galaxies with high-quality optical or \ion{H}{i} data \citep[][]{fraternali-2005-891, heald-2006, heald-2006-b, heald-2007, Zschaechner-2015-a, ho-2016, bizyaev-2017, bizyaev-2022, levy-2019, marasco-2019}. CIG~171 is one of the seven galaxies presenting lag in this paper, however, in \citetalias{sardaneta-2024} was noted that this galaxy does not present \Ha emission outside the stellar disc, so that the lag in rotation may actually be linked to the emission in the thick disc. On the other hand, we discarded the exceptional galaxy in our sample, CIG~906, which exhibited an unusual high value of lag ($95.0\pm24.3$~\kmspc, see Table~\ref{Tab:kin}), likely as a result of its low \Ha emission reflected in the $\arctan$ function fitting to its PVDs.

We studied how the lag in rotation changed with radius finding that all the seven galaxies exhibiting lag displayed a gradient in the lag vs. radius curve. Specifically, four galaxies (CIG~71, 171, 329, 1003) showed a positive gradient with radius, while three galaxies (CIG~201, 416, 847) presented negative gradients. According to \cite{levy-2019}, radial variations in lag may indicate different origins of the extraplanar gas: a decrease in lag with radius suggests an internal origin (e.g., galactic fountains), no radial variation indicates gas accretion from the CGM, and larger lags at larger radii imply cold material accretion from streams. 
However, \cite{levy-2019} also pointed out that since the gravitational potential influences the motion and settling of the gas at equilibrium, it can also mislead the effects of internal and external processes on the observed radial gradients in the lag, making it difficult to distinguish between the two origins. 
This makes challenging to distinguish between internal and external origins just by measuring radial variations of the lag. Indeed, \cite{bizyaev-2022} observed that gas accretion via in-plane inflow can create positive radial gradients at the edges of galaxies.

To determine if the lag is mainly due to an internal or external process we helped our analysis by studying the PVDs along the minor axis (see middle panel of Figure~\ref{Fig:e.g.-c201} and Figures in  Appendix~\ref{Appendix:Maps}).  
PVDs aligned perpendicular to the kinematic major axis provide insight into the kinematic behaviour along the z-axis highlighting its main features and asymmetries. These PVDs indicate whether extraplanar gas rotates slower along the z-axis than in the plane, potentially losing angular momentum. If the gas is in circular rotation, the PVDs will be symmetric with respect to the major axis, while radial motion of the gas inwards or outwards will break this symmetry \citep{fraternali-2006}. For instance, a thinning emission effect observed in  PVDs aligned to the minor axis, observed in galaxies like NGC~891 \citep{fraternali-2006} and the Milky Way \citep{walker-1996}, has suggested a lagging halo, although it also might be a result of mild LoS warp \citep[e.g.][]{swaters-1997, becquaert-1997}. 
On the other hand, in high inclined galaxies ($i \geq 80^{\circ}$)  
%In addition, it has been observed that 
PVDs along minor axis draw a heart-shaped profile indicating the presence of an inflow in the galaxy \citep[e.g.][]{fraternali-2006, rosado-2013}.  
In galaxies with intermediate inclination ($50^{\circ}<i<72^{\circ}$) heart-shaped profiles are commonly observed, appearing in regularly rotating discs. In these galaxies, however, asymmetries in PVDs along minor axis typically indicate non-circular motions, while slow rotation is harder to detect \citep{marasco-2019}.

In this sense, CIG~201 displayed the expected heart-shaped PVDs consistent with its negative lag vs. radius gradient. In contrast, CIG~416 and 847 did not exhibit this characteristic in their minor axis PVDs as expected. These two galaxies, in particular, have in common that the velocity dispersion map shows regions with velocities of 40\kms distributed across the disc rather than concentrated around the brightest \Ha emission knots. Moreover, unlike the distinctive radial velocity map of CIG~201 previously discussed, the radial velocity fields of CIG~416 and 847 have a rigid-body rotation. However, CIG~847 was defined as well isolated with \ion{H}{i} data by \cite{Jones2018},  while CIG~416 may have a dwarf companion \citepalias[see][]{sardaneta-2024}. In this way, there may be additional factors influencing the dynamics of these two galaxies rather than inflowing material.

\begin{figure*}
\centering
\includegraphics[width=1\hsize]{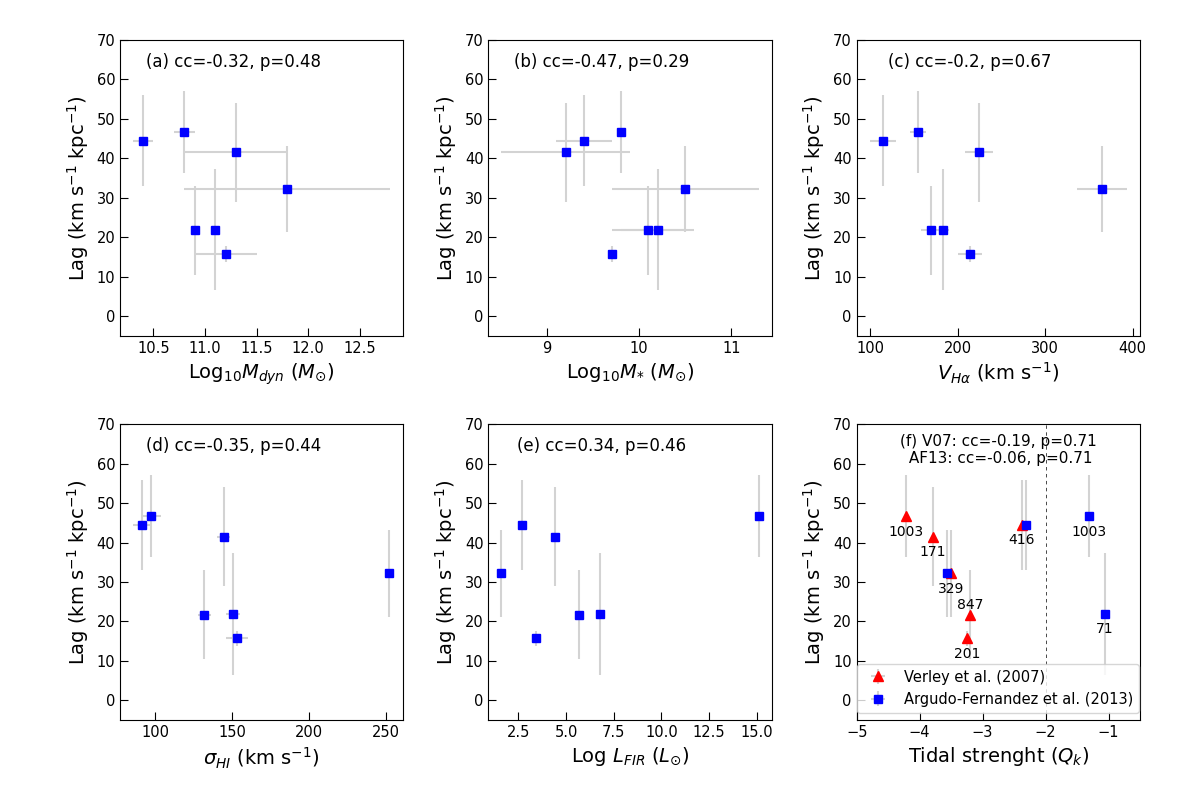}
\caption{
Relationship between the amplitude of the lag in rotation and various physical and environmental properties of the galaxies in our sample: 
(a)~dynamical mass ($M_{\rm dyn}$), 
(b)~stellar mass ($M_{*}$), 
(c)~\Ha maximum rotation velocity ($V_{\rm H\alpha}$), 
(d)~\ion{H}{i} global velocity dispersion ($\sigma_{ \ion{H}{i}}$) from \citet{Jones2018}, 
(e)~FIR luminosity ($F_{\rm FIR}$), and 
(f)~tidal strength ($Q_{k}$) from \citet{verley-2007-i} (red triangles) and \citet{Argudo-F-2013} (blue squares). The trend is evaluated by calculating the correlation coefficient (cc) and the corresponding p-value.
}
\label{fig:lag-properties}
\end{figure*}

In order to understand the physical processes that drive the different types of disc–halo interactions, previous studies, such as \cite{ho-2016}, \citet{bizyaev-2017, bizyaev-2022} and \cite{levy-2019}, investigated the connection between the amplitude of the lag and some  physical properties of the galaxies. Although our sample of seven galaxies with measurable lag is limited, in Figure~\ref{fig:lag-properties}, we examined whether there is a trend between the lag in rotation and some physical characteristics of these seven galaxies by calculating  the correlation coefficient (cc).

First, like \cite{levy-2019}, we did not find any correlation between the lag and the maximum rotation velocity. 
Second, since the FP data provide us with a measure of local velocity dispersion rather than a global one, we used the velocity width of the \ion{H}{i} emission stated by \cite{Jones2018} for the AMIGA sample, which showed a weak negative correlation (cc$\simeq$-0.35) with the lag as reported by \cite{levy-2019}. 
And third, in \citetalias{sardaneta-2024}, we showed that galaxies in our sample have very extended \Ha and UV emission discs relative to the disc traced by the NIR emission. Indeed, when calculating the dynamical mass in this work (see Table~\ref{Tab:kin}), we noticed a difference of one order of magnitude compared to the stellar mass calculated using IR emission\footnote{See Appendix~C from \citetalias{sardaneta-2024} for further details on this calculation.}. A comparison of the lag with both masses revealed a negative and low correlation in both cases, consistent with the findings of \cite{levy-2019}.

The only correlation found by \cite{ho-2016} for SAMI galaxies was for the SFR surface density. This is in agreement with the  diagnostic eDIG diagram of \cite{rossa-2000} which indicated that the prominence of the gaseous haloes could be correlated with tracers of SF in the disc, such as the surface density of FIR emission and dust temperatures. However, in \citetalias{sardaneta-2024}, we showed that not all galaxies with eDIG were located in the defined eDIG area and vice versa. 
Indeed, in Figure~\ref{fig:lag-properties},  we did not find any correlation between the lag and the FIR luminosity, tracer of the SFR, similarly to \cite{bizyaev-2017, bizyaev-2022} and \cite{levy-2019}.

In Figure~\ref{fig:lag-properties} we also investigate the correlation between the lag and the tidal strength parameter, $Q_{k}$. The tidal strength parameter is an estimation of the total gravitational interaction strength that the neighbours produce on the central galaxy with respect to its internal binding forces \citep[][]{Argudo-F-2013}. 
In the frame of AMIGA project, this parameter was obtained first, based on digitized photographic plates from DSS POSS-1 and POSS-2 by \cite{verley-2007, verley-2007-i} and later, on available photometric and spectroscopic data from SDSS-DR9 by \cite{Argudo-F-2013} \citepalias[see][]{sardaneta-2024}. 
Note that this plot excludes CIG~183, limiting it to galaxies whose kinematics show no signs of environmental interaction. 
In Figure~\ref{fig:lag-properties}, there is apparently a trend of negative correlation between the lag and the tidal force measured by \cite{verley-2007, verley-2007-i}, except for CIG~416.  Statistically, if a measurement significantly deviates from the established trend or pattern observed in the data, it may be considered an outlier and potentially discarded. 
A Dixon's test for outliers \citep{nist-statistics} was applied on the dataset giving as a result that CIG~416 is not considered a significant outlier in this dataset    \citep[$Q_{k} = 0.45$, $p{\rm -value}=0.2646$, ][]{R-citation}. In addition, we refer to the isolation review of AMIGA sample using \ion{H}{i} emission data of \cite{Jones2018}, noticing that only 2 galaxies (CIG~329 and 847) out of the 6 galaxies tracing this trend are considered properly isolated. Furthermore, CIG~1003 is no longer within the region of bona fide isolated galaxies in the AMIGA sample after  \cite{Argudo-F-2013} reviewed the tidal strength parameter. 
On the other hand, the fact that \cite{Argudo-F-2013} had confirmed the isolation parameters with more reliable data and that the measure of the tidal strength for CIG~416 is aligned in both surveys, implies that this galaxy cannot simply be excluded, obtaining a negligible correlation coefficient of cc$\simeq$-0.1 for both surveys. A larger sample could resolve this inconsistency.

\section{Discussion}\label{Sec:Disc}

With the aim of understanding the effects of the environment on the origin of the galactic halo, in this paper we studied the \Ha emission line kinematics obtained from FP data of a representative sample of edge-on late-type galaxies ($i \geq 80^{\circ}$) from the CIG catalogue of \cite{Karachentseva-1973}. 
In the following, we discuss our results with those kinematic results obtained from studies on samples of edge-on galaxies using IFU spectroscopy data from SAMI \citep{ho-2016}, MaNGA \citep[][]{bizyaev-2017, bizyaev-2022} and CALIFA \citep{levy-2018, levy-2019} surveys. Henceforth, it is worth considering the selection criteria of each sample to assess whether there is any environmental influence on their results. 
This is, \cite{ho-2016} chosen only edge-on galaxies from SAMI survey presenting regular rotation on their discs, consistent with not undergoing major interactions; 
\cite{bizyaev-2017} included all edge-on MaNGA galaxies, rejecting all galaxies having another object in the field, including stars; 
\cite{levy-2019} claimed edge-on CALIFA galaxies in their sample are not interacting, however, their environmental selection criteria are not clear; 
finally, \cite{bizyaev-2022} was less restrictive on the environment by focusing on selecting edge-on galaxies, even though some of these now included features such as low surface brightness shells and asymmetries. Let us examine how their findings compare to ours.

% \subsection{Velocity fields }

To study the gaseous component, we present \Ha radial velocity fields of the 14 galaxies in our sample. In general, the radial velocity maps indicate that all our galaxies exhibit regular rotation characteristic of disc galaxies. 
On the other hand, \cite{ho-2016} quantified the velocity asymmetry across the galactic major axis on the velocity map to differentiate between extraplanar emissions tracing winds or eDIG, where asymmetry implied galactic winds. Later, \cite{bizyaev-2017} distinguished wind galaxies from those with eDIG as those showing a symmetric decrease in velocity amplitude with vertical distance from the midplane. However, this selection is widespread, since it has been shown that in some of the wind-dominated galaxies, a portion of the pre-existing eDIG remains corotating with the discs while the outflowing gas interacts with the rest of the eDIG \citep[e.g.][]{ho-2016, tomicic-2021, sardaneta-2022, bizyaev-2022}. We qualitatively discuss the asymmetry in our data based on the isovelocities observed in each radial velocity map since the \Ha emission in most galaxies in our sample %CIG~71, 95, 159, 593, 847 and 906, 
is significantly obscured by dust, resulting in fragmented gaseous discs or in a poor detection of \Ha emission. Consequently, the isovelocities in these incomplete maps are mainly traced by distorted lines.

A particular case is CIG~183, where the kinematic and morphological centres 
differ by $\sim$8.5~arcsec. Despite this offset, the \Ha emission disc shows a non-perturbed velocity field. A comparison between the radial velocity map and 
the K-band image might suggest they are two distinct galaxies. However, in \citetalias{sardaneta-2024} it was showed that UV emission covers both ISM 
phases, with FUV emission following the morphology traced by the \Ha emission. 
In any case, this galaxy has a \Ha distribution and kinematics that strongly point to the presence of strong environmental effects. 
On the other hand, three galaxies (CIG~201, 416, and 1003) display extensive gaseous discs, with \Ha emission prevailing over NIR emission. The isovelocities for these galaxies are symmetrical with respect to the minor and major axes. However, while CIG~416 and 1003 exhibit solid-body rotation with parallel and straight velocity contours, CIG~201 shows isovelocities typical of galaxies with steeply rising-to-flat rotation curve which is clearly traced by the PVD's envelope. In \citetalias{sardaneta-2024}, these three galaxies also displayed long filaments and multiple inner emission knots of \Ha and UV emissions. Spectral data would help to determine whether these features originate from an AGN or star-forming regions. In addition, further radio data could provide additional insights into the large-scale kinematics of these targets.

The resolution of GHASP allowed us to compute the dispersion velocity maps for the 14 galaxies in our sample. In general, the velocity dispersion maps are uniform, with values ranging between 20 and 35~\kms across the galaxy.  While the velocity dispersion maps do not exhibit extreme fluctuations, most regions show variations within a range of approximately 15~\kms, suggesting some degree of kinematic uniformity, although  there are some regional variations. This relative uniformity suggests a lack of strong internal or external forces disrupting the gas, which is typical for isolated galaxies. This is consistent with \citep{heald-2006, heald-2006-b, heald-2007}, who found that extraplanar gas in galaxies such as NGC~891, NGC~5775, and NGC~4302 typically exhibits velocity dispersion of $\sigma\leq50$~\kms\, for eDIG.

Six galaxies in our sample (CIG~159, 171, 329, 416, 847, 906) presented some regions of higher values of velocity dispersion ($\sigma\simeq40$~\kms) distributed along the disc. This could imply weak disc-halo interactions, where the velocity dispersion of eDIG arises predominantly from the LoS projection of gas co-rotating with the disc, as noted by \cite{ho-2016}. 
On the other hand, four galaxies (CIG~95, 183, 922, 1003) show high dispersion velocity values around their \Ha brightest knots or along the borders of the old-stellar population disc. In agreement with \cite{levy-2019}, this pattern suggests local turbulence likely driven by star formation activity in the disc. 
Specifically, CIG~593 has very low \Ha emission, so we did not detect important regions with significant velocity dispersion values. 
Lastly, three galaxies (CIG~71, 201, 936) presented high dispersion values in the innermost regions of the galactic disc. In particular, the velocity dispersion map, as the multiwavelength maps (see \citetalias{sardaneta-2024}), of CIG~201 shows a cone shape. \cite{ho-2016} showed that classical wind galaxies exhibit line splitting and high dispersion velocities due to strong disc-halo interactions. Therefore, these features may be indicative of more complex internal dynamics or the presence of galactic winds.

We computed the rotation curves of galaxies in our sample using two different techniques, one of them allowing us to correct the LoS effects due to the inclination. 
The approach to the maximum rotation velocity was verified by comparing the resulting maximum velocities with \ion{H}{i} data from \cite{Jones2018} and further confirmed using B- and K-band Tully-Fisher relationships \citep{epinat-2008-ii, torres-flores-2011} (see Section~\ref{Sec:Results}). This is consistent with \cite{ho-2016}, who also used velocity maps and the stellar mass Tully-Fisher relation for galaxies with limited spatial coverage. In contrast, \cite{levy-2018} used the barycentre method to compute the rotation curves of CALIFA galaxies finding slower \Ha rotation compared to molecular gas, inferring that this difference in velocity was a consequence of the eDIG contribution. It has been shown that the molecular ISM phase traces better the rotation in the inner regions of the galaxy than the gaseous one because of the dust extinction \citep[see e.g.][]{sardaneta-2022}. This highlights the importance of applying LoS corrections to the RCs traced by ionized gas to obtain accurate rotational velocity measurements in edge-on galaxies. In this work, we compute the rotation curves using as well the barycentre method to demonstrate that, in the inner regions of edge-on galaxies, this approach systematically underestimates the true rotational velocity. While previous studies \citep[e.g.,][]{heald-2006, heald-2007, rosado-2013, sardaneta-2022} have shown this effect for individual galaxies in the optical band, we extend this analysis to a sample of 14 galaxies, confirming its systematic approach.

We computed the rotation of the disc at different heights by using the same technique to compute the maximum rotation velocity. We obtained gradients in azimuthal velocity of $32.0\pm10.6$~\kmspc, consistent with previous studies \citep[e.g.][]{fraternali-2005-891, heald-2006, heald-2006-b, heald-2007, Zschaechner-2015-a, ho-2016, bizyaev-2017, bizyaev-2022, levy-2019, marasco-2019}. However, the detection of a lag in rotation may not necessarily indicate the presence of eDIG, since it could rather be related to emission in the thick disc as long as the \Ha emission still originates in the main disc, as observed in CIG~171. Similarly, \cite{levy-2019} observed that five galaxies in their sample showed morphological signs of eDIG but they did not exhibit a measurable lag, suggesting also that lag might not be a definitive marker for eDIG. 
The importance of this kinematic feature lies in its impact on the samples chosen by \cite{bizyaev-2017} and \cite{bizyaev-2022} from MaNGA, focusing on edge-on galaxies with measurable rotation lags to investigate eDIG properties. This leads to uncertainty regarding whether galaxies lacking lags also lack eDIG or if eDIG is present but not detected via kinematic measurements, resulting in a segregation within these samples. 
This complexity highlights the need to conduct extensive studies combining both morphological and kinematic indicators to fully understand the presence and characteristics of eDIG in galaxies.

To explore the physical processes behind disc–halo interactions, studies like \cite{ho-2016}, \citet{bizyaev-2017, bizyaev-2022}, and \cite{levy-2019}, we examined the link between lag amplitude and galaxy properties. In contrast to our results, both \citet{bizyaev-2017, bizyaev-2022} reported a positive correlation between the lag and both the stellar mass and the central velocity dispersion. Specifically, \cite{bizyaev-2022} found a cc=0.45 for the relation between lag and stellar mass, while we observed a cc=-0.47, which is notably different and can be considered low. This discrepancy may be linked to statistical biases due to our smaller sample size compared to the much larger MaNGA sample used by \cite{bizyaev-2022}. Additionally, the galaxies in our sample, as described in \citetalias{sardaneta-2024}, are primarily inhabited by young stars detected through UV emission. This suggests that a more reliable comparison would require an estimation of the stellar mass using spectral energy distribution analysis rather than relying on IR-based calculations.

\cite{bizyaev-2017} argued that this not necessarily indicated that there was an external source of ionization of the gas at high altitudes. However, in isolated galaxies like the ones in our sample, it is unlikely that there is such external source. Therefore, we searched for a relation between the lag and a parameter providing of a picture of the environment around the isolated galaxies. From the analysis to  the correlation between the lag and the tidal strength parameter  \citep[$Q_k$, ][]{verley-2007, verley-2007-i, Argudo-F-2013} a possible negative correlation between the lag and tidal strength emerges. However, this apparent correlation is called into question by possible inconsistencies in galaxy isolation within the AMIGA sample. The review by \cite{Argudo-F-2013} using more reliable data may reaffirm the robustness of the tidal strength parameter but highlights the challenge of interpreting these trends with a limited sample size. Expanding the sample could clarify the apparent inconsistency and strengthen the statistical significance of any potential trends.

In \citetalias{sardaneta-2024}, eDIG traced by \Ha emission was detected vertically in 11 out of 14 galaxies, while extraplanar UV emission was observed in 13 out of 14 galaxies suggesting that star formation extends well beyond the disc defined by the \Ha emission. Despite this high fraction of eDIG, the comparison with the generic sample of inclined spirals of \cite{rossa-2003-i} indicated that the phenomenon is uncorrelated to the immediate galaxy environment. The kinematic findings described above suggest that internal processes alone do not account for the eDIG presence. However, assuming that galaxies in our sample have not suffered any interaction since at least one Gyr, the apparent discrepancies between the presence of eDIG and the kinematic signatures may be reconciled by considering the interaction with the Cicumgalactic Medium (CGM).

The CGM contains significant amounts of gas, metals, and dust, that can influence galaxy evolution through processes like gas accretion and feedback \citep[e.g.][]{stewart-2011, ho-2016}. 
Thus, \cite{bizyaev-2017} suggested that accretion from the intergalactic medium (IGM) might not lead to symmetric velocity lags, whereas interaction with hot halo gas might, this scenario is consistent with our observations of varied lag gradients and regions of high velocity dispersion. In addition, \cite{marasco-2019}, \cite{levy-2019} and \cite{bizyaev-2022} emphasized the role of isotropic gas accretion from the CGM in shaping the observed rotation velocity lags. The influence of the CGM can mask or convolve the effects of internal processes, such as galactic fountains, making it challenging to distinguish between different origins of eDIG based solely on radial variations in the lag. The presence of eDIG in galaxies without clear kinematic signatures of inflow and the varied velocity dispersion maps suggest that interactions with the CGM are a significant factor in generating the observed eDIG and extended UV discs.

\section{Conclusions}\label{Sec:Conc}

In this work, we investigated the kinematic and environmental properties of galaxies in a representative sample of 14 nearby, late-type, highly inclined ($i\geq 80^{\circ}$), isolated galaxies from the CIG catalogue, whose \Ha emission was observed using scanning Fabry-Perot interferometry. The sample was presented previously in  \citetalias{sardaneta-2024}, where a photometric multiwavelength analysis (NIR, optical and UV) was performed finding extraplanar \Ha emission (or eDIG) in 11 out of 14 galaxies, as well as  extraplanar UV emission in 13 out of 14 galaxies. In \citetalias{sardaneta-2024}, we concluded that the phenomenon is uncorrelated to the galaxy environment and that the nature of these haloes in galaxies are subject to a complex interplay of factors, including SF, ionization processes, and possible CGM or ICM contributions. However, the kinematic characteristics of extraplanar components which provide a powerful tool for studying the perturbation they may be undergoing was missing in \citetalias{sardaneta-2024} analysis. Therefore, the complementary analysis we have carried out in this paper revealed several key findings regarding the incidence of eDIG:

\begin{itemize}

\item Radial velocity and velocity dispersion fields and rotation curves:

\begin{enumerate}

%% rotation curves
\item Different techniques were employed to compute the rotation curves, specifically correcting for LoS affects due to inclination. The resulting rotational velocities were validated against \ion{H}{i} data from \cite{Jones2018} and confirmed through B- and K-band Tully-Fisher relationships \citep[][]{epinat-2008-ii, torres-flores-2011}. These comparisons confirmed the accuracy of our measurements and underscored the critical importance of correcting for LoS effects in determining precise rotation curves.

\medskip

\item  For our entire sample of 14 galaxies, the LoS velocity fields exhibit regular rotation, which is typical of disc galaxies. Due to the dust obscuring the \Ha emission and projection effects, in most of our galaxies, we quantitatively observed that most of the rotation curves display solid body shapes.

\medskip

\item  The velocity dispersion maps are generally uniform across the entire disc of each galaxy, ranging from 20 to 35~\kms, which is typical of isolated galaxies without significant shock features from galactic interactions. However, specific patterns were observed: some galaxies exhibited high values across the galactic disc, others around the \Ha brightest knots and some in others in the innermost regions.

\end{enumerate}

\item Lag in rotation:

\begin{enumerate}

\item A rotational lag along the $z$-axis, with an average of approximately $\Delta V/\Delta z=32.0\pm10.6$~\kmspc, was detected in 7 out of 14 galaxies, consistent with previous studies  \citep[][]{fraternali-2005-891, heald-2006, heald-2006-b, heald-2007, Zschaechner-2015-a, ho-2016, bizyaev-2017, bizyaev-2022, levy-2019, marasco-2019}. However, one of these galaxies, CIG~171, showed no emission of extraplanar \Ha in its morphology \citepalias[see][]{sardaneta-2024}. Since morphological signatures of eDIG without the presence of a rotational lag have also been reported in the literature   \citep[see e.g.][]{levy-2019}, we conclude that the lag may not be a definitive marker for eDIG. 

\medskip

\item The galaxies that exhibited a rotational lag showed also a gradient depending on the radius. This suggests that the eDIG either has an internal origin or results from the accretion of cold material from streams. However, since the PVDs parallel to the minor axis did not show specific features of accretion or outflows, we infer that the interaction with the CGM may be the primary factor contributing to the presence of the eDIG.

\medskip

\item Our data suggest a potential correlation between lag and tidal strength. Confirming this relationship requires a larger sample. In agreement with some authors \citep[e.g.][]{levy-2019, bizyaev-2017, bizyaev-2022}, we did not find correlations between the amplitude  of the rotation lag and the physical properties of the galaxies (dynamical mass, stellar mass, maximum rotation velocity, dispersion velocity, FIR luminosity). However, to draw a definitive conclusion, estimating stellar mass using spectral energy distribution is necessary.  %% lag vs. properties

\end{enumerate}

\item Specific cases:

\begin{enumerate}

\item The extended \Ha emission disc of CIG~201 displayed isovelocities typical of galaxies with a steeply rising-to-flat rotation curve, as clearly observed from the rotation curve in the PVD's envelope. Its multiwavelength maps  \citepalias[see][]{sardaneta-2024} and the highest  \Ha velocity dispersion values indicate a cone-shape in the central regions. Given the negative gradient in the lag vs. radius, we used the PVDs parallel to the minor axis to confirm that this galaxy has an inflow.

\end{enumerate}

\end{itemize}

Thus, the presence of eDIG and the kinematic properties observed suggest an interaction with the CGM as a significant factor influencing these galaxies.  
Further research with spectroscopic and high resolution images in different bands would contribute for a better understanding of the interaction of the galactic disc with the CGM and thereby provide insight into the origin of the eDIG. 
In that sense, we are acquiring \texttt{Astrosat-UVIT} \citep{tandon-2017} imagery in the FUV band and long-slit spectroscopic data of our galaxies, to be presented in a forthcoming paper.

\section*{Acknowledgements}
Based on observations taken with the GHASP spectrograph at the Observatoire de Haute Provence (OHP, France), operated by the French CNRS. The authors acknowledge the technical assistance provided by the late Olivier Boissin \Cross\, from LAM and the OHP team before and during the observations, namely the night team: Jean Balcaen, St\'ephane Favard, Jean-Pierre Troncin, Didier Gravallon and the day team led by Fran\c{c}ois Moreau.  
This research has made use of the NASA/IPAC Extragalactic Database (NED), which is operated by the Jet Propulsion Laboratory, California Institute of Technology, under contract with the National Aeronautics and Space Administration. 
This research has made use of the SIMBAD database, operated at CDS, Strasbourg, France. 
M.M.S. thanks the ``Programa de Becas Posdoctorales en la UNAM'' of DGAPA-UNAM. 
M.M.S acknowledges partial support from FONDECYT through grant No~3250253. 
M.M.S. and M.R. acknowledges the project CONACyT~CF-86367. 
I.F.C. acknowledges SIP project 20241455. 
We especially thank Carlos L\'opez-Cob\'a for his helpful comments and assistance in applying the XookSuut 3D model (\url{https://github.com/CarlosCoba/XS3D}) in the context of the Intensity Peak Method (IPM). 
We gratefully acknowledge the anonymous referee for their insightful and constructive comments, which have substantially enhanced the quality and clarity of the manuscript. 
In this work, we used the software: 
\texttt{Astropy}~v5.2.2 \citep{astropy:2013, astropy:2018, astropy:2022}; 
\texttt{Numpy}~v1.24.3 \citep{numpy-harris2020array}; 
\texttt{Matplotlib}~v3.7.1 \citep{matplotlib-Hunter:2007}; 
\texttt{Scipy}~v1.10.1 \citep{2020SciPy-NMeth}; 
\texttt{Lmfit}~v1.2.1 \citep{lmfit-2016ascl.soft06014N}; 
\texttt{Scikit-learn (sklearn)}~v1.3.2 \citep{scikit-learn}; 
\textsc{3D Tilted Ring Fitting Code} \citep[\textsc{TiRiFiC} by][]{jozsa-2007-tirific}; 
\textsc{Python Fully Automated TiRiFiC Code} \citep[\textsc{Py-FAT} by][]{kamphuis-2015-pyFat-i, kamphuis-2024-pyFat-software}.

%%%%%%%%%%%%%%%%%%%%%%%%%%%%%%%%%%%%%%%%%%%%%%%%%%
\section*{Data Availability}
%%The inclusion of a Data Availability Statement is a requirement for articles published in MNRAS. Data Availability Statements provide a standardised format for readers to understand the availability of data underlying the research results described in the article. The statement may refer to original data generated in the course of the study or to third-party data analysed in the article. The statement should describe and provide means of access, where possible, by linking to the data or providing the required accession numbers for the relevant databases or DOIs.
%The data used in this work can be found online on the data servers of \textit{The STScI Digitized Sky Survey}\footnote{\textit{The STScI Digitized Sky Survey} \url{https://archive.stsci.edu/cgi-bin/dss_form}}, \textit{Two Micron All Sky Survey} (2MASS)\footnote{\textit{Two Micron All Sky Survey} (2MASS) \url{https://irsa.ipac.caltech.edu/Missions/2mass.html}} and the \textit{Galaxy Evolution Explorer} (GALEX) \footnote{\textit{Galaxy Evolution Explorer} (GALEX) \url{https://archive.stsci.edu/missions-and-data/galex}}. 
The FP data underlying this article will be shared on reasonable request to the corresponding author.



\bibliographystyle{mnras}
\bibliography{biblio_2024} % if your bibtex file is called example.bib

\begin{thebibliography}{}
\makeatletter
\relax
\def\mn@urlcharsother{\let\do\@makeother \do\$\do\&\do\#\do\^\do\_\do\%\do\~}
\def\mn@doi{\begingroup\mn@urlcharsother \@ifnextchar [ {\mn@doi@}
  {\mn@doi@[]}}
\def\mn@doi@[#1]#2{\def\@tempa{#1}\ifx\@tempa\@empty \href
  {http://dx.doi.org/#2} {doi:#2}\else \href {http://dx.doi.org/#2} {#1}\fi
  \endgroup}
\def\mn@eprint#1#2{\mn@eprint@#1:#2::\@nil}
\def\mn@eprint@arXiv#1{\href {http://arxiv.org/abs/#1} {{\tt arXiv:#1}}}
\def\mn@eprint@dblp#1{\href {http://dblp.uni-trier.de/rec/bibtex/#1.xml}
  {dblp:#1}}
\def\mn@eprint@#1:#2:#3:#4\@nil{\def\@tempa {#1}\def\@tempb {#2}\def\@tempc
  {#3}\ifx \@tempc \@empty \let \@tempc \@tempb \let \@tempb \@tempa \fi \ifx
  \@tempb \@empty \def\@tempb {arXiv}\fi \@ifundefined
  {mn@eprint@\@tempb}{\@tempb:\@tempc}{\expandafter \expandafter \csname
  mn@eprint@\@tempb\endcsname \expandafter{\@tempc}}}

\bibitem[\protect\citeauthoryear{{Amram}, {Boulesteix}, {Georgelin},
  {Georgelin}, {Laval}, {Le Coarer}, {Marcelin}  \& {Rosado}}{{Amram}
  et~al.}{1991}]{amram-1991}
{Amram} P.,  {Boulesteix} J.,  {Georgelin} Y.~M.,  {Georgelin} Y.~P.,  {Laval}
  A.,  {Le Coarer} E.,  {Marcelin} M.,   {Rosado} M.,  1991, The Messenger,
  \href {https://ui.adsabs.harvard.edu/abs/1991Msngr..64...44A} {64, 44}

\bibitem[\protect\citeauthoryear{{Amram}, {Balkowski}, {Boulesteix}, {Cayatte},
  {Marcelin}  \& {Sullivan}}{{Amram} et~al.}{1996}]{amram-1996}
{Amram} P.,  {Balkowski} C.,  {Boulesteix} J.,  {Cayatte} V.,  {Marcelin} M.,
  {Sullivan} W.~T. I.,  1996, \aap, \href
  {https://ui.adsabs.harvard.edu/abs/1996A&A...310..737A} {310, 737}

\bibitem[\protect\citeauthoryear{{Argudo-Fern{\'a}ndez}
  et~al.,}{{Argudo-Fern{\'a}ndez} et~al.}{2013}]{Argudo-F-2013}
{Argudo-Fern{\'a}ndez} M.,  et~al., 2013, \mn@doi [\aap]
  {10.1051/0004-6361/201321326}, \href
  {https://ui.adsabs.harvard.edu/abs/2013A&A...560A...9A} {560, A9}

\bibitem[\protect\citeauthoryear{{Astropy Collaboration} et~al.,}{{Astropy
  Collaboration} et~al.}{2013}]{astropy:2013}
{Astropy Collaboration} et~al., 2013, \mn@doi [\aap]
  {10.1051/0004-6361/201322068}, \href
  {http://adsabs.harvard.edu/abs/2013A%26A...558A..33A} {558, A33}

\bibitem[\protect\citeauthoryear{{Astropy Collaboration} et~al.,}{{Astropy
  Collaboration} et~al.}{2018}]{astropy:2018}
{Astropy Collaboration} et~al., 2018, \mn@doi [\aj] {10.3847/1538-3881/aabc4f},
  \href {https://ui.adsabs.harvard.edu/abs/2018AJ....156..123A} {156, 123}

\bibitem[\protect\citeauthoryear{{Astropy Collaboration} et~al.,}{{Astropy
  Collaboration} et~al.}{2022}]{astropy:2022}
{Astropy Collaboration} et~al., 2022, \mn@doi [\apj]
  {10.3847/1538-4357/ac7c74}, \href
  {https://ui.adsabs.harvard.edu/abs/2022ApJ...935..167A} {935, 167}

\bibitem[\protect\citeauthoryear{{Becquaert} \& {Combes}}{{Becquaert} \&
  {Combes}}{1997}]{becquaert-1997}
{Becquaert} J.~F.,  {Combes} F.,  1997, \aap, \href
  {https://ui.adsabs.harvard.edu/abs/1997A&A...325...41B} {325, 41}

\bibitem[\protect\citeauthoryear{{Bizyaev} et~al.,}{{Bizyaev}
  et~al.}{2017}]{bizyaev-2017}
{Bizyaev} D.,  et~al., 2017, \mn@doi [\apj] {10.3847/1538-4357/aa6979}, \href
  {https://ui.adsabs.harvard.edu/abs/2017ApJ...839...87B} {839, 87}

\bibitem[\protect\citeauthoryear{{Bizyaev}, {Walterbos}, {Chen}, {Drory},
  {Lane}, {Brownstein}  \& {Riffel}}{{Bizyaev} et~al.}{2022}]{bizyaev-2022}
{Bizyaev} D.,  {Walterbos} R. A.~M.,  {Chen} Y.-M.,  {Drory} N.,  {Lane} R.~R.,
   {Brownstein} J.~R.,   {Riffel} R.~A.,  2022, \mn@doi [\mnras]
  {10.1093/mnras/stac1806}, \href
  {https://ui.adsabs.harvard.edu/abs/2022MNRAS.515.1598B} {515, 1598}

\bibitem[\protect\citeauthoryear{{Bregman}}{{Bregman}}{1980}]{bregman-1980}
{Bregman} J.~N.,  1980, \mn@doi [\apj] {10.1086/157776}, \href
  {https://ui.adsabs.harvard.edu/abs/1980ApJ...236..577B} {236, 577}

\bibitem[\protect\citeauthoryear{{Burstein}}{{Burstein}}{1979}]{burstein-1979}
{Burstein} D.,  1979, \mn@doi [\apj] {10.1086/157563}, \href
  {https://ui.adsabs.harvard.edu/abs/1979ApJ...234..829B} {234, 829}

\bibitem[\protect\citeauthoryear{{C{\'a}rdenas-Mart{\'\i}nez} \&
  {Fuentes-Carrera}}{{C{\'a}rdenas-Mart{\'\i}nez} \&
  {Fuentes-Carrera}}{2018}]{cardenas-2018}
{C{\'a}rdenas-Mart{\'\i}nez} N.,  {Fuentes-Carrera} I.,  2018, \mn@doi [\apj]
  {10.3847/1538-4357/aae891}, \href
  {https://ui.adsabs.harvard.edu/abs/2018ApJ...868..141C} {868, 141}

\bibitem[\protect\citeauthoryear{{Catinella}, {Giovanelli}  \&
  {Haynes}}{{Catinella} et~al.}{2006}]{catinella-2006}
{Catinella} B.,  {Giovanelli} R.,   {Haynes} M.~P.,  2006, \mn@doi [\apj]
  {10.1086/500171}, \href
  {https://ui.adsabs.harvard.edu/abs/2006ApJ...640..751C} {640, 751}

\bibitem[\protect\citeauthoryear{{Collins}, {Benjamin}  \& {Rand}}{{Collins}
  et~al.}{2002}]{collins-2002}
{Collins} J.~A.,  {Benjamin} R.~A.,   {Rand} R.~J.,  2002, \mn@doi [\apj]
  {10.1086/342309}, \href
  {https://ui.adsabs.harvard.edu/abs/2002ApJ...578...98C} {578, 98}

\bibitem[\protect\citeauthoryear{{Courteau}}{{Courteau}}{1997}]{courteau-1997}
{Courteau} S.,  1997, \mn@doi [\aj] {10.1086/118656}, \href
  {https://ui.adsabs.harvard.edu/abs/1997AJ....114.2402C} {114, 2402}

\bibitem[\protect\citeauthoryear{{Court{\`e}s}}{{Court{\`e}s}}{1989}]{courtes-1989}
{Court{\`e}s} G.,  1989, in {Tenorio-Tagle} G.,  {Moles} M.,   {Melnick} J.,
  eds, Lecture Notes in Physics, Vol.~350, IAU Colloq. 120: Structure and
  Dynamics of the Interstellar Medium.
Springer, Berlin, Heidelberg, p.~80, \mn@doi{10.1007/BFb0114843}

\bibitem[\protect\citeauthoryear{{Daigle}, {Carignan}, {Amram}, {Hernandez},
  {Chemin}, {Balkowski}  \& {Kennicutt}}{{Daigle}
  et~al.}{2006a}]{daigle-2006-i}
{Daigle} O.,  {Carignan} C.,  {Amram} P.,  {Hernandez} O.,  {Chemin} L.,
  {Balkowski} C.,   {Kennicutt} R.,  2006a, \mn@doi [\mnras]
  {10.1111/j.1365-2966.2006.10002.x}, \href
  {https://ui.adsabs.harvard.edu/abs/2006MNRAS.367..469D} {367, 469}

\bibitem[\protect\citeauthoryear{{Daigle}, {Carignan}, {Hernandez}, {Chemin}
  \& {Amram}}{{Daigle} et~al.}{2006b}]{daigle-2006-fp}
{Daigle} O.,  {Carignan} C.,  {Hernandez} O.,  {Chemin} L.,   {Amram} P.,
  2006b, \mn@doi [\mnras] {10.1111/j.1365-2966.2006.10216.x}, \href
  {https://ui.adsabs.harvard.edu/abs/2006MNRAS.368.1016D} {368, 1016}

\bibitem[\protect\citeauthoryear{{Dettmar}}{{Dettmar}}{1993}]{dettmar-1993}
{Dettmar} R.~J.,  1993, Reviews in Modern Astronomy, \href
  {https://ui.adsabs.harvard.edu/abs/1993RvMA....6...33D} {6, 33}

\bibitem[\protect\citeauthoryear{{Drew}, {Casey}, {Burnham}, {Hung}, {Kassin},
  {Simons}  \& {Zavala}}{{Drew} et~al.}{2018}]{Drew-2018ApJ...869...58D}
{Drew} P.~M.,  {Casey} C.~M.,  {Burnham} A.~D.,  {Hung} C.-L.,  {Kassin} S.~A.,
   {Simons} R.~C.,   {Zavala} J.~A.,  2018, \mn@doi [\apj]
  {10.3847/1538-4357/aaedbf}, \href
  {https://ui.adsabs.harvard.edu/abs/2018ApJ...869...58D} {869, 58}

\bibitem[\protect\citeauthoryear{{Epinat} et~al.,}{{Epinat}
  et~al.}{2008a}]{benoit-2008}
{Epinat} B.,  et~al., 2008a, \mn@doi [\mnras]
  {10.1111/j.1365-2966.2008.13422.x}, \href
  {http://adsabs.harvard.edu/abs/2008MNRAS.388..500E} {388, 500}

\bibitem[\protect\citeauthoryear{{Epinat} et~al.,}{{Epinat}
  et~al.}{2008b}]{epinat-2008-i}
{Epinat} B.,  et~al., 2008b, \mn@doi [\mnras]
  {10.1111/j.1365-2966.2008.13422.x}, \href
  {https://ui.adsabs.harvard.edu/abs/2008MNRAS.388..500E} {388, 500}

\bibitem[\protect\citeauthoryear{{Epinat}, {Amram}  \& {Marcelin}}{{Epinat}
  et~al.}{2008c}]{epinat-2008-ii}
{Epinat} B.,  {Amram} P.,   {Marcelin} M.,  2008c, \mn@doi [\mnras]
  {10.1111/j.1365-2966.2008.13796.x}, \href
  {https://ui.adsabs.harvard.edu/abs/2008MNRAS.390..466E} {390, 466}

\bibitem[\protect\citeauthoryear{{Epinat}, {Amram}, {Balkowski}  \&
  {Marcelin}}{{Epinat} et~al.}{2010}]{epinat-2010}
{Epinat} B.,  {Amram} P.,  {Balkowski} C.,   {Marcelin} M.,  2010, \mn@doi
  [\mnras] {10.1111/j.1365-2966.2009.15688.x}, \href
  {https://ui.adsabs.harvard.edu/abs/2010MNRAS.401.2113E} {401, 2113}

\bibitem[\protect\citeauthoryear{{Eskridge} et~al.,}{{Eskridge}
  et~al.}{2002}]{Eskridge-2002}
{Eskridge} P.~B.,  et~al., 2002, \mn@doi [\apjs] {10.1086/342340}, \href
  {http://adsabs.harvard.edu/abs/2002ApJS..143...73E} {143, 73}

\bibitem[\protect\citeauthoryear{{Flores-Fajardo}, {Morisset}, {Stasi{\'n}ska}
  \& {Binette}}{{Flores-Fajardo} et~al.}{2011}]{FloresFajardo-2011}
{Flores-Fajardo} N.,  {Morisset} C.,  {Stasi{\'n}ska} G.,   {Binette} L.,
  2011, \mn@doi [\mnras] {10.1111/j.1365-2966.2011.18848.x}, \href
  {https://ui.adsabs.harvard.edu/abs/2011MNRAS.415.2182F} {415, 2182}

\bibitem[\protect\citeauthoryear{{Fraternali} \& {Binney}}{{Fraternali} \&
  {Binney}}{2006}]{fraternali-2006}
{Fraternali} F.,  {Binney} J.~J.,  2006, \mn@doi [\mnras]
  {10.1111/j.1365-2966.2005.09816.x}, \href
  {https://ui.adsabs.harvard.edu/abs/2006MNRAS.366..449F} {366, 449}

\bibitem[\protect\citeauthoryear{{Fraternali} \& {Binney}}{{Fraternali} \&
  {Binney}}{2008}]{fraternali-binney-2008}
{Fraternali} F.,  {Binney} J.~J.,  2008, \mn@doi [\mnras]
  {10.1111/j.1365-2966.2008.13071.x}, \href
  {https://ui.adsabs.harvard.edu/abs/2008MNRAS.386..935F} {386, 935}

\bibitem[\protect\citeauthoryear{{Fraternali}, {Oosterloo}  \&
  {Sancisi}}{{Fraternali} et~al.}{2004}]{fraternali-2004}
{Fraternali} F.,  {Oosterloo} T.,   {Sancisi} R.,  2004, \mn@doi [\aap]
  {10.1051/0004-6361:20040529}, \href
  {https://ui.adsabs.harvard.edu/abs/2004A&A...424..485F} {424, 485}

\bibitem[\protect\citeauthoryear{{Fraternali}, {Oosterloo}, {Sancisi}  \&
  {Swaters}}{{Fraternali} et~al.}{2005a}]{fraternali-2005}
{Fraternali} F.,  {Oosterloo} T.~A.,  {Sancisi} R.,   {Swaters} R.,  2005a, in
  {Braun} R.,  ed.,  Astronomical Society of the Pacific Conference Series Vol.
  331, Extra-Planar Gas. p.~239 (\mn@eprint {arXiv} {astro-ph/0410375}),
  \mn@doi{10.48550/arXiv.astro-ph/0410375}

\bibitem[\protect\citeauthoryear{{Fraternali}, {Oosterloo}, {Sancisi}  \&
  {Swaters}}{{Fraternali} et~al.}{2005b}]{fraternali-2005-891}
{Fraternali} F.,  {Oosterloo} T.~A.,  {Sancisi} R.,   {Swaters} R.,  2005b, in
  {Braun} R.,  ed.,  Astronomical Society of the Pacific Conference Series Vol.
  331, Extra-Planar Gas. p.~239 (\mn@eprint {arXiv} {astro-ph/0410375})

\bibitem[\protect\citeauthoryear{{Fuentes-Carrera} et~al.,}{{Fuentes-Carrera}
  et~al.}{2004}]{fuentesc-2004}
{Fuentes-Carrera} I.,  et~al., 2004, \mn@doi [\aap]
  {10.1051/0004-6361:20034190}, \href
  {http://adsabs.harvard.edu/abs/2004A%26A...415..451F} {415, 451}

\bibitem[\protect\citeauthoryear{{Fuentes-Carrera}, {Rosado}, {Amram}, {Salo}
  \& {Laurikainen}}{{Fuentes-Carrera} et~al.}{2007}]{fuentes-2007}
{Fuentes-Carrera} I.,  {Rosado} M.,  {Amram} P.,  {Salo} H.,   {Laurikainen}
  E.,  2007, \mn@doi [\aap] {10.1051/0004-6361:20077071}, \href
  {https://ui.adsabs.harvard.edu/abs/2007A&A...466..847F} {466, 847}

\bibitem[\protect\citeauthoryear{{Garc{\'{\i}}a-Ruiz}, {Sancisi}  \&
  {Kuijken}}{{Garc{\'{\i}}a-Ruiz} et~al.}{2002}]{garcia-ruiz-2002}
{Garc{\'{\i}}a-Ruiz} I.,  {Sancisi} R.,   {Kuijken} K.,  2002, \mn@doi [\aap]
  {10.1051/0004-6361:20020976}, \href
  {http://adsabs.harvard.edu/abs/2002A%26A...394..769G} {394, 769}

\bibitem[\protect\citeauthoryear{{Garrido}, {Marcelin}, {Amram}  \&
  {Boulesteix}}{{Garrido} et~al.}{2002}]{garrido-2002}
{Garrido} O.,  {Marcelin} M.,  {Amram} P.,   {Boulesteix} J.,  2002, \mn@doi
  [\aap] {10.1051/0004-6361:20020479}, \href
  {https://ui.adsabs.harvard.edu/abs/2002A&A...387..821G} {387, 821}

\bibitem[\protect\citeauthoryear{{Giovanelli} \& {Haynes}}{{Giovanelli} \&
  {Haynes}}{2002}]{giovalnelli-haynes-2002}
{Giovanelli} R.,  {Haynes} M.~P.,  2002, \mn@doi [\apjl] {10.1086/341368},
  \href {https://ui.adsabs.harvard.edu/abs/2002ApJ...571L.107G} {571, L107}

\bibitem[\protect\citeauthoryear{{G{\'o}mez-L{\'o}pez}
  et~al.,}{{G{\'o}mez-L{\'o}pez} et~al.}{2019}]{gomezl-2019}
{G{\'o}mez-L{\'o}pez} J.~A.,  et~al., 2019, \mn@doi [\aap]
  {10.1051/0004-6361/201935869}, \href
  {https://ui.adsabs.harvard.edu/abs/2019A&A...631A..71G} {631, A71}

\bibitem[\protect\citeauthoryear{{Gonz{\'a}lez-D{\'\i}az}
  et~al.,}{{Gonz{\'a}lez-D{\'\i}az} et~al.}{2024a}]{gonzalezDiaz-2024-betis-i}
{Gonz{\'a}lez-D{\'\i}az} R.,  et~al., 2024a, \mn@doi [\aap]
  {10.1051/0004-6361/202348453}, \href
  {https://ui.adsabs.harvard.edu/abs/2024A&A...687A..20G} {687, A20}

\bibitem[\protect\citeauthoryear{{Gonz{\'a}lez-D{\'\i}az}, {Rosales-Ortega}  \&
  {Galbany}}{{Gonz{\'a}lez-D{\'\i}az}
  et~al.}{2024b}]{gonzalezDiaz-2024-betis-ii}
{Gonz{\'a}lez-D{\'\i}az} R.,  {Rosales-Ortega} F.~F.,   {Galbany} L.,  2024b,
  \mn@doi [\aap] {10.1051/0004-6361/202451240}, \href
  {https://ui.adsabs.harvard.edu/abs/2024A&A...691A..25G} {691, A25}

\bibitem[\protect\citeauthoryear{{Gooch}}{{Gooch}}{1996}]{gooch-1996}
{Gooch} R.,  1996, in {Jacoby} G.~H.,  {Barnes} J.,  eds,  Astronomical Society
  of the Pacific Conference Series Vol. 101, Astronomical Data Analysis
  Software and Systems V. p.~80

\bibitem[\protect\citeauthoryear{Harris et~al.,}{Harris
  et~al.}{2020}]{numpy-harris2020array}
Harris C.~R.,  et~al., 2020, \mn@doi [Nature] {10.1038/s41586-020-2649-2}, 585,
  357

\bibitem[\protect\citeauthoryear{{Heald}, {Rand}, {Benjamin}, {Collins}  \&
  {Bland-Hawthorn}}{{Heald} et~al.}{2006a}]{heald-2006}
{Heald} G.~H.,  {Rand} R.~J.,  {Benjamin} R.~A.,  {Collins} J.~A.,
  {Bland-Hawthorn} J.,  2006a, \mn@doi [\apj] {10.1086/497902}, \href
  {http://adsabs.harvard.edu/abs/2006ApJ...636..181H} {636, 181}

\bibitem[\protect\citeauthoryear{{Heald}, {Rand}, {Benjamin}  \&
  {Bershady}}{{Heald} et~al.}{2006b}]{heald-2006-b}
{Heald} G.~H.,  {Rand} R.~J.,  {Benjamin} R.~A.,   {Bershady} M.~A.,  2006b,
  \mn@doi [\apj] {10.1086/505464}, \href
  {https://ui.adsabs.harvard.edu/abs/2006ApJ...647.1018H} {647, 1018}

\bibitem[\protect\citeauthoryear{{Heald}, {Rand}, {Benjamin}  \&
  {Bershady}}{{Heald} et~al.}{2007}]{heald-2007}
{Heald} G.~H.,  {Rand} R.~J.,  {Benjamin} R.~A.,   {Bershady} M.~A.,  2007,
  \mn@doi [\apj] {10.1086/518087}, \href
  {https://ui.adsabs.harvard.edu/\#abs/2007ApJ...663..933H} {663, 933}

\bibitem[\protect\citeauthoryear{{Hern{\'a}ndez-Toledo}, {V{\'a}zquez-Mata},
  {Mart{\'\i}nez-V{\'a}zquez}, {Choi}  \& {Park}}{{Hern{\'a}ndez-Toledo}
  et~al.}{2010}]{hernandez-toledo-2010}
{Hern{\'a}ndez-Toledo} H.~M.,  {V{\'a}zquez-Mata} J.~A.,
  {Mart{\'\i}nez-V{\'a}zquez} L.~A.,  {Choi} Y.-Y.,   {Park} C.,  2010, \mn@doi
  [\aj] {10.1088/0004-6256/139/6/2525}, \href
  {https://ui.adsabs.harvard.edu/abs/2010AJ....139.2525H} {139, 2525}

\bibitem[\protect\citeauthoryear{{Ho} et~al.,}{{Ho} et~al.}{2016}]{ho-2016}
{Ho} I.~T.,  et~al., 2016, \mn@doi [\mnras] {10.1093/mnras/stw017}, \href
  {https://ui.adsabs.harvard.edu/\#abs/2016MNRAS.457.1257H} {457, 1257}

\bibitem[\protect\citeauthoryear{Hunter}{Hunter}{2007}]{matplotlib-Hunter:2007}
Hunter J.~D.,  2007, \mn@doi [Computing in Science \& Engineering]
  {10.1109/MCSE.2007.55}, 9, 90

\bibitem[\protect\citeauthoryear{{Jarrett}, {Chester}, {Cutri}, {Schneider}  \&
  {Huchra}}{{Jarrett} et~al.}{2003}]{jarrett-2003}
{Jarrett} T.~H.,  {Chester} T.,  {Cutri} R.,  {Schneider} S.~E.,   {Huchra}
  J.~P.,  2003, \mn@doi [\aj] {10.1086/345794}, \href
  {https://ui.adsabs.harvard.edu/abs/2003AJ....125..525J} {125, 525}

\bibitem[\protect\citeauthoryear{{Jones} et~al.,}{{Jones}
  et~al.}{2017}]{jones-manga-2017}
{Jones} A.,  et~al., 2017, \mn@doi [\aap] {10.1051/0004-6361/201629802}, \href
  {https://ui.adsabs.harvard.edu/abs/2017A&A...599A.141J} {599, A141}

\bibitem[\protect\citeauthoryear{{Jones} et~al.,}{{Jones}
  et~al.}{2018}]{Jones2018}
{Jones} M.~G.,  et~al., 2018, \mn@doi [\aap] {10.1051/0004-6361/201731448},
  \href {https://ui.adsabs.harvard.edu/abs/2018A&A...609A..17J} {609, A17}

\bibitem[\protect\citeauthoryear{{Joye} \& {Mandel}}{{Joye} \&
  {Mandel}}{2003}]{ds9}
{Joye} W.~A.,  {Mandel} E.,  2003, in {Payne} H.~E.,  {Jedrzejewski} R.~I.,
  {Hook} R.~N.,  eds,  Astronomical Society of the Pacific Conference Series
  Vol. 295, Astronomical Data Analysis Software and Systems XII. p.~489

\bibitem[\protect\citeauthoryear{{J{\'o}zsa}, {Kenn}, {Klein}  \&
  {Oosterloo}}{{J{\'o}zsa} et~al.}{2007}]{jozsa-2007-tirific}
{J{\'o}zsa} G.~I.~G.,  {Kenn} F.,  {Klein} U.,   {Oosterloo} T.~A.,  2007,
  \mn@doi [\aap] {10.1051/0004-6361:20066164}, \href
  {https://ui.adsabs.harvard.edu/abs/2007A&A...468..731J} {468, 731}

\bibitem[\protect\citeauthoryear{{Kamphuis}}{{Kamphuis}}{2024}]{kamphuis-2024-pyFat-software}
{Kamphuis} P.,  2024, {pyFAT: Python Fully Automated TiRiFiC}, Astrophysics
  Source Code Library, record ascl:2407.002

\bibitem[\protect\citeauthoryear{{Kamphuis}, {Peletier}, {Dettmar}, {van der
  Hulst}, {van der Kruit}  \& {Allen}}{{Kamphuis} et~al.}{2007}]{kamphuis-2007}
{Kamphuis} P.,  {Peletier} R.~F.,  {Dettmar} R.~J.,  {van der Hulst} J.~M.,
  {van der Kruit} P.~C.,   {Allen} R.~J.,  2007, \mn@doi [\aap]
  {10.1051/0004-6361:20066989}, \href
  {https://ui.adsabs.harvard.edu/abs/2007A&A...468..951K} {468, 951}

\bibitem[\protect\citeauthoryear{{Kamphuis}, {J{\'o}zsa}, {Oh}, {Spekkens},
  {Urbancic}, {Serra}, {Koribalski}  \& {Dettmar}}{{Kamphuis}
  et~al.}{2015}]{kamphuis-2015-pyFat-i}
{Kamphuis} P.,  {J{\'o}zsa} G.~I.~G.,  {Oh} S. .~H.,  {Spekkens} K.,
  {Urbancic} N.,  {Serra} P.,  {Koribalski} B.~S.,   {Dettmar} R.~J.,  2015,
  \mn@doi [\mnras] {10.1093/mnras/stv1480}, \href
  {https://ui.adsabs.harvard.edu/abs/2015MNRAS.452.3139K} {452, 3139}

\bibitem[\protect\citeauthoryear{{Karachentsev}, {Makarov}, {Karachentseva}  \&
  {Melnyk}}{{Karachentsev} et~al.}{2011}]{Karachentsev-2011}
{Karachentsev} I.~D.,  {Makarov} D.~I.,  {Karachentseva} V.~E.,   {Melnyk}
  O.~V.,  2011, \mn@doi [Astrophysical Bulletin] {10.1134/S1990341311010019},
  \href {https://ui.adsabs.harvard.edu/abs/2011AstBu..66....1K} {66, 1}

\bibitem[\protect\citeauthoryear{{Karachentseva}}{{Karachentseva}}{1973}]{Karachentseva-1973}
{Karachentseva} V.~E.,  1973, Soobshcheniya Spetsial'noj Astrofizicheskoj
  Observatorii, \href {http://adsabs.harvard.edu/abs/1973SoSAO...8....3K} {8}

\bibitem[\protect\citeauthoryear{{Karachentseva}, {Mitronova}, {Melnyk}  \&
  {Karachentsev}}{{Karachentseva} et~al.}{2010}]{Karachentseva-2009}
{Karachentseva} V.~E.,  {Mitronova} S.~N.,  {Melnyk} O.~V.,   {Karachentsev}
  I.~D.,  2010, \mn@doi [Astrophysical Bulletin] {10.1134/S1990341310010013},
  \href {https://ui.adsabs.harvard.edu/abs/2010AstBu..65....1K} {65, 1}

\bibitem[\protect\citeauthoryear{{Kormendy} \& {Illingworth}}{{Kormendy} \&
  {Illingworth}}{1982}]{kormendy-1982}
{Kormendy} J.,  {Illingworth} G.,  1982, \mn@doi [\apj] {10.1086/159923}, \href
  {http://adsabs.harvard.edu/abs/1982ApJ...256..460K} {256, 460}

\bibitem[\protect\citeauthoryear{{Kourkchi}, {Tully}, {Neill}, {Seibert},
  {Courtois}  \& {Dupuy}}{{Kourkchi} et~al.}{2019}]{kourkchi-2019}
{Kourkchi} E.,  {Tully} R.~B.,  {Neill} J.~D.,  {Seibert} M.,  {Courtois}
  H.~M.,   {Dupuy} A.,  2019, \mn@doi [\apj] {10.3847/1538-4357/ab4192}, \href
  {https://ui.adsabs.harvard.edu/abs/2019ApJ...884...82K} {884, 82}

\bibitem[\protect\citeauthoryear{{Lee}, {Irwin}, {Dettmar}, {Cunningham},
  {Golla}  \& {Wang}}{{Lee} et~al.}{2001}]{lee-2001}
{Lee} S.~W.,  {Irwin} J.~A.,  {Dettmar} R.~J.,  {Cunningham} C.~T.,  {Golla}
  G.,   {Wang} Q.~D.,  2001, \mn@doi [\aap] {10.1051/0004-6361:20011046}, \href
  {https://ui.adsabs.harvard.edu/abs/2001A&A...377..759L} {377, 759}

\bibitem[\protect\citeauthoryear{{Lee} et~al.,}{{Lee} et~al.}{2017}]{lee-2017}
{Lee} B.,  et~al., 2017, \mn@doi [\mnras] {10.1093/mnras/stw3162}, \href
  {https://ui.adsabs.harvard.edu/abs/2017MNRAS.466.1382L} {466, 1382}

\bibitem[\protect\citeauthoryear{{Lequeux}}{{Lequeux}}{1983}]{lequeux-1983}
{Lequeux} J.,  1983, \aap, \href
  {https://ui.adsabs.harvard.edu/abs/1983A&A...125..394L} {125, 394}

\bibitem[\protect\citeauthoryear{{Levy} et~al.,}{{Levy}
  et~al.}{2018}]{levy-2018}
{Levy} R.~C.,  et~al., 2018, \mn@doi [\apj] {10.3847/1538-4357/aac2e5}, \href
  {https://ui.adsabs.harvard.edu/abs/2018ApJ...860...92L} {860, 92}

\bibitem[\protect\citeauthoryear{{Levy} et~al.,}{{Levy}
  et~al.}{2019}]{levy-2019}
{Levy} R.~C.,  et~al., 2019, \mn@doi [\apj] {10.3847/1538-4357/ab2ed4}, \href
  {https://ui.adsabs.harvard.edu/abs/2019ApJ...882...84L} {882, 84}

\bibitem[\protect\citeauthoryear{{Li}, {Fraternali}, {Marasco}, {Trager},
  {Pezzulli}, {Mancera Pi{\~n}a}  \& {Verheijen}}{{Li}
  et~al.}{2023}]{li-fraternali-marasco-2023}
{Li} A.,  {Fraternali} F.,  {Marasco} A.,  {Trager} S.~C.,  {Pezzulli} G.,
  {Mancera Pi{\~n}a} P.~E.,   {Verheijen} M. A.~W.,  2023, \mn@doi [\mnras]
  {10.1093/mnras/stad129}, \href
  {https://ui.adsabs.harvard.edu/abs/2023MNRAS.520..147L} {520, 147}

\bibitem[\protect\citeauthoryear{{Marasco} \& {Fraternali}}{{Marasco} \&
  {Fraternali}}{2011}]{marasco-fraternali-2011A&A...525A.134M}
{Marasco} A.,  {Fraternali} F.,  2011, \mn@doi [\aap]
  {10.1051/0004-6361/201015508}, \href
  {https://ui.adsabs.harvard.edu/abs/2011A&A...525A.134M} {525, A134}

\bibitem[\protect\citeauthoryear{{Marasco}, {Fraternali}  \&
  {Binney}}{{Marasco} et~al.}{2012}]{marasco-2012}
{Marasco} A.,  {Fraternali} F.,   {Binney} J.~J.,  2012, \mn@doi [\mnras]
  {10.1111/j.1365-2966.2011.19771.x}, \href
  {https://ui.adsabs.harvard.edu/abs/2012MNRAS.419.1107M} {419, 1107}

\bibitem[\protect\citeauthoryear{{Marasco} et~al.,}{{Marasco}
  et~al.}{2019}]{marasco-2019}
{Marasco} A.,  et~al., 2019, \mn@doi [\aap] {10.1051/0004-6361/201936338},
  \href {https://ui.adsabs.harvard.edu/abs/2019A&A...631A..50M} {631, A50}

\bibitem[\protect\citeauthoryear{{Marino} et~al.,}{{Marino}
  et~al.}{2011}]{marino-2011-mnras}
{Marino} A.,  et~al., 2011, \mn@doi [\mnras]
  {10.1111/j.1365-2966.2010.17684.x}, \href
  {https://ui.adsabs.harvard.edu/abs/2011MNRAS.411..311M} {411, 311}

\bibitem[\protect\citeauthoryear{{Mihalas} \& {Binney}}{{Mihalas} \&
  {Binney}}{1981}]{mihalas}
{Mihalas} D.,  {Binney} J.,  1981, {Galactic astronomy: Structure and
  kinematics /2nd edition/}.
W. H. Freeman, San Francisco

\bibitem[\protect\citeauthoryear{{Miller} \& {Veilleux}}{{Miller} \&
  {Veilleux}}{2003}]{miller-2003}
{Miller} S.~T.,  {Veilleux} S.,  2003, \mn@doi [\apj] {10.1086/375620}, \href
  {http://adsabs.harvard.edu/abs/2003ApJ...592...79M} {592, 79}

\bibitem[\protect\citeauthoryear{{Mo}, {van den Bosch}  \& {White}}{{Mo}
  et~al.}{2010}]{mo}
{Mo} H.,  {van den Bosch} F.~C.,   {White} S.,  2010, {Galaxy Formation and
  Evolution}.
Cambridge University Press, Cambridge

\bibitem[\protect\citeauthoryear{{Moiseev}}{{Moiseev}}{2014}]{moiseev-2014}
{Moiseev} A.~V.,  2014, \mn@doi [Astrophysical Bulletin]
  {10.1134/S1990341314010015}, \href
  {http://adsabs.harvard.edu/abs/2014AstBu..69....1M} {69, 1}

\bibitem[\protect\citeauthoryear{{Moiseev} \& {Egorov}}{{Moiseev} \&
  {Egorov}}{2008}]{moiseev-2008}
{Moiseev} A.~V.,  {Egorov} O.~V.,  2008, \mn@doi [Astrophysical Bulletin]
  {10.1134/S1990341308020089}, \href
  {https://ui.adsabs.harvard.edu/abs/2008AstBu..63..181M} {63, 181}

\bibitem[\protect\citeauthoryear{NIST/SEMATECH}{NIST/SEMATECH}{2023}]{nist-statistics}
NIST/SEMATECH 2023, {e-Handbook of Statistical Methods},
  \mn@doi{10.18434/M32189}, \url {http://www.itl.nist.gov/div898/handbook/}

\bibitem[\protect\citeauthoryear{{Newville}, {Stensitzki}, {Allen}, {Rawlik},
  {Ingargiola}  \& {Nelson}}{{Newville}
  et~al.}{2016}]{lmfit-2016ascl.soft06014N}
{Newville} M.,  {Stensitzki} T.,  {Allen} D.~B.,  {Rawlik} M.,  {Ingargiola}
  A.,   {Nelson} A.,  2016, {Lmfit: Non-Linear Least-Square Minimization and
  Curve-Fitting for Python}, Astrophysics Source Code Library, record
  ascl:1606.014

\bibitem[\protect\citeauthoryear{{Osterbrock} \& {Ferland}}{{Osterbrock} \&
  {Ferland}}{2006}]{osterbrock-2006}
{Osterbrock} D.~E.,  {Ferland} G.~J.,  2006, {Astrophysics of gaseous nebulae
  and active galactic nuclei}, 2 edn.
University Science Books, Sausalito, CA

\bibitem[\protect\citeauthoryear{Pedregosa et~al.,}{Pedregosa
  et~al.}{2011}]{scikit-learn}
Pedregosa F.,  et~al., 2011, Journal of Machine Learning Research, 12, 2825

\bibitem[\protect\citeauthoryear{{Persic}, {Salucci}  \& {Stel}}{{Persic}
  et~al.}{1996}]{persic-1996}
{Persic} M.,  {Salucci} P.,   {Stel} F.,  1996, \mn@doi [\mnras]
  {10.1093/mnras/278.1.27}, \href
  {https://ui.adsabs.harvard.edu/abs/1996MNRAS.281...27P} {281, 27}

\bibitem[\protect\citeauthoryear{{Peters}, {van der Kruit}, {Allen}  \&
  {Freeman}}{{Peters} et~al.}{2017}]{peters-2017}
{Peters} S.~P.~C.,  {van der Kruit} P.~C.,  {Allen} R.~J.,   {Freeman} K.~C.,
  2017, \mn@doi [\mnras] {10.1093/mnras/stw2101}, \href
  {https://ui.adsabs.harvard.edu/abs/2017MNRAS.464...65P} {464, 65}

\bibitem[\protect\citeauthoryear{{R Core Team}}{{R Core
  Team}}{2021}]{R-citation}
{R Core Team} 2021, R: A Language and Environment for Statistical Computing.
R Foundation for Statistical Computing, Vienna, Austria, \url
  {https://www.R-project.org/}

\bibitem[\protect\citeauthoryear{{Rampazzo} et~al.,}{{Rampazzo}
  et~al.}{2016}]{rampazzo-2016}
{Rampazzo} R.,  et~al., 2016, in {D'Onofrio} M.,  {Rampazzo} R.,   {Zaggia} S.,
   eds,  Astrophysics and Space Science Library Vol. 435, From the Realm of the
  Nebulae to Populations of Galaxies. p.~381,
  \mn@doi{10.1007/978-3-319-31006-0_5}

\bibitem[\protect\citeauthoryear{{Rampazzo}, {Omizzolo}, {Uslenghi},
  {Rom{\'a}n}, {Mazzei}, {Verdes-Montenegro}, {Marino}  \& {Jones}}{{Rampazzo}
  et~al.}{2020}]{Rampazzo2020}
{Rampazzo} R.,  {Omizzolo} A.,  {Uslenghi} M.,  {Rom{\'a}n} J.,  {Mazzei} P.,
  {Verdes-Montenegro} L.,  {Marino} A.,   {Jones} M.~G.,  2020, \mn@doi [\aap]
  {10.1051/0004-6361/202038156}, \href
  {https://ui.adsabs.harvard.edu/abs/2020A&A...640A..38R} {640, A38}

\bibitem[\protect\citeauthoryear{{Rampazzo} et~al.,}{{Rampazzo}
  et~al.}{2021}]{rampazzo-2021}
{Rampazzo} R.,  et~al., 2021, \mn@doi [Journal of Astrophysics and Astronomy]
  {10.1007/s12036-021-09690-x}, \href
  {https://ui.adsabs.harvard.edu/abs/2021JApA...42...31R} {42, 31}

\bibitem[\protect\citeauthoryear{{Rand}}{{Rand}}{1996}]{rand-1996}
{Rand} R.~J.,  1996, \mn@doi [\apj] {10.1086/177184}, \href
  {http://adsabs.harvard.edu/abs/1996ApJ...462..712R} {462, 712}

\bibitem[\protect\citeauthoryear{{Rand}}{{Rand}}{1998}]{rand-1998}
{Rand} R.~J.,  1998, \mn@doi [\pasa] {10.1071/AS98106}, \href
  {http://adsabs.harvard.edu/abs/1998PASA...15..106R} {15, 106}

\bibitem[\protect\citeauthoryear{{Rand}}{{Rand}}{2000}]{rand-2000}
{Rand} R.~J.,  2000, \mn@doi [\apjl] {10.1086/312756}, \href
  {https://ui.adsabs.harvard.edu/abs/2000ApJ...537L..13R} {537, L13}

\bibitem[\protect\citeauthoryear{{Rautio}, {Watkins}, {Comer{\'o}n}, {Salo},
  {D{\'\i}az-Garc{\'\i}a}  \& {Janz}}{{Rautio} et~al.}{2022}]{rautio-2022}
{Rautio} R.~P.~V.,  {Watkins} A.~E.,  {Comer{\'o}n} S.,  {Salo} H.,
  {D{\'\i}az-Garc{\'\i}a} S.,   {Janz} J.,  2022, \mn@doi [\aap]
  {10.1051/0004-6361/202142440}, \href
  {https://ui.adsabs.harvard.edu/abs/2022A&A...659A.153R} {659, A153}

\bibitem[\protect\citeauthoryear{{Rela{\~n}o} \& {Beckman}}{{Rela{\~n}o} \&
  {Beckman}}{2005}]{relano-beckman-2005}
{Rela{\~n}o} M.,  {Beckman} J.~E.,  2005, \mn@doi [\aap]
  {10.1051/0004-6361:20041708}, \href
  {https://ui.adsabs.harvard.edu/abs/2005A&A...430..911R} {430, 911}

\bibitem[\protect\citeauthoryear{{Repetto}, {Rosado}, {Gabbasov}  \&
  {Fuentes-Carrera}}{{Repetto} et~al.}{2010}]{rep}
{Repetto} P.,  {Rosado} M.,  {Gabbasov} R.,   {Fuentes-Carrera} I.,  2010,
  \mn@doi [\aj] {10.1088/0004-6256/139/4/1600}, \href
  {http://adsabs.harvard.edu/abs/2010AJ....139.1600R} {139, 1600}

\bibitem[\protect\citeauthoryear{{Reynolds}}{{Reynolds}}{1984}]{reynolds-1984}
{Reynolds} R.~J.,  1984, in NASA Conference Publication. p.~97

\bibitem[\protect\citeauthoryear{{Reynolds}}{{Reynolds}}{1993}]{reynolds-1993}
{Reynolds} R.~J.,  1993, in {Holt} S.~S.,  {Verter} F.,  eds,  American
  Institute of Physics Conference Series Vol. 278, Back to the Galaxy. pp
  156--165, \mn@doi{10.1063/1.44005}

\bibitem[\protect\citeauthoryear{{Rosado}, {Ghosh}  \&
  {Fuentes-Carrera}}{{Rosado} et~al.}{2008}]{rosado-2008}
{Rosado} M.,  {Ghosh} K.~K.,   {Fuentes-Carrera} I.,  2008, \mn@doi [\aj]
  {10.1088/0004-6256/136/1/212}, \href
  {https://ui.adsabs.harvard.edu/abs/2008AJ....136..212R} {136, 212}

\bibitem[\protect\citeauthoryear{{Rosado}, {Gabbasov}, {Repetto},
  {Fuentes-Carrera}, {Amram}, {Martos}  \& {Hernandez}}{{Rosado}
  et~al.}{2013}]{rosado-2013}
{Rosado} M.,  {Gabbasov} R.~F.,  {Repetto} P.,  {Fuentes-Carrera} I.,  {Amram}
  P.,  {Martos} M.,   {Hernandez} O.,  2013, \mn@doi [\aj]
  {10.1088/0004-6256/145/5/135}, \href
  {http://adsabs.harvard.edu/abs/2013AJ....145..135R} {145, 135}

\bibitem[\protect\citeauthoryear{{Rossa} \& {Dettmar}}{{Rossa} \&
  {Dettmar}}{2000}]{rossa-2000}
{Rossa} J.,  {Dettmar} R.-J.,  2000, \aap, \href
  {http://adsabs.harvard.edu/abs/2000A%26A...359..433R} {359, 433}

\bibitem[\protect\citeauthoryear{{Rossa} \& {Dettmar}}{{Rossa} \&
  {Dettmar}}{2003a}]{rossa-2003-i}
{Rossa} J.,  {Dettmar} R.-J.,  2003a, \mn@doi [\aap]
  {10.1051/0004-6361:20030615}, \href
  {http://adsabs.harvard.edu/abs/2003A%26A...406..493R} {406, 493}

\bibitem[\protect\citeauthoryear{{Rossa} \& {Dettmar}}{{Rossa} \&
  {Dettmar}}{2003b}]{rossa-2003-ii}
{Rossa} J.,  {Dettmar} R.-J.,  2003b, \mn@doi [\aap]
  {10.1051/0004-6361:20030698}, \href
  {http://adsabs.harvard.edu/abs/2003A%26A...406..505R} {406, 505}

\bibitem[\protect\citeauthoryear{{Sabater}, {Verdes-Montenegro}, {Leon}, {Best}
   \& {Sulentic}}{{Sabater} et~al.}{2012}]{sabater-2012}
{Sabater} J.,  {Verdes-Montenegro} L.,  {Leon} S.,  {Best} P.,   {Sulentic} J.,
   2012, \mn@doi [\aap] {10.1051/0004-6361/201118692}, \href
  {https://ui.adsabs.harvard.edu/abs/2012A&A...545A..15S} {545, A15}

\bibitem[\protect\citeauthoryear{{Sancisi} \& {Allen}}{{Sancisi} \&
  {Allen}}{1979}]{sancisi-1979}
{Sancisi} R.,  {Allen} R.~J.,  1979, \aap, \href
  {http://adsabs.harvard.edu/abs/1979A%26A....74...73S} {74, 73}

\bibitem[\protect\citeauthoryear{{Sardaneta}, {Rosado}  \&
  {S{\'a}nchez-Cruces}}{{Sardaneta} et~al.}{2020}]{sardaneta-2020}
{Sardaneta} M.~M.,  {Rosado} M.,   {S{\'a}nchez-Cruces} M.,  2020, \mn@doi
  [\rmxaa] {10.22201/ia.01851101p.2020.56.01.09}, \href
  {https://ui.adsabs.harvard.edu/abs/2020RMxAA..56...71S} {56, 71}

\bibitem[\protect\citeauthoryear{{Sardaneta} et~al.,}{{Sardaneta}
  et~al.}{2022}]{sardaneta-2022}
{Sardaneta} M.~M.,  et~al., 2022, \mn@doi [\aap] {10.1051/0004-6361/202142270},
  \href {https://ui.adsabs.harvard.edu/abs/2022A&A...659A..45S} {659, A45}

\bibitem[\protect\citeauthoryear{{Sardaneta}, {Amram}, {Rampazzo}, {Rosado},
  {S{\'a}nchez-Cruces}, {Fuentes-Carrera}  \& {Ghosh}}{{Sardaneta}
  et~al.}{2024}]{sardaneta-2024}
{Sardaneta} M.~M.,  {Amram} P.,  {Rampazzo} R.,  {Rosado} M.,
  {S{\'a}nchez-Cruces} M.,  {Fuentes-Carrera} I.,   {Ghosh} S.,  2024, \mn@doi
  [\mnras] {10.1093/mnras/stae043}, \href
  {https://ui.adsabs.harvard.edu/abs/2024MNRAS.528.2145S} {528, 2145}

\bibitem[\protect\citeauthoryear{Sharp \& Bland-Hawthorn}{Sharp \&
  Bland-Hawthorn}{2010}]{Sharp-2010}
Sharp R.~G.,  Bland-Hawthorn J.,  2010, \mn@doi [The Astrophysical Journal]
  {10.1088/0004-637X/711/2/818}, 711, 818

\bibitem[\protect\citeauthoryear{{Sil'chenko}, {Moiseev}, {Smirnova}  \&
  {Uklein}}{{Sil'chenko} et~al.}{2023}]{silchenko-2023}
{Sil'chenko} O.,  {Moiseev} A.~V.,  {Smirnova} A.,   {Uklein} R.,  2023,
  \mn@doi [Galaxies] {10.3390/galaxies11060119}, \href
  {https://ui.adsabs.harvard.edu/abs/2023Galax..11..119S} {11, 119}

\bibitem[\protect\citeauthoryear{{Sofue} \& {Rubin}}{{Sofue} \&
  {Rubin}}{2001}]{sofue-2001}
{Sofue} Y.,  {Rubin} V.,  2001, \mn@doi [\araa]
  {10.1146/annurev.astro.39.1.137}, \href
  {http://adsabs.harvard.edu/abs/2001ARA%26A..39..137S} {39, 137}

\bibitem[\protect\citeauthoryear{{Sofue}, {Tutui}, {Honma}  \&
  {Tomita}}{{Sofue} et~al.}{1997}]{sofue-1997AJ....114.2428S}
{Sofue} Y.,  {Tutui} Y.,  {Honma} M.,   {Tomita} A.,  1997, \mn@doi [\aj]
  {10.1086/118657}, \href
  {https://ui.adsabs.harvard.edu/abs/1997AJ....114.2428S} {114, 2428}

\bibitem[\protect\citeauthoryear{{Sofue}, {Tomita}, {Tutui}, {Honma}  \&
  {Takeda}}{{Sofue} et~al.}{1998}]{sofue-1998}
{Sofue} Y.,  {Tomita} A.,  {Tutui} Y.,  {Honma} M.,   {Takeda} Y.,  1998,
  \mn@doi [\pasj] {10.1093/pasj/50.5.427}, \href
  {http://adsabs.harvard.edu/abs/1998PASJ...50..427S} {50, 427}

\bibitem[\protect\citeauthoryear{{Stewart}, {Kaufmann}, {Bullock}, {Barton},
  {Maller}, {Diemand}  \& {Wadsley}}{{Stewart} et~al.}{2011}]{stewart-2011}
{Stewart} K.~R.,  {Kaufmann} T.,  {Bullock} J.~S.,  {Barton} E.~J.,  {Maller}
  A.~H.,  {Diemand} J.,   {Wadsley} J.,  2011, \mn@doi [\apj]
  {10.1088/0004-637X/738/1/39}, \href
  {https://ui.adsabs.harvard.edu/abs/2011ApJ...738...39S} {738, 39}

\bibitem[\protect\citeauthoryear{{Swaters}, {Sancisi}  \& {van der
  Hulst}}{{Swaters} et~al.}{1997}]{swaters-1997}
{Swaters} R.~A.,  {Sancisi} R.,   {van der Hulst} J.~M.,  1997, \mn@doi [\apj]
  {10.1086/304958}, \href {http://adsabs.harvard.edu/abs/1997ApJ...491..140S}
  {491, 140}

\bibitem[\protect\citeauthoryear{{Takamiya} \& {Sofue}}{{Takamiya} \&
  {Sofue}}{2002}]{takamiya-2002}
{Takamiya} T.,  {Sofue} Y.,  2002, \mn@doi [\apjl] {10.1086/343028}, \href
  {https://ui.adsabs.harvard.edu/abs/2002ApJ...576L..15T} {576, L15}

\bibitem[\protect\citeauthoryear{{Tandon} et~al.,}{{Tandon}
  et~al.}{2017}]{tandon-2017}
{Tandon} S.~N.,  et~al., 2017, \mn@doi [\aj] {10.3847/1538-3881/aa8451}, \href
  {https://ui.adsabs.harvard.edu/abs/2017AJ....154..128T} {154, 128}

\bibitem[\protect\citeauthoryear{{Tomi{\v{c}}i{\'c}}
  et~al.,}{{Tomi{\v{c}}i{\'c}} et~al.}{2021}]{tomicic-2021}
{Tomi{\v{c}}i{\'c}} N.,  et~al., 2021, \mn@doi [\apj]
  {10.3847/1538-4357/ac230e}, \href
  {https://ui.adsabs.harvard.edu/abs/2021ApJ...922..131T} {922, 131}

\bibitem[\protect\citeauthoryear{{Torres-Flores}, {Epinat}, {Amram}, {Plana}
  \& {Mendes de Oliveira}}{{Torres-Flores} et~al.}{2011}]{torres-flores-2011}
{Torres-Flores} S.,  {Epinat} B.,  {Amram} P.,  {Plana} H.,   {Mendes de
  Oliveira} C.,  2011, \mn@doi [\mnras] {10.1111/j.1365-2966.2011.19169.x},
  \href {https://ui.adsabs.harvard.edu/abs/2011MNRAS.416.1936T} {416, 1936}

\bibitem[\protect\citeauthoryear{{Tully} \& {Fisher}}{{Tully} \&
  {Fisher}}{1977}]{tully-fisher-1977}
{Tully} R.~B.,  {Fisher} J.~R.,  1977, \aap, \href
  {https://ui.adsabs.harvard.edu/abs/1977A&A....54..661T} {54, 661}

\bibitem[\protect\citeauthoryear{{Urrejola-Mora}, {G{\'o}mez}, {Torres-Flores},
  {Amram}, {Epinat}, {Monachesi}, {Marinacci}  \& {de
  Oliveira}}{{Urrejola-Mora} et~al.}{2022}]{UrrejolaM-2022ApJ...935...20U}
{Urrejola-Mora} C.,  {G{\'o}mez} F.~A.,  {Torres-Flores} S.,  {Amram} P.,
  {Epinat} B.,  {Monachesi} A.,  {Marinacci} F.,   {de Oliveira} C.~M.,  2022,
  \mn@doi [\apj] {10.3847/1538-4357/ac78ec}, \href
  {https://ui.adsabs.harvard.edu/abs/2022ApJ...935...20U} {935, 20}

\bibitem[\protect\citeauthoryear{{Valotto} \& {Giovanelli}}{{Valotto} \&
  {Giovanelli}}{2004}]{valotto-2004}
{Valotto} C.,  {Giovanelli} R.,  2004, \mn@doi [\aj] {10.1086/421359}, \href
  {https://ui.adsabs.harvard.edu/abs/2004AJ....128..115V} {128, 115}

\bibitem[\protect\citeauthoryear{{Veilleux}, {Bland-Hawthorn}  \&
  {Cecil}}{{Veilleux} et~al.}{1999a}]{veilleux-1999-ii}
{Veilleux} S.,  {Bland-Hawthorn} J.,   {Cecil} G.,  1999a, \mn@doi [\aj]
  {10.1086/301095}, \href
  {https://ui.adsabs.harvard.edu/abs/1999AJ....118.2108V} {118, 2108}

\bibitem[\protect\citeauthoryear{{Veilleux}, {Bland-Hawthorn}, {Cecil}, {Tully}
   \& {Miller}}{{Veilleux} et~al.}{1999b}]{veilleux-1999}
{Veilleux} S.,  {Bland-Hawthorn} J.,  {Cecil} G.,  {Tully} R.~B.,   {Miller}
  S.~T.,  1999b, \mn@doi [\apj] {10.1086/307453}, \href
  {http://adsabs.harvard.edu/abs/1999ApJ...520..111V} {520, 111}

\bibitem[\protect\citeauthoryear{{Veilleux}, {Cecil}  \&
  {Bland-Hawthorn}}{{Veilleux} et~al.}{2005}]{veilleux-2005}
{Veilleux} S.,  {Cecil} G.,   {Bland-Hawthorn} J.,  2005, \mn@doi [\araa]
  {10.1146/annurev.astro.43.072103.150610}, \href
  {https://ui.adsabs.harvard.edu/abs/2005ARA&A..43..769V} {43, 769}

\bibitem[\protect\citeauthoryear{{Verdes-Montenegro}, {Sulentic}, {Lisenfeld},
  {Leon}, {Espada}, {Garcia}, {Sabater}  \& {Verley}}{{Verdes-Montenegro}
  et~al.}{2005}]{verdesm-2005}
{Verdes-Montenegro} L.,  {Sulentic} J.,  {Lisenfeld} U.,  {Leon} S.,  {Espada}
  D.,  {Garcia} E.,  {Sabater} J.,   {Verley} S.,  2005, \mn@doi [\aap]
  {10.1051/0004-6361:20042280}, \href
  {http://adsabs.harvard.edu/abs/2005A%26A...436..443V} {436, 443}

\bibitem[\protect\citeauthoryear{{Verheijen}}{{Verheijen}}{2001}]{verheijen-2001}
{Verheijen} M. A.~W.,  2001, \mn@doi [\apj] {10.1086/323887}, \href
  {https://ui.adsabs.harvard.edu/abs/2001ApJ...563..694V} {563, 694}

\bibitem[\protect\citeauthoryear{{Verley} et~al.,}{{Verley}
  et~al.}{2007a}]{verley-2007-i}
{Verley} S.,  et~al., 2007a, \mn@doi [\aap] {10.1051/0004-6361:20077307}, \href
  {https://ui.adsabs.harvard.edu/abs/2007A&A...470..505V} {470, 505}

\bibitem[\protect\citeauthoryear{{Verley} et~al.,}{{Verley}
  et~al.}{2007b}]{verley-2007}
{Verley} S.,  et~al., 2007b, \mn@doi [\aap] {10.1051/0004-6361:20077481}, \href
  {http://adsabs.harvard.edu/abs/2007A%26A...472..121V} {472, 121}

\bibitem[\protect\citeauthoryear{Virtanen et~al.,}{Virtanen
  et~al.}{2020}]{2020SciPy-NMeth}
Virtanen P.,  et~al., 2020, \mn@doi [Nature Methods]
  {10.1038/s41592-019-0686-2}, \href {https://rdcu.be/b08Wh} {17, 261}

\bibitem[\protect\citeauthoryear{{Vollmer}}{{Vollmer}}{2009}]{vollmer-2009}
{Vollmer} B.,  2009, \mn@doi [\aap] {10.1051/0004-6361/200911892}, \href
  {https://ui.adsabs.harvard.edu/abs/2009A&A...502..427V} {502, 427}

\bibitem[\protect\citeauthoryear{{Walker}, {Mihos}  \& {Hernquist}}{{Walker}
  et~al.}{1996}]{walker-1996}
{Walker} I.~R.,  {Mihos} J.~C.,   {Hernquist} L.,  1996, \mn@doi [\apj]
  {10.1086/176956}, \href
  {https://ui.adsabs.harvard.edu/abs/1996ApJ...460..121W} {460, 121}

\bibitem[\protect\citeauthoryear{{Yuan} \& {Zhu}}{{Yuan} \&
  {Zhu}}{2004}]{yuan-2004}
{Yuan} Q.-r.,  {Zhu} C.-x.,  2004, \mn@doi [\caa]
  {10.1016/S0275-1062(04)90015-X}, \href
  {https://ui.adsabs.harvard.edu/abs/2004ChA&A..28..127Y} {28, 127}

\bibitem[\protect\citeauthoryear{{Zhao}, {Sun}, {Shen}, {Liu}, {Zhou}  \&
  {Ji}}{{Zhao} et~al.}{2021}]{Zhao-2021ApJ...913..111Z}
{Zhao} Q.,  {Sun} L.,  {Shen} L.,  {Liu} G.,  {Zhou} H.,   {Ji} T.,  2021,
  \mn@doi [\apj] {10.3847/1538-4357/abf4de}, \href
  {https://ui.adsabs.harvard.edu/abs/2021ApJ...913..111Z} {913, 111}

\bibitem[\protect\citeauthoryear{{Zschaechner}, {Rand}  \&
  {Walterbos}}{{Zschaechner} et~al.}{2015}]{Zschaechner-2015-a}
{Zschaechner} L.~K.,  {Rand} R.~J.,   {Walterbos} R.,  2015, \mn@doi [\apj]
  {10.1088/0004-637X/799/1/61}, \href
  {https://ui.adsabs.harvard.edu/abs/2015ApJ...799...61Z} {799, 61}

\makeatother
\end{thebibliography}

%\begin{thebibliography}{99}
%\end{thebibliography}

%%%%%%%%%%%%%%%%%%%%%%%%%%%%%%%%%%%%%%%%%%%%%%%%%%

%%%%%%%%%%%%%%%%% APPENDICES %%%%%%%%%%%%%%%%%%%%%

\appendix


\newpage

\section{Individual notes} \label{Sec:individual-notes}

\paragraph*{\textbf{CIG\,71 (UGC\,1391).}}
Fig.~\ref{Fig:c71}: %The monochromatic \Ha map shows asymmetry towards the south, while the morphological and kinematic centre and PA remain consistent. 
The ionized gas shows a noticeable asymmetry, being shifted towards the south with respect to the infrared emission. However, the morphological and kinematic centre and PA of the galaxy are consistent. 
Extrapolating the arctan function up to $D_{25_{\rm B}}$, we found the same maximum rotation velocity in \Ha and \ion{H}{i} emissions within a difference of $\pm3$~\kms. According to \citetalias{sardaneta-2024}, the extraplanar component of this galaxy reaches a vertical distance of 1.2~kpc, lagging in rotation by $-21.9$~\kms. CIG~71 exhibits high dispersion velocity values of up to $\sim$45~\kms southwards of the morphological centre, mainly within the stellar disc and not in the southern gaseous knot, as one might expect.

\paragraph*{\textbf{CIG\,95 (UGC\,1733).}}
Fig.~\ref{Fig:c95}: According to \citetalias{sardaneta-2024}, the ionized gas of this galaxy extends radially up to 10.4~kpc. Here, it displays a symmetrical radial velocity field. The morphological and kinematic centres align within $\sim1$~arcsec, and the morphological and kinematic PA agree within $\sim2^{\circ}$. While the observed data indicates an increasing rotation curve, likely peaking before $D_{25_{\rm B}}$, the best-fitting matches the maximum velocity of the \ion{H}{i} within $\pm5$~\kms. \citetalias{sardaneta-2024} reported an extraplanar component reaching a vertical distance of up to 1.8~kpc, however it does not show a lag in rotation along the vertical axis. CIG~95 exhibits a bright \Ha knot at the southern border of the stellar disc, with dispersion velocity values reaching up to $\sim$40\kms.

\paragraph*{\textbf{CIG\,159 (UGC\,3326).}}
Fig.~\ref{Fig:c159}: The photometric and kinematic centres exhibit a notable discrepancy of $\sim15$~arcsec. The morphological and kinematic PA align within $\sim1.5^{\circ}$, although in \citetalias{sardaneta-2024}, the ionized gas disc is reported as warped. The radial velocity field reveals a symmetric and increasing rotation curve in the innermost regions computed using the barycentre method. Due to the low brightness emission, the maximum rotation velocity computed with the envelope tracing method appears lower, but it matches the \ion{H}{i} rotation velocity within $\pm3.5$~\kms. The thinness of the \Ha disc prevented the measurement of the lag in rotation along the $z-$axis. Some regions distributed across the disc display dispersion velocities $\leq$40~\kms.

\paragraph*{\textbf{CIG\,171 (UGC\,3474).}}
Fig.~\ref{Fig:c171}: In \citetalias{sardaneta-2024}, the asymmetry of the ionized gas disc in this galaxy was attributed to dust extinction. Despite this, sufficient \Ha emission enables the computation of the rotation curve using the envelope tracing method. The morphological and kinematic centre and PA are in agreement, with the PVDs indicating an increasing trend of the rotation curve. However, in cases like CIG~171, the barycentre method can yield inaccurate asymmetrical rotation curves. The maximum rotation velocity from \ion{H}{i} emission is also undervalued by $\sim$55~\kms. While no extraplanar emission along the vertical axis is observed \citepalias[see][]{sardaneta-2024}, the extended ionized gas emission allows detection of a rotation lag of $\sim-41.5$~\kms. Uniformly distributed regions across the disc display dispersion velocities $\sim$35-40~\kms.

\paragraph*{\textbf{CIG\,183 (UGC\,3791).}} 
Fig.~\ref{Fig:c183}: In \citetalias{sardaneta-2024}, it was shown that the \Ha monochromatic map of CIG~183, displaying an asymmetric disc displaced as a whole to the northeast  beyond the 2.2$\mu$m emission disc, aligns with the locus of the young stellar population traced by the NUV and FUV emissions. 
In this paper, the \Ha radial velocity field is symmetrical with respect to the kinematic centre, located $\sim15$~arcsec northwards of the photometric centre. 
In addition, there is a significant divergence of $\sim8.5$~arcsec between the morphological and kinematic PAs. 
The rotation curve is symmetrical, and the maximum rotation velocities in \Ha and  \ion{H}{i} maximum rotation velocities agree, differing by $\sim5$~\kms. 
With the PA almost parallel to the north, the \Ha emission disc appears deformed and bent into a C-shape pointing westward. This morphology  suggests that the gas on the eastern side is less compressed, resulting in faster rotation compared to the more compressed gas in the west from PVDs. The \Ha distribution and kinematics of CIG\,183 strongly point towards the presence of strong environmental effects. 
High dispersion velocity values of $\sim$40~\kms are mainly located on the northern side of the \Ha emission disc, outside the disc of the old stellar population.

\paragraph*{\textbf{CIG\,201 (UGC\,3979).}} Fig.~\ref{Fig:e.g.-c201}:  
The significantly extended \Ha emission disc exhibits a regular velocity field, with the kinematic and photometric centres and PAs in agreement. Both the barycentre and envelope tracing methods yield similar rotation velocities at $D_{25_{B}}$, but diverge by $\sim$20~\kms from the maximum rotation velocity measured in \ion{H}{i}.  Because of the dust extinction, in the inner regions, the barycentre method  underestimates the rotation velocities, rendering the  envelope tracing method  more accurate in that region. In contrast, the arctan fitting used in the envelope tracing method enhances the reliability of the barycentre method beyond $D_{25_{B}}$, which indicates an increasing rotation in that region for CIG~201. 
Along the vertical ($z$-axis) direction, the extended emission shows a rotation lag of $-15.7\pm2.0$~\kms. However, in \citetalias{sardaneta-2024}, the multiwavelength morphology of CIG~201 suggested that the galactic disc may host a galactic wind. This hypothesis is supported here by a cone-shaped pattern observed in the \Ha dispersion velocity values of $\sim$50~\kms in the innermost regions of the galactic disc. Further spectral data are necessary to confirm this scenario.

\paragraph*{\textbf{CIG\,329 (UGC\,5010).}} Fig.~\ref{Fig:c329}:  
The ionized gas disc exhibits a ring structure due to dust extinction of a stellar bar \citepalias[see][]{sardaneta-2024}, rotating symmetrically with respect to the minor axis. The photometric and kinematic centres and PAs coincide. The central points from the central PVD, used to compute the rotation curve with the envelope tracing method, indicate an inflection point towards a flat rotation curve near the old stellar population disc (i.e., the radius of the 3$\sigma$ surface brightness level of the 2MASS K${s}$-band image), coinciding with the maximum rotation velocity measured in \ion{H}{i}. However, the \Ha rotation curve computed with the barycentre method and the best-fitting  of the arctan function show an increasing solid body rotation, reaching a higher maximum rotation velocity by $\sim60$~\kms at $D{25_{B}}$. We adopt the maximum rotation velocity in \Ha since the barycentre method provides a reliable rotation curve in the outermost region. According to \citetalias{sardaneta-2024}, the filament-like extraplanar component of this galaxy extends to a vertical distance of 1.7~kpc, lagging in rotation by $-32.2\pm11.0$~\kms. Regions distributed across the disc display dispersion velocities of $\sim$40~\kms.

\paragraph*{\textbf{CIG\,416 (UGC\,5642).}} Fig.~\ref{Fig:c416}:  
In \citetalias{sardaneta-2024}, the \Ha emission disc was described as asymmetric,  filamentary, wide and radially extended, featuring a tail-shape at the western tip and several detached ionized gas clouds. The morphological and kinematic centres, located between the two brightest \Ha knots, coincide, while the position angles (PAs) align within $1^{\circ}$. Despite the asymmetries in the \Ha emission disc, the radial velocity field exhibits symmetry with respect to the kinematic centre. The \Ha rotation curves display differences in the inner disc, consistent with the galaxy's inclination, however, the maximum rotation velocity measured in \Ha in all cases matches that of \ion{H}{i}. Along the vertical ($z$-axis) direction, the extended emission reveals a regular rotation lag of $-44.5\pm11.5$~\kms. Uniformly distributed regions across the disc exhibit dispersion velocities of approximately $\sim$35~\kms. There is no particular feature of radial velocity or velocity dispersion at the brightest \Ha knots.

\paragraph*{\textbf{CIG\,593 (UGC\,8598).}} Fig.~\ref{Fig:c593}:  
As outlined in \citetalias{sardaneta-2024}, we detected only a few faint \Ha clouds tracing an asymmetric and truncated ionized gas distribution displaced northwards with respect to the minor axis of the old stellar population disc. Indeed, the kinematic centre is located at the northern edge of the old stellar population disc, approximately vary by $4.5^{\circ}$. Nevertheless, the radial velocity field shows regular rotation. From a PVD computed with a virtual slit of 5~pixels ($\sim$3.4~arcsec), we were able to compute the \Ha rotation curve indicating a maximum rotation velocity that agrees with the one at \ion{H}{i} by 10~\kms. 
The asymmetric rotation curve computed with the barycentre method indicates that the receding side, comprising the clouds belonging to the extraplanar gas, rotates faster than the approaching side. Therefore, from the distribution of the young stellar population \citepalias[see figure~E8 from][]{sardaneta-2024}, the dynamic mass might be undervalued here. There are no regions across the disc exhibiting significant dispersion velocity values.

\paragraph*{\textbf{CIG\,847 (UGC\,11132).}} Fig.~\ref{Fig:c847}:  
The \Ha emission displays a radially extended, asymmetric, and warped disc, with greater prominence and brightness towards the south compared to the north along the major axis \citepalias[see][]{sardaneta-2024}. Although the \Ha emission peak is located towards the south-east from the NIR photometric centre, the kinematic and morphological centres match, with the PAs varying only by $1.6^{\circ}$. 
Within the old stellar population disc, the velocity field remains uniform, while perturbations are observed in the radially extended regions, as evidenced by asymmetries in the barycentre rotation curve. The maximum rotation velocity measured at \Ha using different methods aligns closely with that measured at \ion{H}{i}, differing by only $\sim$11~\kms. 
Because of the galactic centre is located to the west of the  \Ha emission disc, particularly for CIG~847, we measured a rotation lag  of $-21.7\pm11.3$~\kms eastwards. 
The presence of a rotation lag is often associated with a significant amount of extended gas along the vertical axis. However, CIG~847 has only 0.3~kpc of extended ionized gas along the vertical axis  \citepalias[see][]{sardaneta-2024}, 
suggesting that the rotation lag may not solely result from the eDIG but could be an intrinsic feature of the galaxy's dynamics. Several regions across the disc show dispersion velocities of approximately 40~\kms, while the dispersion velocity at the \Ha emission peak remains relatively unchanging at $\sim$25~\kms.

\paragraph*{\textbf{CIG\,906 (UGC\,11723).}}  Fig.~\ref{Fig:c906}:  
In \citetalias{sardaneta-2024}, only faint \Ha clouds resembling knots were identified along the galactic major axis. In this paper, we observed that these clouds trace a rotating \Ha disc, indicating a general rotational motion. Consequently, the rotation curves exhibit a solid-body rotation pattern in the innermost region. Nevertheless, as we move towards the outermost regions, the rotational velocity of the farthest clouds from the kinematic centre decreases, resulting in an asymmetrical barycentre rotation curve. The arctan fitting aligns with the solid-body rotation pattern, reaching a maximum rotation velocity at \Ha lower than the one measured at \ion{H}{i} by approximately $\sim20$~\kms at $D_{25_{\rm B}}$. The morphological and kinematic centres show a variation of approximately $\sim5$~arcsec, while the PAs vary by only $2^{\circ}$. 
A rotation lag is typically associated with an extension of gas along the vertical axis (or the eDIG).  However, while CIG\,906 does not show \Ha extraplanar emission along the vertical component, we measured a median rotation lag in that direction of $-95.0\pm24.3$~\kms. 
This suggests that the rotation lag might be an inherent characteristic of the galaxy's dynamics. Finally, the dispersion velocity map is rather uniform with a value $\sim20$~\kms along the \Ha emission disc.

\paragraph*{\textbf{CIG\,922 (UGC\,11785).}} Fig.~\ref{Fig:c922}:  
Although the \Ha emission disc exhibits asymmetry and warping \citepalias[see also][]{sardaneta-2024}, CIG~922 displays a solid-body-like rotation, observed in both the isovelocities on the radial velocity field and the rotation curves. 
This solid-body-like rotation pattern results in an increasing curve that exceeds the maximum rotation velocity measured at \ion{H}{i} by 45~\kms at $D_{25_{\rm B}}$. 
Due to the warping of the \Ha emission disc, the PVDs parallel to the major axis appear asymmetric, with the gas rotating slower northwards and faster southwards compared to the gas along the major axis. 
However, consistent with a solid-body rotation, neither the gas in the disc nor the eDIG show a different rotation along the vertical axis. 
The morphological and kinematic centres align within $\sim5$~arcsec, while the position angles vary by $1.6^{\circ}$. Additionally, the dispersion velocity map shows values of $25-40$~\kms in  and around the brightest \Ha emission knots.

\paragraph*{\textbf{CIG\,936 (UGC\,11859).}} Fig.~\ref{Fig:c936}:  
The K${s}$-band and \Ha emissions are concentrated within the central 30~arcsec of this galaxy, which has an optical diameter of $D_{25_{\rm B}}=185.4$~arcsec \citepalias[see][]{sardaneta-2024}. The \Ha monochromatic image reveals that the warm gas is compressed at the centre within the old stellar population disc, forming a cone shape. 
Although the radial velocity field shows symmetry with respect to the minor axis and, the kinematic and morphological centres agree within $\sim$2~arcsec, the PAs differ by $\sim5^{\circ}$. 
The \Ha emitting gas detected enabled the computation of the rotation curve in the inner region. Extrapolating the best-fitting  of the arctan function up to $D_{25_{\rm B}}$ revealed a maximum rotation velocity in \Ha higher than that in \ion{H}{i} by $\sim$18~\kms. 
Due to the low quantity of ionized gas detected in this galaxy, a reliable lag in rotation along the vertical axis was not detected. 
Two regions displaying dispersion velocities of $\leq$40~\kms are located on both sides of the kinematic centre along the major axis.

\paragraph*{\textbf{CIG\,1003 (UGC\,12304).}} Fig.~\ref{Fig:c1003}:  
The \Ha emission appears compressed within the galactic disc, displaying three bright knots: one at the centre and the other two symmetrically positioned $\sim$15~arcsec apart \citepalias[see][]{sardaneta-2024}. Despite asymmetry along the minor axis, the radial velocity map suggests a solid-body-like rotation across the disc. Thus, while the barycentre rotation curve increases, the envelope tracing method yields a flat rotation curve. The maximum rotation velocity at \Ha exceeds that measured at \ion{H}{i} by approximately 40~\kms. The kinematic and morphological centres agree, with the PAs differing by only $1.7^{\circ}$. 
Notably, the \Ha emission map reveals a prominent filament surrounding the central knot, resembling a cone-shaped structure pointing southwest. That filament and the multiwavelength morphology of CIG~1003 suggested that the galactic disc may host a galactic wind  \citepalias[see][]{sardaneta-2024}. 
Along the vertical axis, extended \Ha emission exhibits a rotation lag of $-46.7\pm10.4$~\kms. 
In contrast, the dispersion velocity map of CIG~1003 shows mostly uniform values across the \Ha disc with a median of $20$~\kms, while values around the brightest \Ha emission knots are as high as  $\sim 35$~\kms. Therefore, additional spectral data are required to confirm the existence of a galactic wind.

%\newpage
\section{Variations in PVD symmetry}\label{Appendix:pvd_symmetry}

\begin{figure*}
\centering
\includegraphics[width=\hsize]{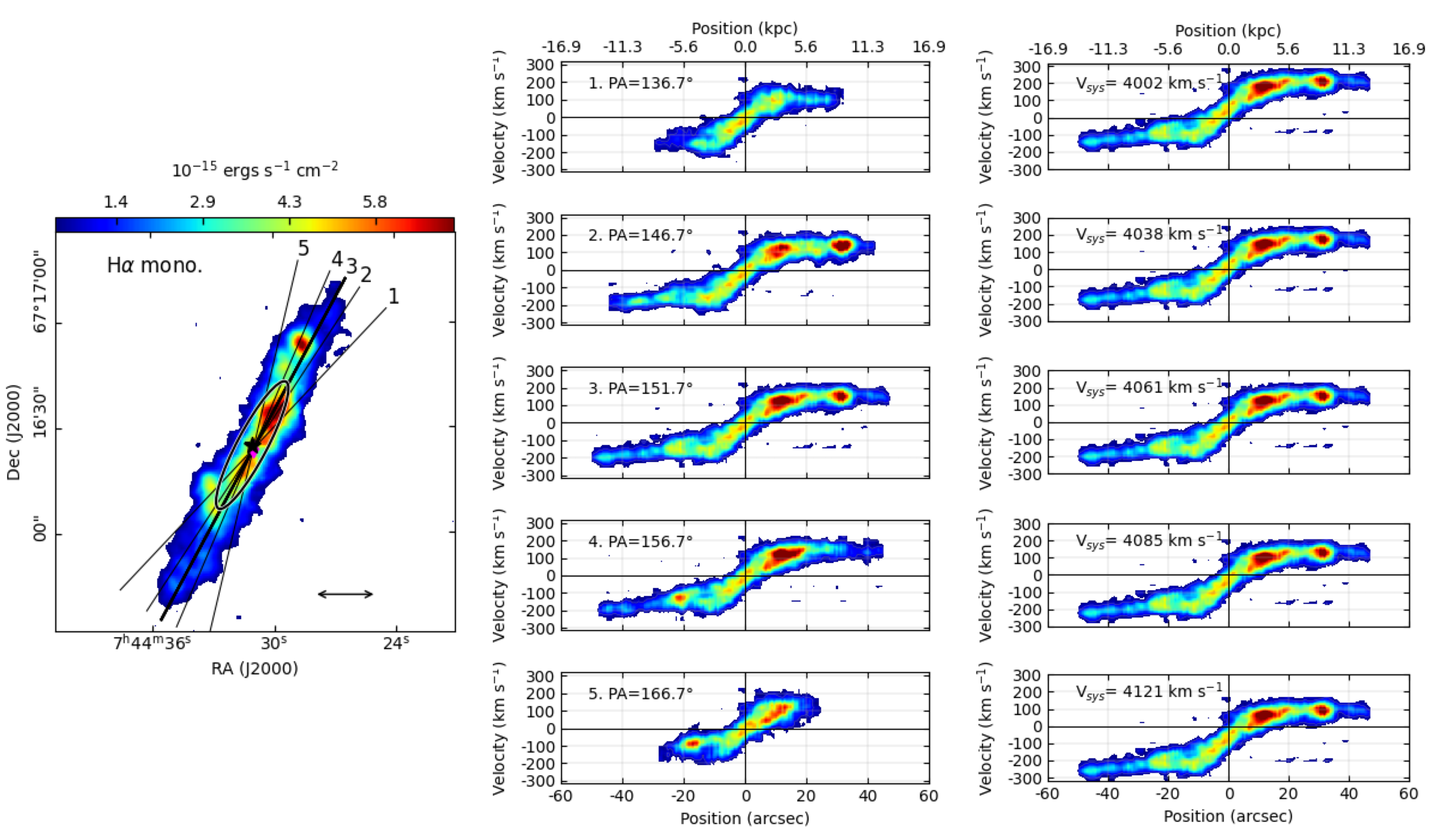}
\caption{Position-velocity diagrams of the galaxy CIG~201. 
Center: The position angle (PA) is varied from $\pm5^{\circ}$ to $\pm15^{\circ}$ relative to the PVD along the major axis (PA=151.7$^{\circ}$). The pseudo-slits are overlaid and numbered from lowest to highest PA on the \Ha monochromatic map (left), with the corresponding PVD number indicated. Right: The heliocentric velocity is varied from the systemic velocity reported by NED ($V_{\rm sys}$=4061~\kms).
}
\label{Fig:pvds_params}
\end{figure*}

Variations in the kinematic parameters such as centre, position angle (PA), and systemic velocity ($V_{\rm sys}$), result in asymmetric PVDs that can distort the representation of gas motions within the galaxy,   
making their interpretation more challenging. 
As an example, Figure~\ref{Fig:pvds_params} shows the PVDs along the major axis of the galaxy CIG~201. It can be observed that varying the PA results in shorter PVDs with asymmetries along the position axis ($x$-axis). Similarly, modifying the ($V_{\rm sys}$) alters the symmetry of the PVD along the velocity axis ($y$-axis).  For this study, we selected PVDs that exhibit the most symmetric profiles in both axes, as these most accurately reflect the kinematic parameters of the galaxies.

\section{Intensity peak method}\label{Appendix:intensity-peak}

To compute the rotation velocity ($V_{\mathrm{rot}}$) of the galaxy with the intensity peak method (IPM) method, it is necessary to proceed by assuming that a galaxy is formed by a well-defined infinitesimally thin disc, that the system rotates around an axis perpendicular to the galactic plane inclined at an angle $i$ with respect to the plane of the sky and that it is dominated by rotational motions around an axis perpendicular to the galactic plane \citep[e.g.][]{mihalas}. Assuming an axial symmetry around the galactic centre, the velocity of the pixels that are at the same distance from the kinematic centre along the major axis can be measured by dividing the galaxy on the sky-plane into rings. 
The rotation velocity $V_{\mathrm{rot}}(r)$ for a ring of radius $r$ is given as the azimuthal average over the ring of the deprojected velocities:
\begin{equation}
V_{\mathrm{rot}}(r,\,\theta)=\frac{V_{\parallel}(r,\,\theta)-V_{sys}}{\cos\theta\sin i}\label{Eq:ipm}
\end{equation}
where $r$ and $\theta$ are the radial and angular coordinates in the plane of the galaxy and $V_{\parallel}(r,\,\theta)$ is the velocity along the Line-of-Sight (LoS) extracted from each pixel of the FP velocity field 
\citep[e.g.][]{fuentesc-2004, epinat-2008-ii, cardenas-2018, sardaneta-2020, sardaneta-2022}. 

Although the IPM is known to be inefficient for high-inclined galaxies due to its non-infinitesimal nature and LoS effects, we deliberately calculated it to provide a glimpse of the appropriate rotation curve. In addition, we aimed to illustrate the limitations of this method and show why it fails in these cases. We computed the RC using the IPM with the \textsc{ADHOCw} software which computes the rotation velocity of the galaxies considering 
the barycentre of the H$\alpha$ profile observed in each pixel within an angular sector along the kinematic major axis \citep[e.g.][]{fuentesc-2004,rep, sardaneta-2020}. In Figure~\ref{Fig:e.g.-c201}, the kinematic major axis and the aperture of the angular sector are traced on the radial velocity map with gray lines, and the resulting rotation curve of the galaxy from the IPM is displayed with filled and empty squares for the receding and approaching side, respectively. The error bars represent pixel value dispersion within an angular sector along the major axis.

%\newpage

\section{Validation of rotational lag measurement}\label{App:tirific}

In previous studies, the vertical rotation lag has been measured from the maximum rotation velocities as a function of height above the disc plane, as done by \citet{marasco-fraternali-2011A&A...525A.134M} using \ion{H}{i} data in the Milky Way, and by \citet{levy-2019} using \Ha emission from CALIFA galaxies. 
In this work, we follow the same principle but using  \Ha emission line data cubes obtained with FP interferometry. As described in Section~\ref{Sec:Lag}, we extracted PVDs at different heights and computed the maximum rotation velocity at each height using the envelope tracing method (ETM) and, finally, we estimated the velocity gradient along the $z$-axis via linear regression of the rotation velocity at a fixed radius. To assess the reliability of this method,  we present in this appendix a mock data cube of the galaxy presented in the main body of the paper (CIG~201, Figure~\ref{Fig:e.g.-c201}), a representative high-inclined galaxy of our sample, designed to test whether the ETM can recover the known observed kinematic inputs under  controlled conditions.

To construct the mock data cube, we used the \textsc{3D Tilted Ring Fitting Code} \citep[\textsc{TiRiFiC};][]{jozsa-2007-tirific, kamphuis-2015-pyFat-i, kamphuis-2024-pyFat-software}, designed to model gas emission in rotating galaxies using the coined \textit{tilted-ring model}. In this scheme, each ring can have its own geometry (position, inclination) and kinematic parameters (rotation velocity, dispersion), allowing the three-dimensional structure of thin discs with a certain vertical thickness to be represented. 
In edge-on galaxies, where the line of sight intersects the disc continuously, several disc elements contribute simultaneously to the observed signal at each position, so beam smearing and projection effects can be overcome. Moreover, when the vertical thickness of the disc is less than one tenth of the beam size, \textsc{TiRiFiC} overestimates this parameter, as it becomes unresolved spatially \citep{kamphuis-2015-pyFat-i}. While the model accounts for vertical extent, the derived thickness must be interpreted with caution, as it is constrained by the spatial resolution of the data. 
Given these challenges in modelling highly inclined galaxies, in this work we constructed a mock data cube by manually setting the physical parameters in \textsc{TiRiFiC}, and fine-tuned the model by visually comparing the emission, radial velocity and velocity dispersion maps with the observed data, following the approach described by \citet{jozsa-2007-tirific}.

To assemble the mock data cube of the galaxy CIG~201, we provided its known physical input parameters. The template cube reproduces the GHASP resolution and physical dimensions of 512$\times$512~pixels, 44~channels, and 11.5~\kms spectral step (Table~\ref{Tab:obs}). We defined 12 concentric rings reaching a maximum radius of 51~arcsec ($\sim$11.9~kpc), corresponding to the major axis extent of the \Ha disc observed in CIG~201. According to Hubble’s formula, the inclination $i$ can be estimated from the observed axis ratio $q = b/a$ as 
\begin{equation}
\cos^2(i) = \frac{(q^2 - q_0^2)}{1 - q_0^2},      
\end{equation}\label{Eq:hubble}
where we adopted an intrinsic axial ratio of $q_0 = 0.11 \pm 0.03$, as reported by \citet{yuan-2004} for highly inclined disc galaxies and a measured minor axis of $b\approx8.5$~arcsec ($\sim$2.4 kpc), yielding an inclination of $i = 82.5^\circ$, consistent with the geometry of the CIG~201 ionised gas disc. To represent a thick disc that, for our purposes, includes both the midplane gas and a diffuse extraplanar component (eDIG), the model includes two superposed concentric discs (\texttt{NDISKS}=2) with the same inclination ($i=82.5^{\circ}$) and position angle (PA=151.7$^{\circ}$). We adopted an exponential radial surface–brightness profile (\texttt{SBR$_i$}) for each of the two components, namely $\mathrm{B}_1(R)$ and $\mathrm{B}_2(R)$.  
%The second disc (representing the eDIG) was normalised ring by ring as \(\mathrm{B}_2(R)=0.2\,\mathrm{B}_1(R)\). 
The second disc (representing the eDIG) was normalised ring by ring as $\mathrm{B}_2(R)=0.2\,\mathrm{B}_1(R)$. The vertical density, $\mathrm{H(Z_1)}$ and $\mathrm{H(Z_2)}$, in both components follows a \(\mathrm{sech}^2\) profile (\texttt{LTYPE}\,=\,2) with scale heights \(Z_{1}=2.5''\) and \(Z_{2}=3.5''\). In \textsc{TiRiFiC}, the 3D emissivity is separable \citep{jozsa-2007-tirific}, \(J(R,\phi,Z)=B(R,\phi)\,H(Z)\). 
Furthermore, the local vertical profile is the linear sum of two vertical profiles weighted by their radial brightness: 
$I(z,R)=\mathrm{B}_1(R)H(Z_{1})\,+\mathrm{B}_2(R)\,H(Z_{2})$. 
With these settings, the vertical thickness (FWHM) of the combined profile is $\simeq$5~arcsec ($\sim 1.5$~kpc).

The 44 spectral channels of the template cube allowed us to assign a maximum rotation velocity of $\sim$213~\kms\, to the disc, consistent with that measured for CIG~201. We adopted a constant velocity dispersion of 25~\kms, and introduced a linear decrease in the rotational velocity with height in the second disc (\texttt{DVRO}~$\simeq$~4.5~\kms~arcsec$^{-1}$), equivalent to a vertical lag of $\sim$16.0\kmspc, in agreement with observational estimates for CIG~201 (see Table~\ref{Tab:kin}). The input parameters are listed in Table~\ref{Tab:tirific}.

\begin{table}%[htbp]
\caption{Input parameters of the mock data cube}
\begin{center}
\begin{tabular}{lc}
\hline
Parameter & Value \\ 
\hline
Dimensions & 512 $\times$ 512~pix $\times$ 44~spectral channels \\ 
Number of discs (\texttt{NDISKS}) & 2 \\
Velocity channel width & 11.5 \kms \\
Pixel scale & 0.68 arcsec pix$^{-1}$ \\
Number of rings  & 12 \\ 
Maximum radius &  51.0~arcsec ($\sim$ 84~pix, 11.9 kpc) \\ 
Inclination ($i$) & 82.5$^{\circ}$ \\ 
Position Angle (PA) & 157$^{\circ}$ \\ 
Heliocentric velocity (V$_{\rm sys}$)  & 4061~\kms \\ 
Maximum rotation velocity (V$_{\rm rot}$) & 213~\kms \\ 
Velocity dispersion ($\sigma_{\rm obs}$) & 25~\kms \\ 
Vertical disc scale height (disc, $Z_{1}$) & 2.5 arcsec ($\sim$0.7 kpc) \\
Vertical disc scale height (eDIG, $Z_{2}$) & 3.5 arcsec ($\sim$1.0 kpc) \\
%Vertical thickness (FWHM) & $\sim$5.0 arcsec ($\sim$1.5 kpc) \\ 
Vertical density profile (\texttt{LTYPE}) & sech$^2$ \\
Rotational lag gradient (\texttt{DVRO}) & 4.5~\kms~arcsec$^{-1}$ ($\sim$16 \kmspc) \\ 
\hline
\end{tabular}
\end{center}
\label{Tab:tirific}
\end{table}

\begin{figure*}
\centering
 \includegraphics[width=\hsize]{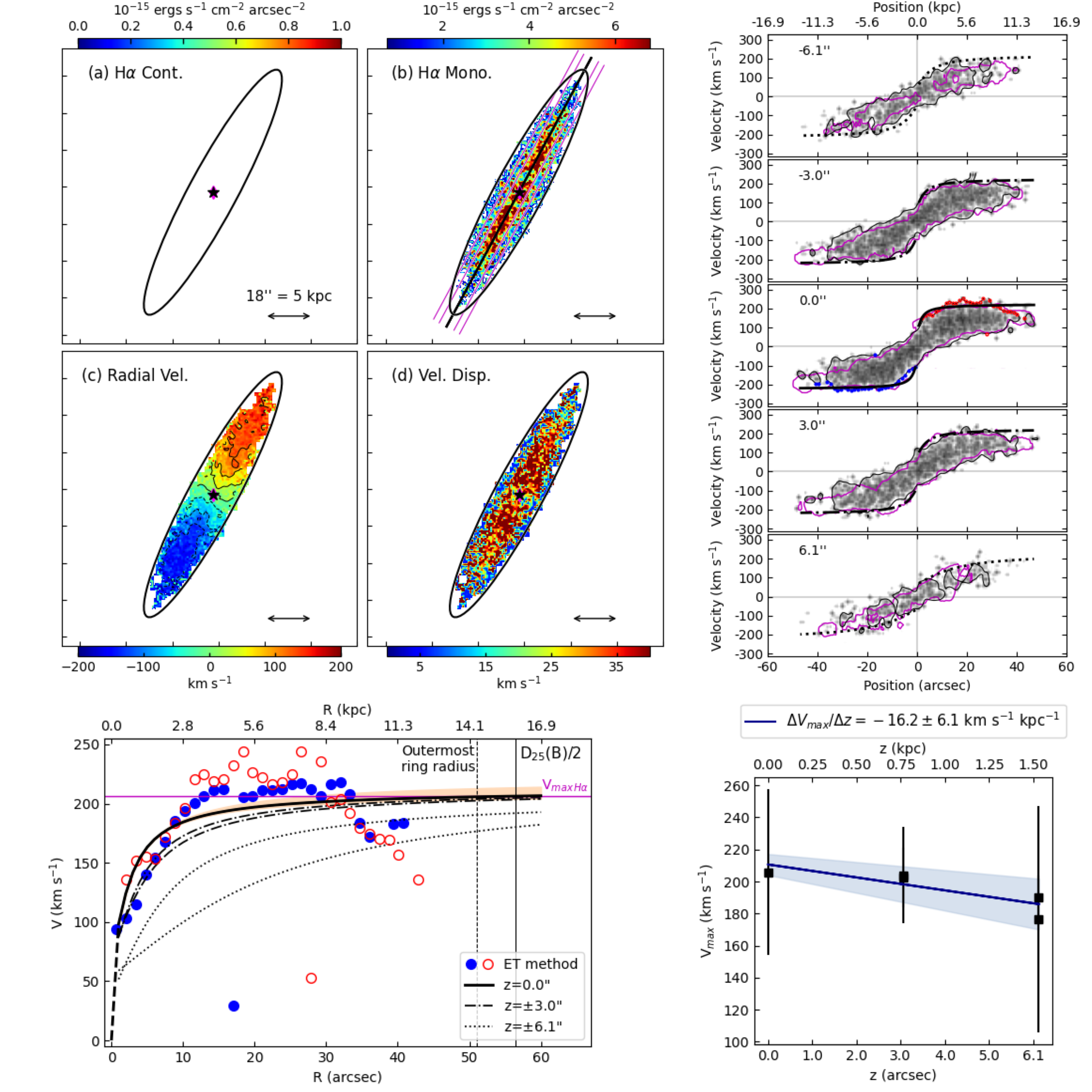}
\caption{
Validation of the vertical lag extraction method using a synthetic \Ha data cube generated with \textsc{TiRiFiC}, configured to match the spatial and spectral properties of GHASP observations. 
Top left: 
(a)~\Ha~continuum (not included in the model), 
(b)~monochromatic, 
(c)~radial velocity and 
(d)~velocity dispersion maps 
derived from the mock data cube by fitting a Gaussian profile at each spatial pixel. 
The black ellipse outlines the outermost ring of the modelled ionized gas disc, corresponding to a measured major axis $a \simeq 51.0$~arcsec ($\sim$11.9~kpc) and minor axis $b \simeq 8.5$~arcsec ($\sim$2.4~kpc) of the CIG~201 \Ha emission. 
Top right: 
PVDs along the major axis at the same heights from the kinematic major axis than those of the galaxy CIG 201 ($z = 0$, $\pm$3.0, and $\pm$6.1 arcsec, see Figure~\ref{Fig:e.g.-c201}). 
Pseudo-slits are traced on the synthetic \Ha monochromatic map. 
For direct comparison, the contours corresponding to the intensity threshold used to define the envelope of the observed PVDs of CIG~201 are superimposed (in magenta) on the synthetic ones. 
The major-axis PVD shows the $I_{env}$ derived using $\eta = 0.3$, represented by filled blue circles for the receding side and empty red circles for the approaching side, these points were used to fit the $\arctan$ function (solid line). 
The $I_{env}$ using $\eta = 0.3$ at different heights from the kinematic major axis are plotted: dash-dotted lines correspond to heights of $\pm 3.0$ arcsec and dotted lines to heights of $\pm 6.1$ arcsec. 
Bottom left: 
rotation curves derived from the ET method (blue filled circles for approaching, red empty circles for receding). 
For the ET method, the inclination-corrected $\arctan$ fitting to I$_{\rm env}$ (with $\eta=0.3$) for the major-axis PVD is shown as a solid line, while the shaded area corresponds to the range $\eta = [0.2, 0.5]$. To measure the lag in rotation, inclination-corrected dashed and dotted lines from the upper panels are also included. 
Bottom right: 
OLS fit of the relation between maximum rotation velocities at different heights and the height ($z$), with the confidence region shaded.
The linear fit gave as a result a vertical lag of $\Delta V/\Delta z = -16.2 \pm 6.1$~\kmspc, consistent with the input model value. 
This test confirms that the method reliably recovers the imposed velocity gradient under controlled, idealized conditions. 
}
\label{Fig:tirific}
\end{figure*}

Figure~\ref{Fig:tirific} shows the results of applying the ET method to the mock data cube. The PVDs extracted at different disc heights, as well as the rotation curves and linear fit used to estimate the vertical lag, are displayed in the right and bottom panels.  
To allow a direct comparison, we overlaid the contours of the observed PVDs of CIG~201 onto the synthetic ones. These contours trace the intensity threshold used to define the envelope, as described in the relation~\ref{equation:I_env}. Additionally, we derived the monochromatic, radial velocity, and velocity dispersion maps by fitting a Gaussian function to the spectral profile of each pixel. As the model includes only \Ha emission and no continuum source was added, the continuum map is empty and was not considered in the analysis. A slight Gaussian smoothing ($\sigma=0.5$~pix) was applied to the maps to enhance the visibility of coherent structures.

With a disc thickness of $q_0=0.11$ and an inclination of $i=82.5^\circ$, projection effects strongly affect the observed vertical distribution. The intrinsic vertical structure becomes apparently only near  the outer edge, where the disc plane no longer projects over the vertical structure. These projection effects can be observed in both the monochromatic and the velocity dispersion map. Although the model assumes a constant intrinsic velocity dispersion of 25~\kms\ across the disc (see Table~\ref{Tab:tirific}), the resulting velocity dispersion map shows spatial variations, particularly high values ($\sim$35~\kms) in the inner regions. This is a consequence of integrating the emission along the line of sight in a highly inclined disc, where the observed profile results from LOS integration. Near the centre, line profiles are broadened due to the superposition of velocities from multiple layers along the LOS, increasing the values of the input dispersion. In contrast, at larger radii or higher vertical positions, the profiles tend to be narrower, resulting in lower measured dispersions.

We extracted five PVDs at vertical positions equivalent to those used for CIG~201 (see Figure~\ref{Fig:e.g.-c201}) above and below the mid-plane.  Using the ETM, we recovered the input maximum rotation velocity of $V_{{\rm rot}}\sim$213~\kms. For consistency, the vertical lag was measured at the B-band optical radius of CIG~201, $D_{25}(B)/2 \simeq 54.0\,{\rm arcsec}$ ($\sim$15.2~kpc, see Table~\ref{Tab:kin}). 
Velocities were corrected for inclination. The vertical lag was then calculated as the slope of a linear fit to $V_{\rm max}$ as a function of $z$. The method successfully recovers a vertical velocity gradient of $\Delta V/\Delta z=-16.2\pm6.1$~\kmspc, consistent with the input value. This result confirms that our approach can reliably retrieve vertical lags under controlled conditions.

\section{Arctan fitting}

Table~\ref{Tab:atan} lists the parameters resulting from fitting the $\arctan$ function $f(x)= A\tan^{-1}(x/x_0)+c$ to the envelope intensity (I$_{\rm env}$, using $\eta=0.3$) of the PVD along the kinematic major axis of the 14 galaxies in our sample. 
\begin{table}%[htbp]
\caption{Parameters for the theoretical rotation curve: $f(x)= A\tan^{-1}(x/x0)+c$.}
\begin{center}
\begin{tabular}{cccccccc}
\hline
CIG Name & A   & $x_{0}$   & c &   $R^{2}$ \\ 
\hline
71 & 121.7 $\pm$  5.1 & 5.6 $\pm$  0.6 &  …  $\pm$   …  & 0.80 \\ 
95 & 94.3 $\pm$  10.0 & 11.5 $\pm$  2.9 &  …  $\pm$  …  & 0.89 \\ 
159 & 192.3 $\pm$  21.2 & 27.8 $\pm$  5.2 &  …  $\pm$   …  & 0.86 \\ 
171 & 158.9 $\pm$  14.7 & 32.9 $\pm$  6.3 & 56.3 $\pm$  5.4 & 0.93 \\ 
183 & 95.3 $\pm$  24.5 & 13.6 $\pm$  5.1 & 40.6 $\pm$  2.9 & 0.81 \\ 
201 & 72.1 $\pm$  9.0 & 3.4 $\pm$  0.9 & 111.2 $\pm$  15.2 & 0.75 \\ 
329 & 279.7 $\pm$  22.5 & 23.5 $\pm$  4.0 & 8.5 $\pm$  4.7 & 0.98 \\ 
416 & 37.6 $\pm$  9.8 & 1.7 $\pm$  1.0 & 59.1 $\pm$  15.4 & 0.60 \\ 
593 & 165.9 $\pm$  8.9 & 17.3 $\pm$  1.8 & 22.2 $\pm$  2.1 & 0.95 \\ 
847 & 115.5 $\pm$  7.2 & 3.2 $\pm$  0.8 &  …  $\pm$  …   & 0.69 \\ 
906 & 134.9 $\pm$  11.9 & 12.6 $\pm$  2.1 &  …  $\pm$  …  &   0.82 \\ 
922 & 130.6 $\pm$  8.6 & 15.7 $\pm$  2.0 & 38.2 $\pm$  1.9 & 0.97 \\ 
936 & 104.4 $\pm$  24.1 & 4.5 $\pm$  2.6 &  …  $\pm$  …   & 0.64 \\ 
1003 & 110.4 $\pm$  5.8 & 3.2 $\pm$  0.7 &  …  $\pm$  …   & 0.65 \\ 
\hline
\end{tabular}
\end{center}
\label{Tab:atan}
\end{table}

\section{Individual maps, PVDs and RCs}\label{Appendix:Maps}

\begin{figure*}
\centering
\includegraphics[width=\hsize]{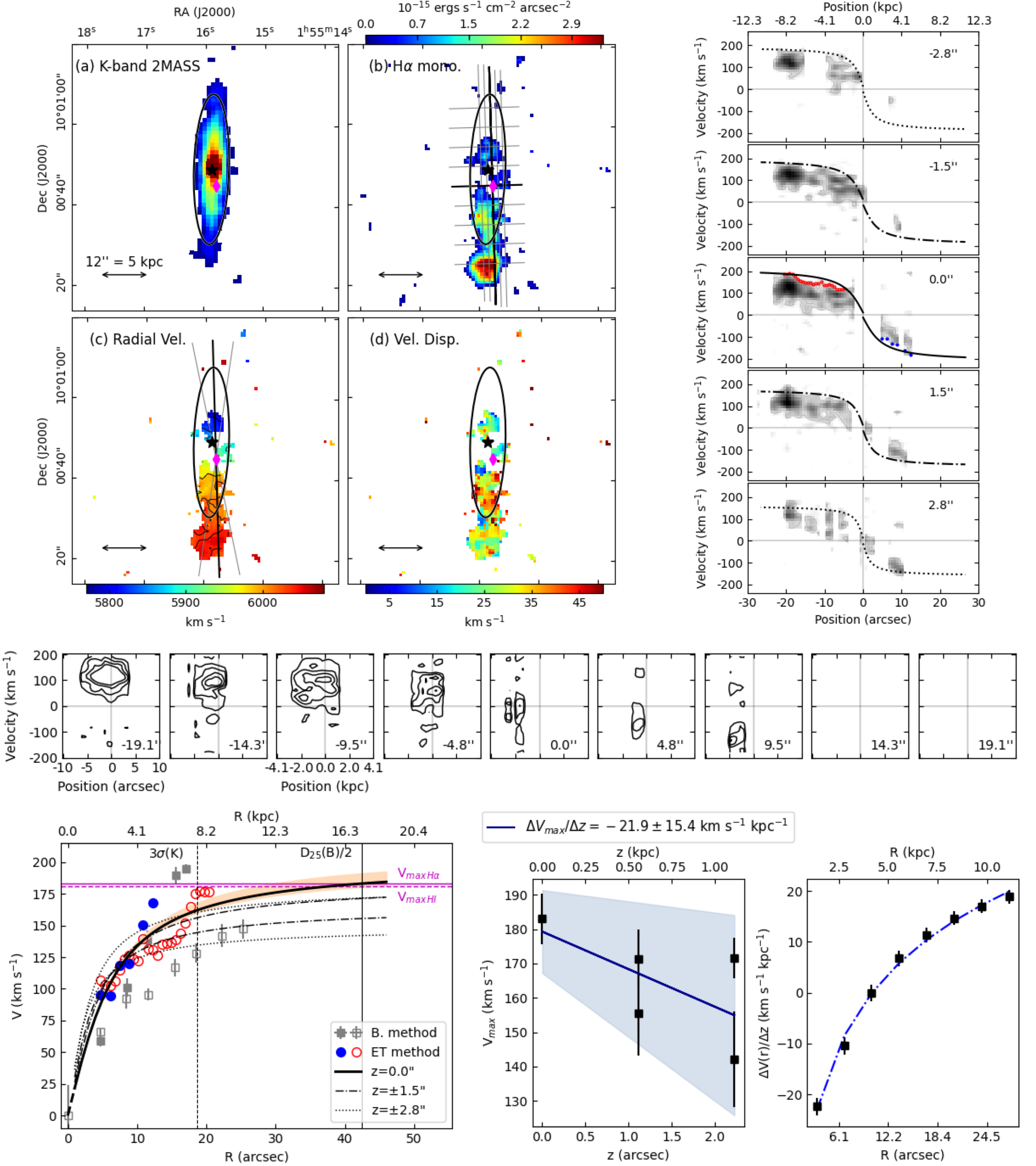}
\caption{CIG~71 (UGC~1391). Same as Fig.~\ref{Fig:e.g.-c201}.
}
\label{Fig:c71}
\end{figure*}

\begin{figure*}
\centering
\includegraphics[width=\hsize]{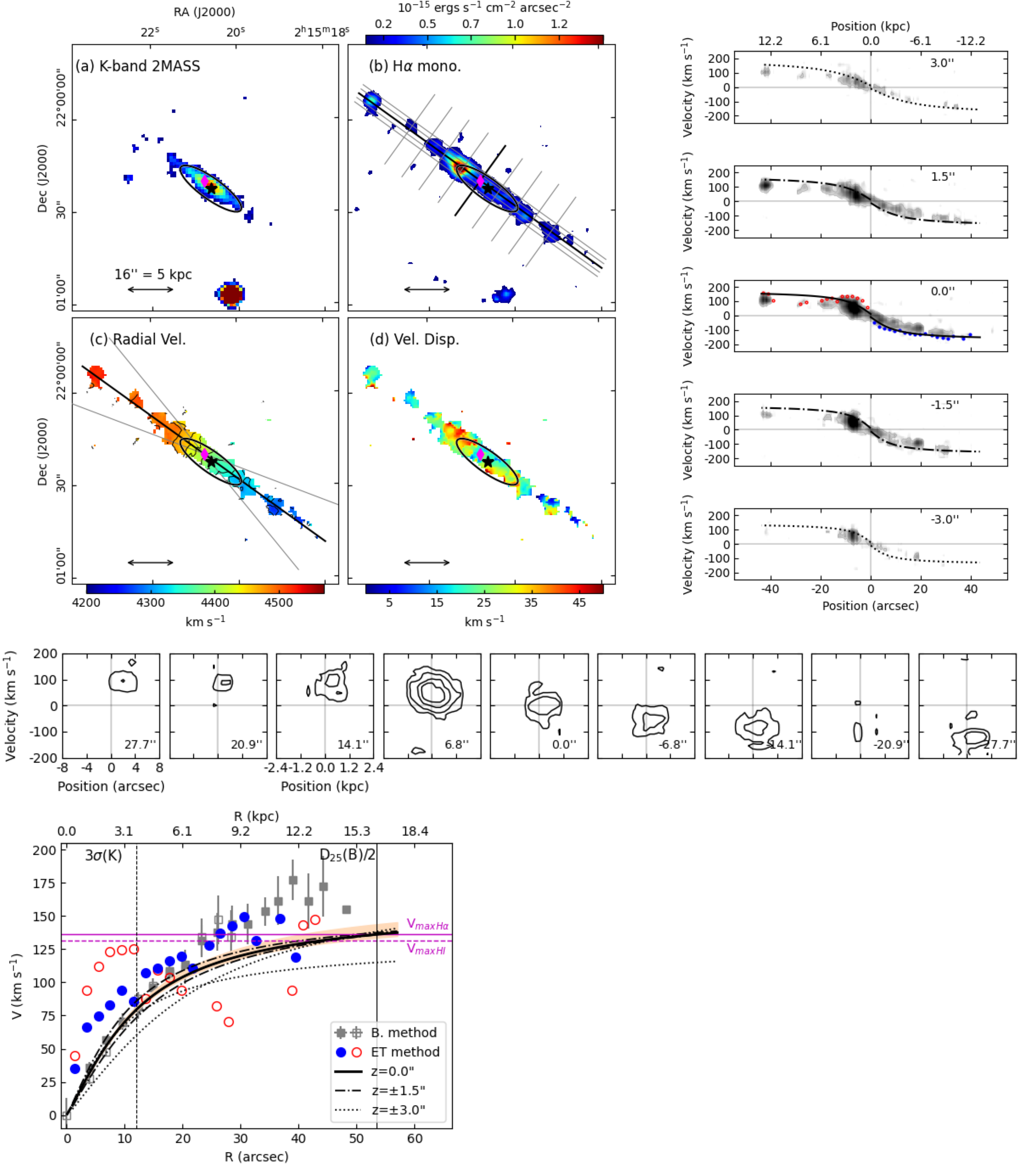}
\caption{CIG~95 (UGC~1733). Same as Fig.~\ref{Fig:e.g.-c201}.
}
\label{Fig:c95}
\end{figure*}

\begin{figure*}
\centering
\includegraphics[width=\hsize]{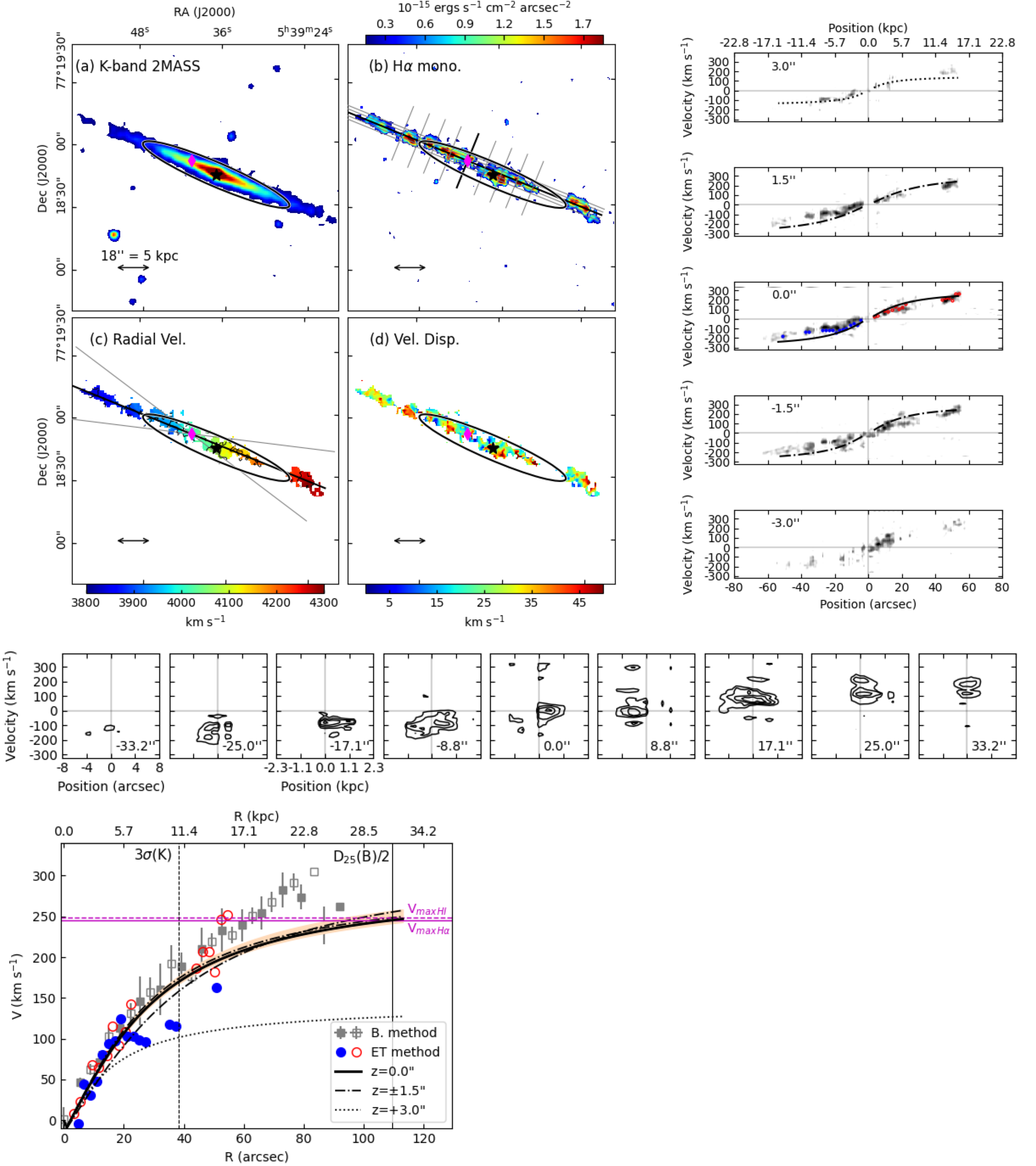}
\caption{CIG~159 (UGC~3326). Same as Fig.~\ref{Fig:e.g.-c201}.
}
\label{Fig:c159}
\end{figure*}

\begin{figure*}
\centering
\includegraphics[width=\hsize]{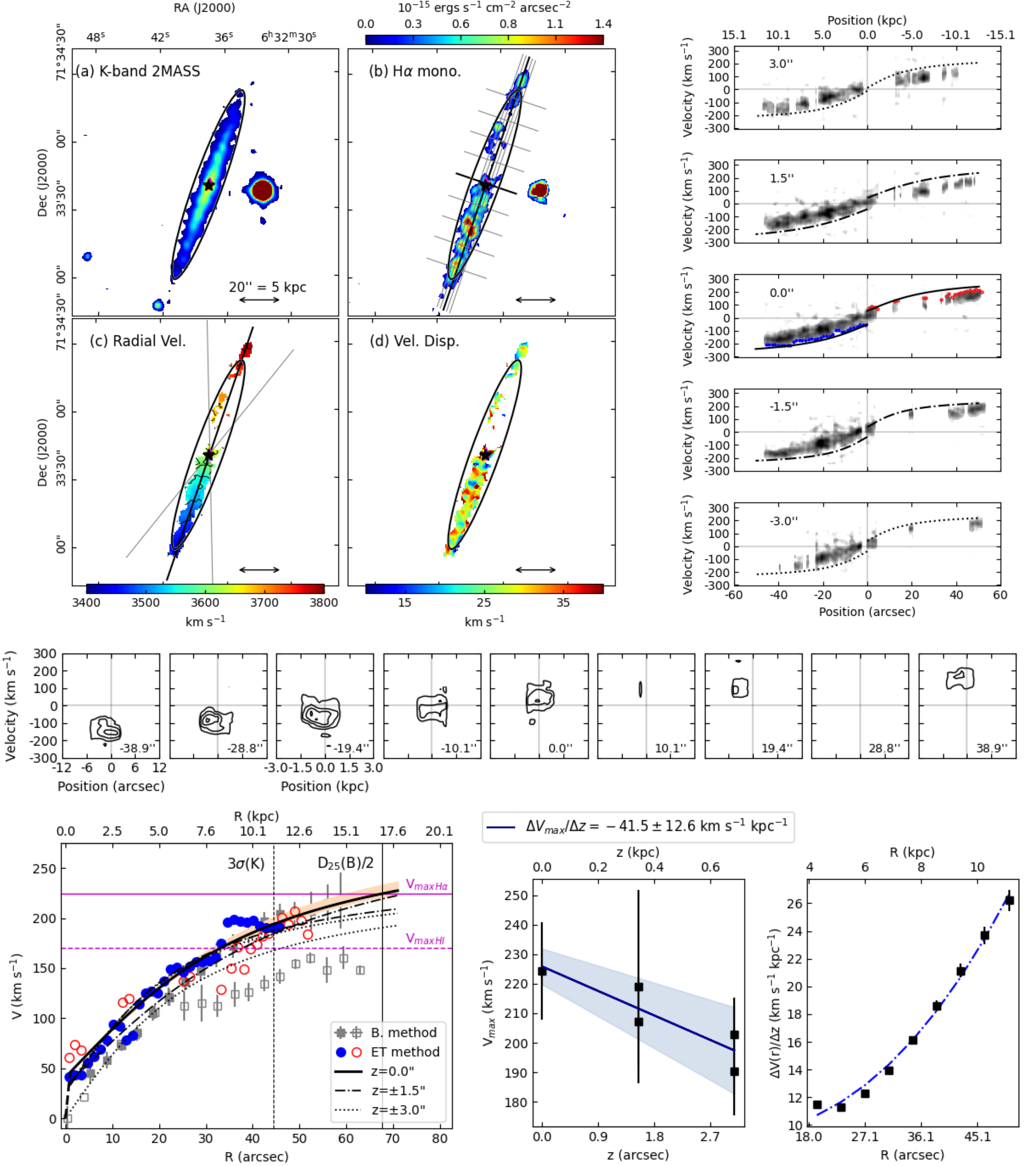}
\caption{CIG~171 (UGC~3474). Same as Fig.~\ref{Fig:e.g.-c201}.
}
\label{Fig:c171}
\end{figure*}

\begin{figure*}
\centering
\includegraphics[width=\hsize]{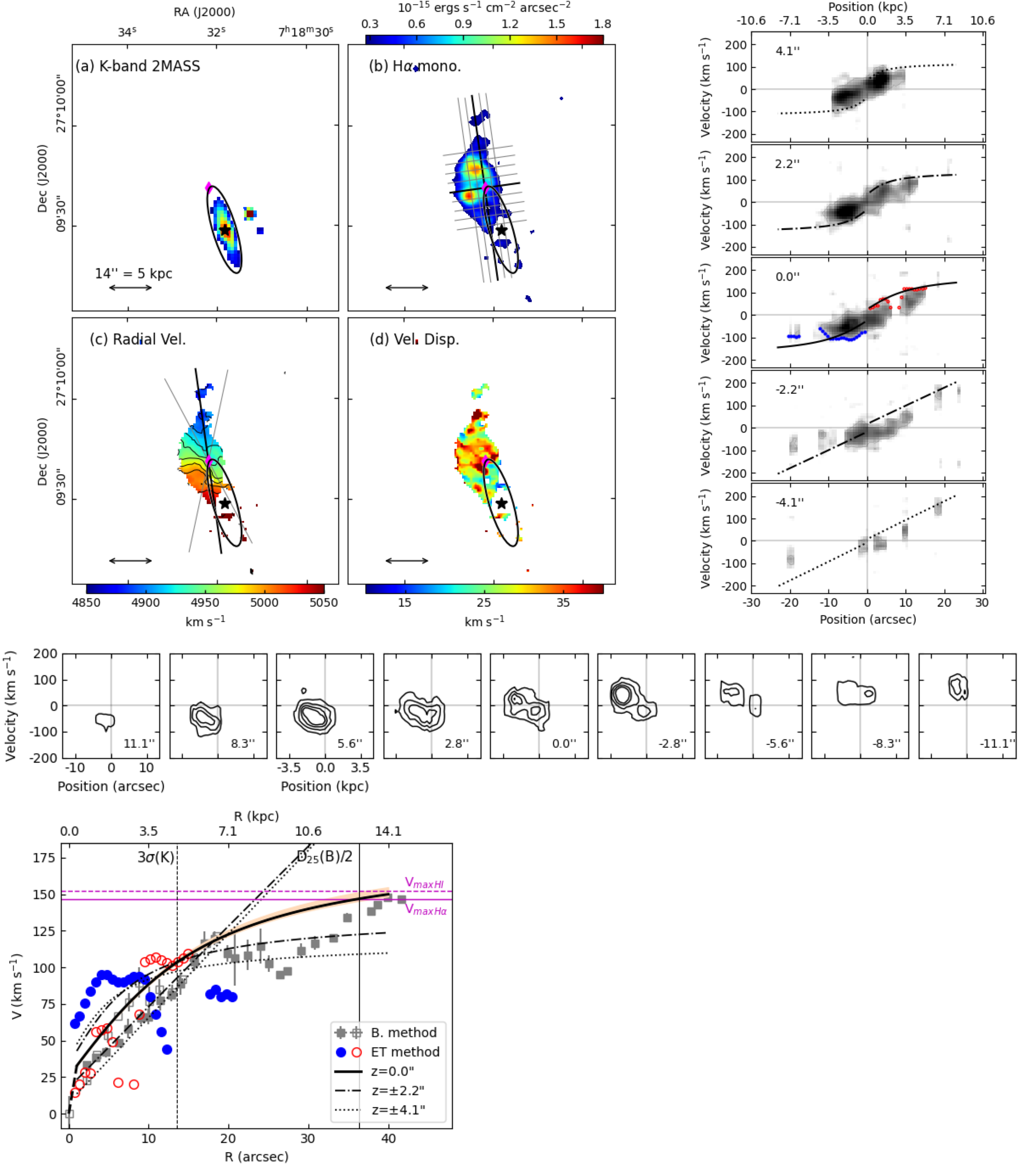}
\caption{CIG~183 (UGC~3791). Same as Fig.~\ref{Fig:e.g.-c201}.
}
\label{Fig:c183}
\end{figure*}

\begin{figure*}
\centering
\includegraphics[width=\hsize]{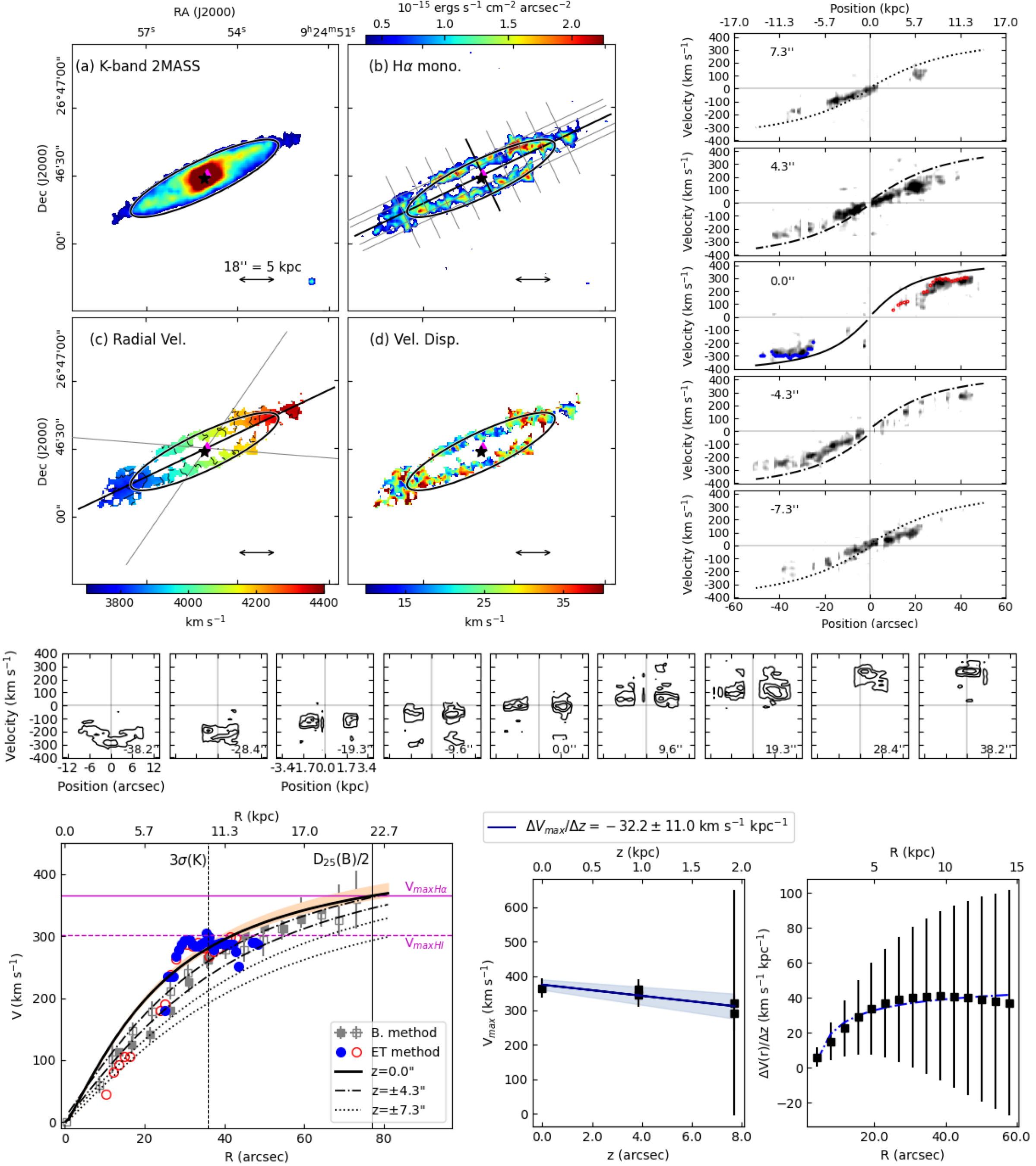}
\caption{CIG~329 (UGC~5010). Same as Fig.~\ref{Fig:e.g.-c201}.
}
\label{Fig:c329}
\end{figure*}

\begin{figure*}
\centering
\includegraphics[width=\hsize]{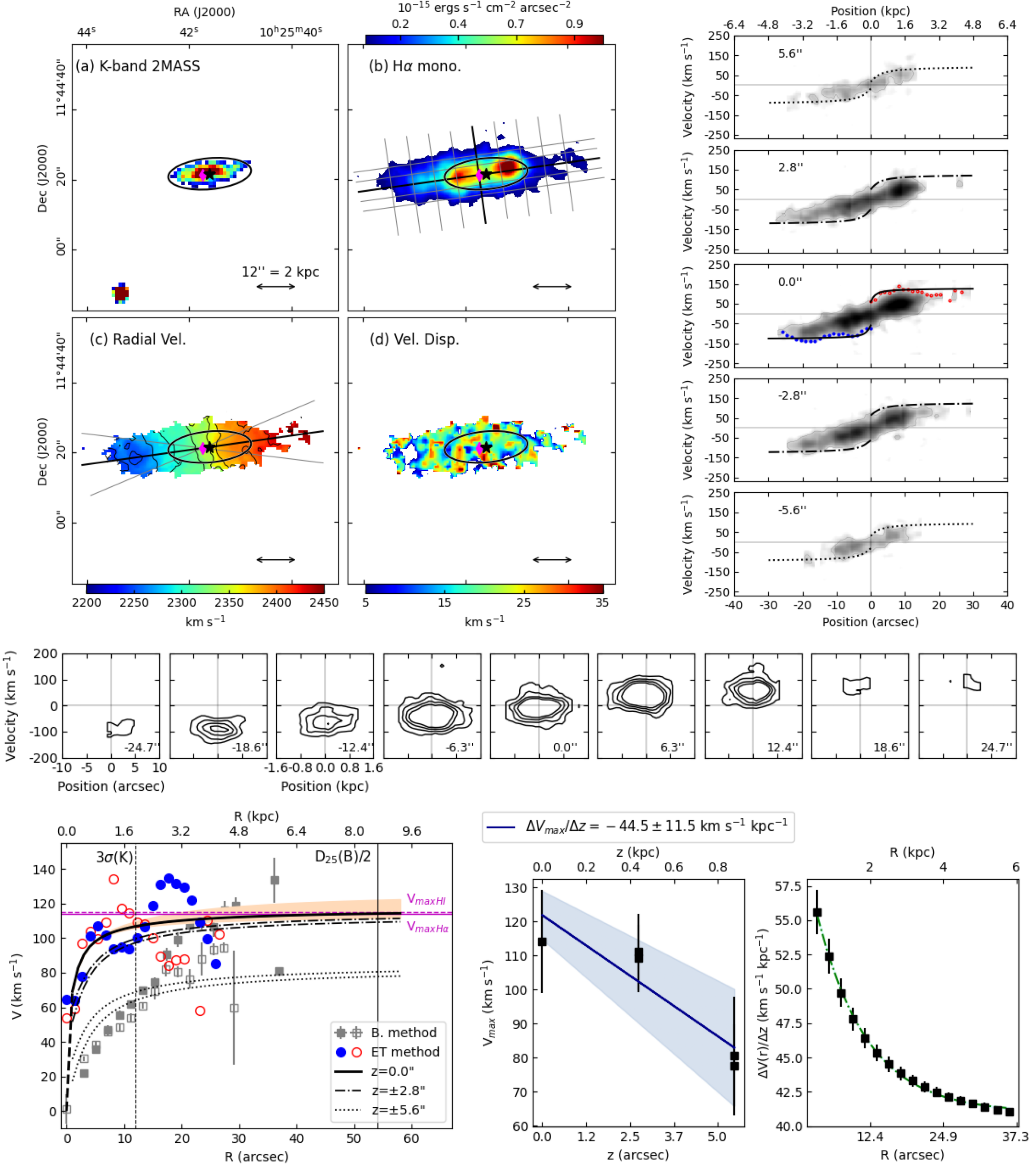}
\caption{CIG~416 (UGC~5642). Same as Fig.~\ref{Fig:e.g.-c201}.
}
\label{Fig:c416}
\end{figure*}

\begin{figure*}
\centering
\includegraphics[width=1\hsize]{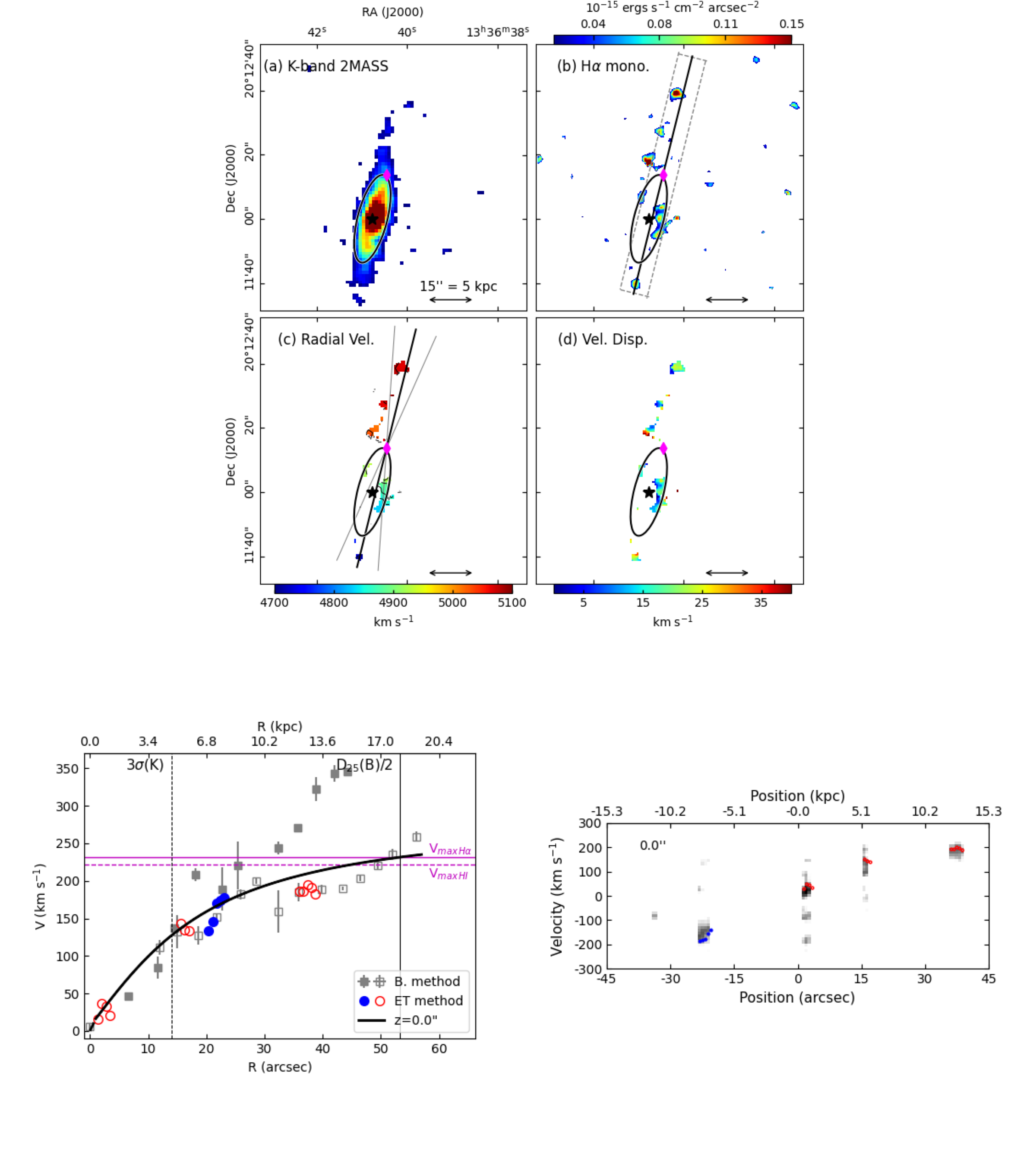}
\caption{CIG~593 (UGC~8598). Similar to Fig.~\ref{Fig:e.g.-c201}. However, because of the limited \Ha-emitting gas in this galaxy CIG\,593, there was computed only one PVD using a wider virtual slit of 5~pixels ($\sim$3.4~arcsec) centred along the kinematic major axis (gray dashed lines). For the same reason, PVDs along the minor axis were also not extracted. 
}
\label{Fig:c593}
\end{figure*}

\begin{figure*}
\centering
\includegraphics[width=\hsize]{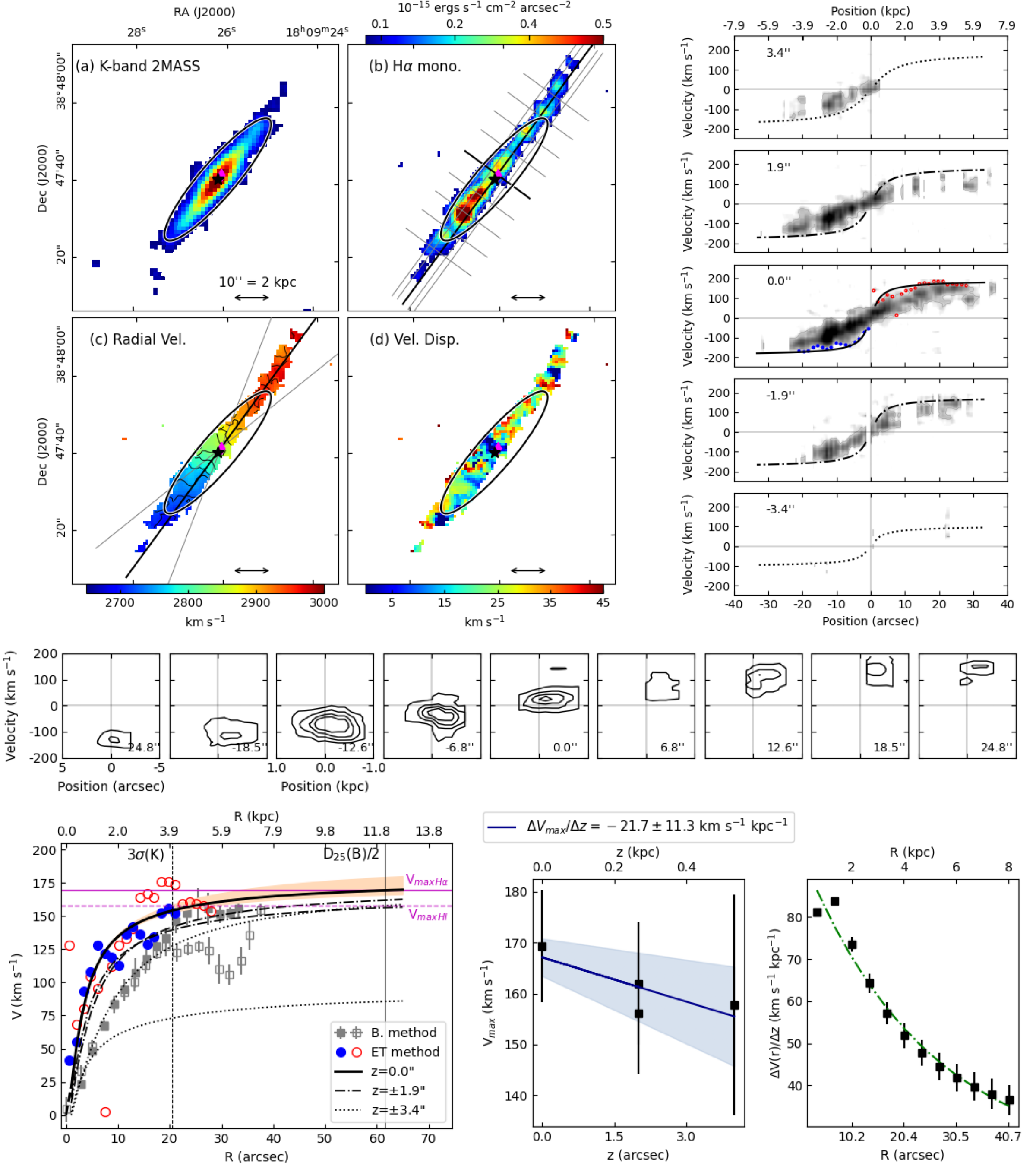}
\caption{CIG~847 (UGC~11132). Same as Fig.~\ref{Fig:e.g.-c201}.
}
\label{Fig:c847}
\end{figure*}

\begin{figure*}
\centering
\includegraphics[width=\hsize]{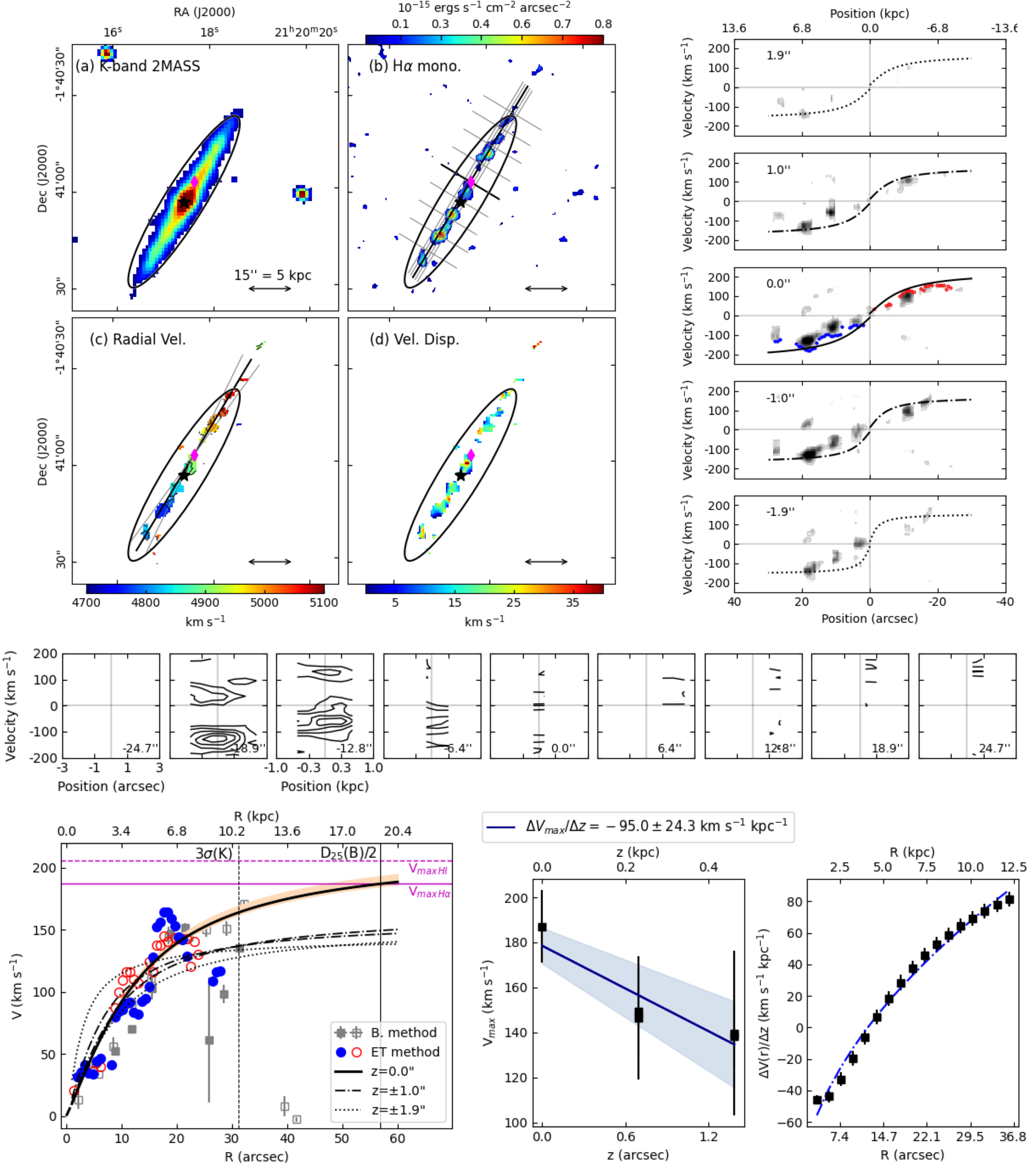}
\caption{CIG~906 (UGC~11723). Same as Fig.~\ref{Fig:e.g.-c201}.
}
\label{Fig:c906}
\end{figure*}

\begin{figure*}
\centering
\includegraphics[width=\hsize]{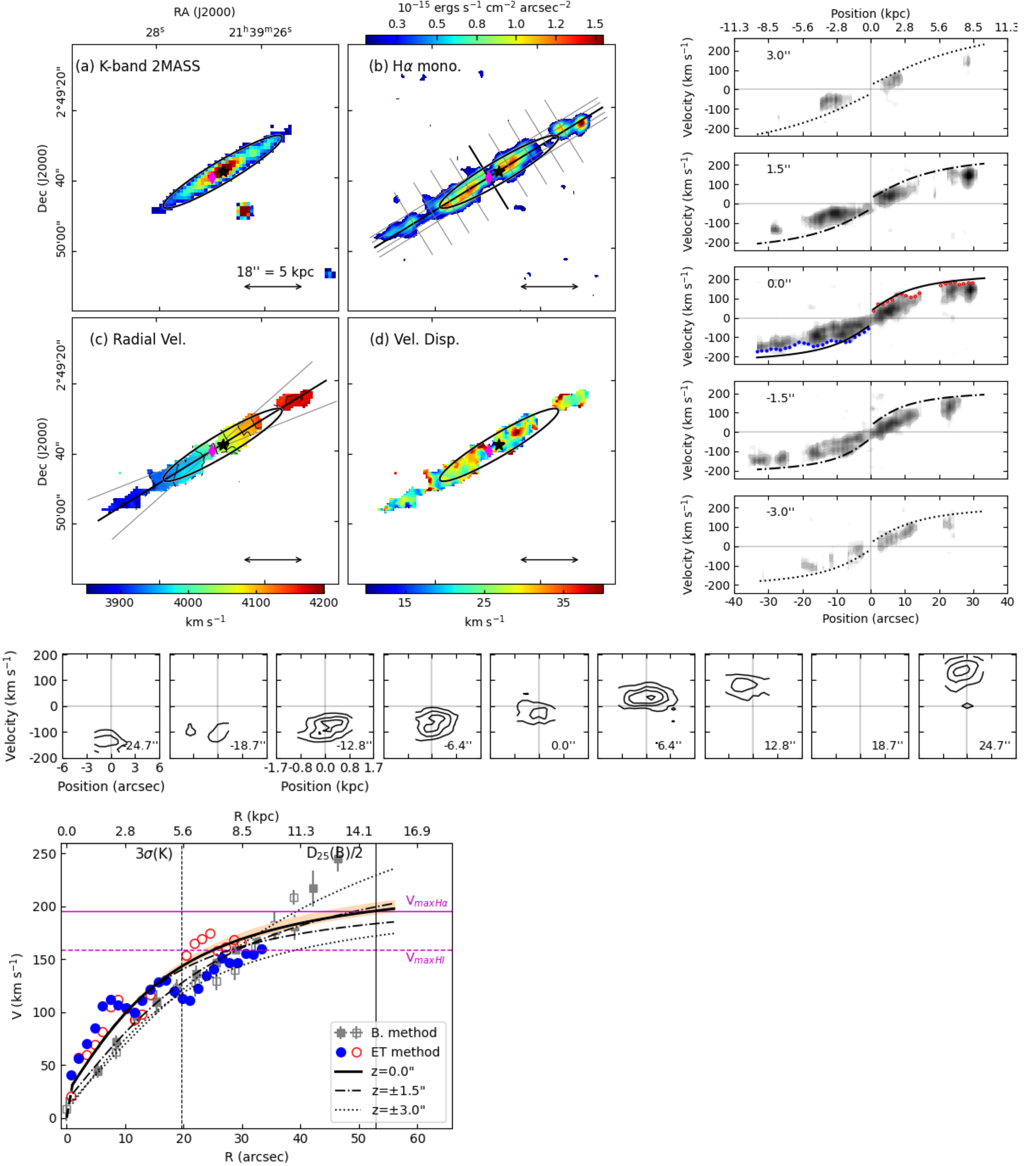}
\caption{CIG~922 (UGC~11785). Same as Fig.~\ref{Fig:e.g.-c201}.
}
\label{Fig:c922}
\end{figure*}

\begin{figure*}
\centering
\includegraphics[width=\hsize]{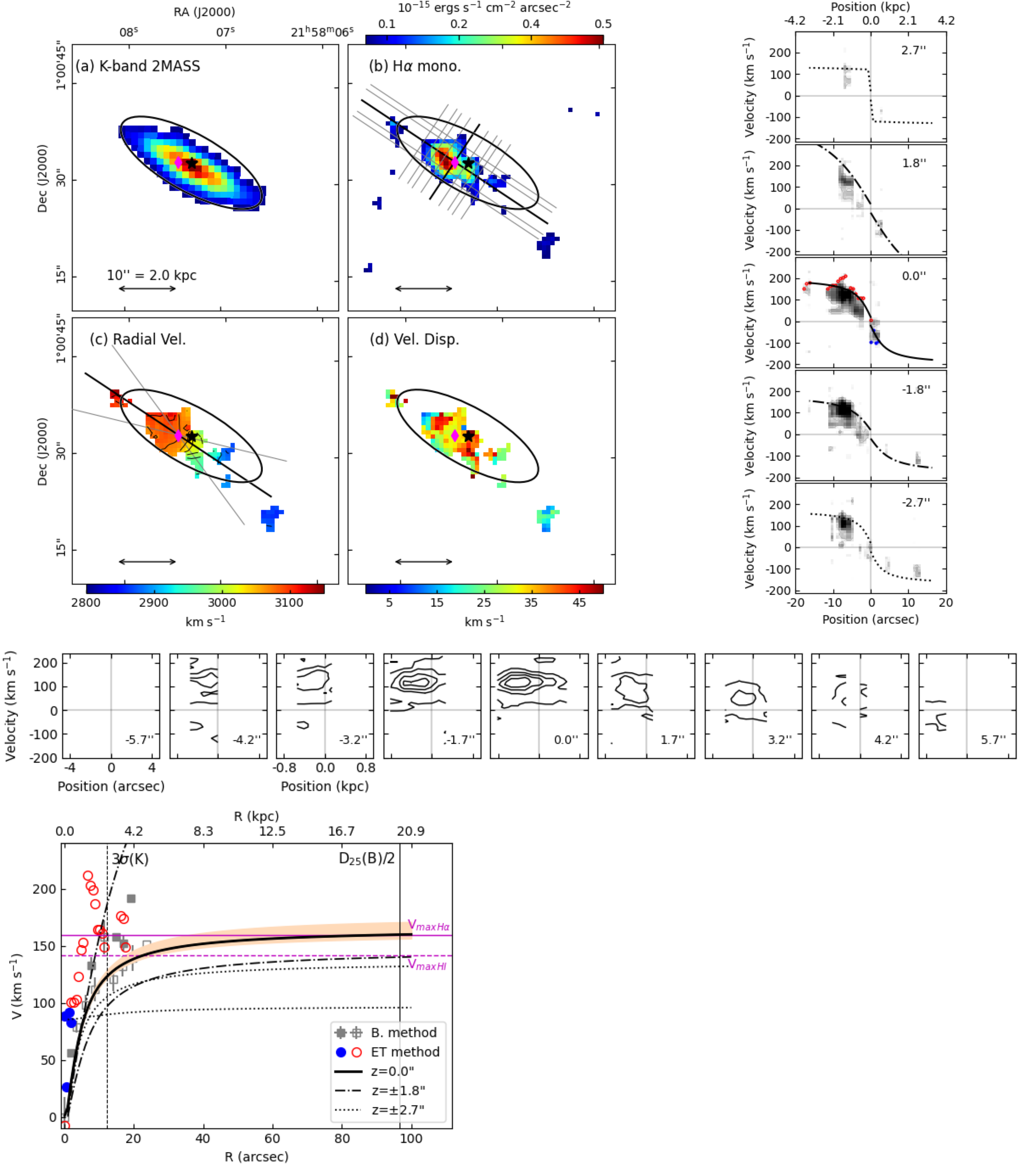}
\caption{CIG~936 (UGC~11859). Same as Fig.~\ref{Fig:e.g.-c201}.
}
\label{Fig:c936}
\end{figure*}

\begin{figure*}
\centering
\includegraphics[width=\hsize]{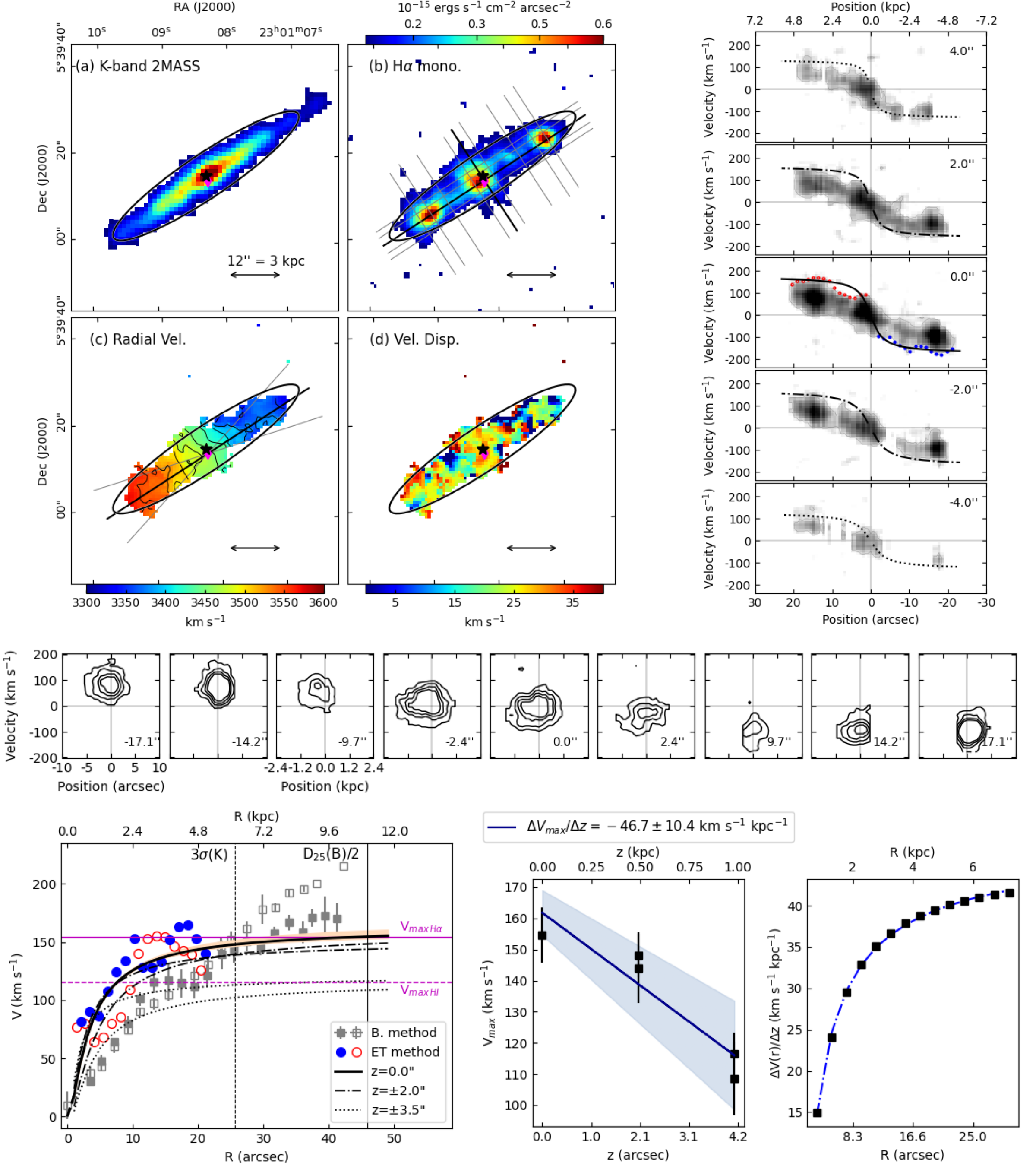}
\caption{CIG~1003 (UGC~12304). Same as Fig.~\ref{Fig:e.g.-c201}.
}
\label{Fig:c1003}
\end{figure*}

In the following section, we present the results of the kinematic analysis for each galaxy in our sample. Figures~\ref{Fig:c71} to~\ref{Fig:c1003} share a common layout, illustrated in Figure~\ref{Fig:e.g.-c201}, and include the maps and diagrams used to extract the rotation curves and estimate the vertical lag. 
The top-left panels show: (a)~the 2MASS~\textit{K$_s$}-band image, (b)~the \Ha monochromatic map, (c)~the radial velocity map, and (d)~the velocity dispersion map. The 3$\sigma$ isophote from the $K$-band image is overlaid and used to define the boundary of the old stellar disc, while the photometric and kinematic centres are marked with a star and a diamond, respectively. 
To quantify the vertical rotation lag, we extracted five PVDs parallel to the major axis at different vertical heights for each galaxy. These are shown in the top-right panels, except in the case of CIG~593, which exhibits poor ionized gas emission and is excluded from this analysis. The middle row of panels shows PVDs along the minor axis to assess the vertical distribution and symmetry of the ionized gas, with the locations of the pseudo-slits indicated on the \Ha map.

In the bottom-left panels, we compare the rotation curves derived from both the IPM and the ETM. The inclination-corrected $\arctan$ fits to the envelope are included as solid lines, with shaded areas marking the variation across the $\eta = [0.2, 0.5]$ range. The inclination-corrected envelopes from the five vertical cuts are also included. Following the method applied in \cite{torres-flores-2011}, we adopted as reference the rotation velocity measured at the optical radius, defined as $D_{25}(B)/2$, to avoid underestimating the true maximum velocity due to rising rotation curves. The envelopes of emission in each PVD were traced using the ET method with a threshold of $\eta = 0.3$ and corrected for inclination. 

The bottom-middle panels display the OLS linear fit of the rotation velocity versus vertical height, used to estimate the average vertical lag, while the bottom-right panels show the lag as a function of radius, which may offer insights into the origin and structure of the extraplanar component.

%

%%%%%%%%%%%%%%%%%%%%%%%%%%%%%%%%%%%%%%%%%%%%%%%%%%

% Don't change these lines
\bsp	% typesetting comment
\label{lastpage}
\end{document}